\documentclass[11pt]{article}
\usepackage{latexsym}
\usepackage{amsmath,amsfonts,amsbsy,amssymb}
\usepackage[dvips]{graphicx,epsfig}

 \csname
@addtoreset\endcsname{equation}{section}

\def\mso{\mathfrak{so}}

\def\msl{\mathfrak{sl}}
\def\msp{\mathfrak{sp}}

\def\mhs{\mathfrak{hs}}

\def\mg{\mathfrak{g}}
\def\mh{\mathfrak{h}}
\def\mD{\mathfrak{D}}

\def\Real{{\mathbb R}}
\def\Comp{{\mathbb C}}
\def\integ{{\mathbb Z}}

\def\bec{\begin{center}}
\def\ec{\end{center}}
\def\a{\alpha} \def\ad{\dot{\a}} 

\def\b{\beta}  \def\bd{\dot{\b}} 
\def\c{\gamma} \def\cd{\dot{\c}}
\def\C{\Gamma}
\def\d{\delta} \def\dd{\dot{\d}}
\def\D{\Delta}
\def\e{\epsilon} 
\def\ve{\varepsilon}

\def\k{\kappa}
\def\vark{\varkappa}
\def\l{\lambda}
\def\L{\Lambda}
\def\m{\mu}
\def\n{\nu}
\def\r{\rho}
\def\s{\sigma}
\def\S{\Sigma}
\def\t{\tau}
\def\th{\theta}

\def\y{\eta}
\def\z{\zeta}
\def\O{\Omega}
\def\o{\omega}

\def\sb{{\bar\s}}

\def\yb{{\bar y}}
\def\zb{{\bar z}}

\def\nn{\nonumber}
\newcommand{\eq}[1]{(\ref{#1})}

\def\be{\begin{equation}}
\def\ee{\end{equation}}
\def\bea{\begin{eqnarray}}
\def\eea{\end{eqnarray}}
\def\ba{\begin{array}}
\def\ea{\end{array}}

\def\ft#1#2{{\textstyle{{\scriptstyle #1}
\over {\scriptstyle #2}}}}

\def\ket#1{|#1\rangle}
\def\bra#1{\langle#1|}
\def\scs#1{\section{\bf{\sc \large #1}}}
\def\scss#1{\subsection{\bf{\sc  #1}}}
\def\scsss#1{\subsubsection{\bf{\sc \small #1}}}

\def\ad{\dot\alpha}
\def\bd{\dot\beta}
\def\sb{\bar\sigma}

\begin{document}



\vspace{8pt}

\begin{center}


{\Large\sc Families of Exact Solutions to Vasiliev's 4D Equations \\[10pt] with Spherical, Cylindrical and Biaxial Symmetry}


\vspace{55pt}

C a r l o~~~~I a z e o l l a \\[15pt] 
{\it Dipartimento di Fisica, Universit\`a di Bologna \\
 and \\
 INFN, Sezione di Bologna\\
via Irnerio 46, I-40126 Bologna, Italy}\\[25pt]

P e r ~~~~ S u n d e l l\footnote{F.R.S.-FNRS Researcher with an Ulysse Incentive Grant for Mobility in Scientific Research.}\\[15pt]
{\it Service de M\'ecanique et Gravitation\\
Universit\'e de Mons\\
Place du Parc, 20, 7000 Mons, Belgium }\\[25pt]


\vspace{10pt} {\sc\large Abstract}\end{center}

We provide Vasiliev's four-dimensional bosonic higher-spin gravities with six families of exact solutions admitting two commuting Killing vectors.
Each family contains a subset of generalized Petrov Type-D solutions in which one of the two $\mso(2)$ symmetries enhances to either $\mso(3)$ or $\mso(2,1)$. 
In particular, the spherically symmetric solutions are static and we expect one of them to be gauge-equivalent to the extremal Didenko-Vasiliev solution \cite{Didenko:2009td}.
The solutions activate all spins and can be characterized either via generalized electric and magnetic charges defined asymptotically in weak-field regions or via the values of fully higher-spin gauge-invariant observables given by on-shell closed zero-forms.
The solutions are obtained by combining the gauge-function method with separation of variables in twistor space via expansion of the Weyl zero-form in Di-Rac supersingleton projectors times deformation parameters in a fashion that is suggestive of a generalized electromagnetic duality.

 \setcounter{page}{1}


\pagebreak

\tableofcontents

\newpage


\scs{Introduction}\label{sec:in}


\scss{Motivations}

The importance of higher-spin gravities \cite{vasiliev,Vasiliev:2003ev,Vasiliev:1999ba,Bekaert:2005vh} --- both in themselves, as some of the few known consistent interacting gauge field theories, and as systems of intermediate complexity between ordinary gauge theories and string field theories --- is currently becoming more widely appreciated. 
All known models in four dimensions and above consist of an infinite tower of gauge fields, essentially tied to dressings of the minimal-bosonic models consisting perturbatively of symmetric-tensor gauge fields of even ranks including a physical scalar \cite{Vasiliev:2003ev,Sezgin:2003pt}. 
Such infinite towers resemble the leading Regge trajectory of string theories collapsed to critical masses given in units of a finite cosmological constant.
Indeed, just like string field theories, higher-spin gravities admit a formulation, found by Vasiliev \cite{vasiliev}, in terms of master fields depending on commuting spacetime coordinates and internal oscillators, and interacting via star-product algebras. 
However, their algebraic structures are simple enough that equations of motion based on a gauge principle --- including generalized spacetime symmetries --- can be spelled out explicitly in a background-independent fashion, as for ordinary (super)gravities.
Corresponding off-shell formulations have been proposed recently in \cite{Boulanger:2011dd} and related issues concerning globally-defined formulations on-shell and off-shell have been studied in \cite{Sezgin:2011hq}.
Moreover, the properties of discretized strings in Anti-de Sitter spacetime \cite{Engquist:2005yt} (motivated by the semi-classical results of \cite{Gubser:2002tv,Kruczenski:2004wg}) suggest that higher-spin gravities are sub-sectors of particular tensionless limits of closed string field theories. 
Higher-spin gravities are thus tractable models for studying large-curvature effects in stringy completions of ordinary gravities.

In particular, this opens a new window on holography in regimes where the boundary theories are weakly coupled and the bulk theories contain higher-spin gravities (and that are hence complementary to the more widely studied dual pairs involving strongly-coupled boundary theories and string theories with low-energy effective gravity descriptions on the bulk side).
In this regime, the massless higher-spin fields correspond to the bilinears in free fields on the boundary, which is a manifestation of the Flato--Fronsdal theorem \cite{Flato:1978qz}.
An important special case is the holographic correspondence between three-dimensional O(N)-vector models and four-dimensional higher-spin gravities \cite{Sezgin:2002rt,Klebanov:2002ja}, for which supersymmetries are not essential and the absence of boundary gauge symmetries implies simplified 1/N expansions.
At the level of three-point functions, this correspondence was verified in the case of scalar self-couplings \cite{Sezgin:2003pt} and more recently for general couplings in \cite{Giombi:2009wh,Giombi:2010vg}. 
Moreover, possessing the full bulk field equations allows for more direct and 
detailed studies of holography using, for example, exact renormalization-group equations \cite{Douglas:2010rc} or bilocal fields \cite{Koch:2010cy}.

Vasiliev's equations \cite{vasiliev,Vasiliev:2003ev} (see \cite{Vasiliev:1999ba,Bekaert:2005vh,Iazeolla:2008bp} for reviews) provide a fully nonlinear framework for higher-spin gravities. 
They encode the classical dynamics of a highly complicated system ---  in which infinitely many fields of all spins are coupled through higher-derivative interaction vertices --- into a combination of zero-curvature constraints, for suitable master fields with simple higher-spin gauge transformations, and algebraic constraints, that actually describe a deformed $\star$-product algebra. 
This elegant description is achieved within the unfolded formulation of dynamics \cite{Vasiliev:1988xc,Vasiliev:1988sa,Vasiliev:1999ba,Bekaert:2005vh}, which is a generalization based on differential algebras of the covariant Hamiltonian formulation of dynamics.
The resulting unfolded systems consist of finitely many fundamental differential forms, which are the aforementioned master fields, living on extensions of spacetime referred to as correspondence spaces. 
Locally, these are products $T^\ast{\cal X}\times {\cal T}$, where ${\cal X}$ contains spacetime, and ${\cal T}$ is a non-commutative twistor space in the case of four-dimensional spacetime.
The unexpected simplicity of the equations resides in that all spacetime component fields required for the unfolded formulation are packed away into the master fields in such a way that contractions of the coordinates of ${\cal T}$, controlled by the deformed $\star$-product algebra, reconstruct generally covariant albeit non-local interactions in spacetime\footnote{Barring the issue of topologically nontrivial configurations on ${\cal T}$, which we shall address later in Section \ref{Sec:defosc} and Appendix \ref{App:D}, projecting out ${\cal T}$ yields a full set of infinitely many component fields on spacetime that can be expressed on-shell in terms of a set of dynamical fields and their derivatives (see Appendix \ref{weakfields}). 
In particular, there is always a dynamical metric tensor that one can treat non-perturbatively while treating all other dynamical fields as weak. 
More precisely, the generally-covariant dynamical equations take the form of standard kinetic terms with critical mass terms \cite{Sezgin:2002ru}, equated to source terms that admit a double expansion in weak fields and derivatives.
At every fixed order in weak fields, the derivative expansions do \emph{not} terminate, at least not in the naively defined basis of dynamical fields (see for example \cite{Kristiansson:2003xx,Boulanger:2008tg,Bekaert:2010hw} and references therein).
Moreover, the derivatives are given in units of the cosmological constant and, as a consequence, the higher-derivative interaction terms are huge on-shell for unitarizable fluctuation fields belonging to lowest-weight spaces. Thus, an important open problem is to specify admissible boundary conditions, possibly related to a weakened notion of perturbative locality \cite{Bekaert:2010hw}.}.
In this sense, the gauge principle based on higher-spin symmetries leads to a departure from the more familiar framework of Einstein gravity with perturbative stringy corrections into a radically different realm involving interesting new phenomena, already at the classical level, for which we would like to gain more intuition.  

Given the aforementioned non-localities, exact solutions of Vasiliev's equations are of great interest as they can provide important insights on both the physics and the geometry underlying such an unconventional physical regime.  
To find examples, one can conveniently solve the zero-curvature constraints locally on ${\cal X}$ using gauge functions; the local degrees of freedom are thus encoded into suitable fibre elements that solve the remaining deformed-oscillator problem on ${\cal T}$ \cite{Vasiliev:1990bu,Bolotin:1999fa}. 
In this precise sense, Vasiliev's unfolded equations naturally map the original dynamical problem in spacetime to an arguably more tractable problem in the fibre space -- a property that has not only proved useful for finding exact solutions \cite{Didenko:2006zd,Sezgin:2005pv,Iazeolla:2007wt} but also greatly simplified the computations of three-point functions in \cite{Giombi:2010vg}. 

\scss{Known exact solutions to higher-spin gravity}\label{Sec1.2}

Besides the anti-de Sitter vacuum solution, a number of non-trivial solutions have been found in recent years. 
In $2+1$ dimensions, where higher-spin fields do not propagate, a class of vacuum solutions have been constructed in models containing non-trivial matter sectors \cite{Prokushkin:1998bq} by making use of a specific Ansatz for the deformed oscillators. More recently, the BTZ black-hole has been embedded into Vasiliev's three-dimensional higher-spin gravity \cite{Didenko:2006zd} and black-hole-like solutions to a certain Chern-Simons higher-spin gravity have appeared \cite{Gutperle:2011kf,Ammon:2011nk}.
In $3+1$ dimensions, the first non-trivial example of an exact solution was found in \cite{Sezgin:2005pv} by using the gauge-function method and adapting the aforementioned Ansatz for deformed oscillators to the four-dimensional case.
In various four-dimensional spacetime signatures, further classes of exact solutions were presented in \cite{Iazeolla:2007wt}, including algebraically special generalizations of type-D\footnote{This terminology refers to Petrov's invariant classification of the Weyl tensors \cite{Petrov:2000bs,PenroseRindler}, based on the algebraic properties of the latter at any spacetime point; for further details, related notation and conventions, see Appendix \ref{App:conv}. } gravitational instantons, with all higher-spin fields turned on, and some new vacuum solutions describing topologically non-trivial field configurations on ${\cal Z}$.
Finally, in \cite{Didenko:2009td}, Didenko and Vasiliev have given a solution that in many respects corresponds to an extremal generalization of the Schwarzschild solution to AdS gravity.

In the latter solution, characterized by a single deformation parameter $M$, spherically symmetric modes for each spin are switched on in a coherent fashion. 
The spin-$s$ Weyl tensors depend on the radial coordinate $r$ of the spherically symmetric coordinate system of $AdS_4$ as $C^{(s)}\propto r^{-s-1}$, are all of (generalized) Petrov-type D \cite{Petrov:2000bs}, and are built in terms of covariant derivatives of a time-like $AdS_4$ Killing vector. 
Asymptotically, at spatial infinity where $r\rightarrow \infty$, the different spins decouple and one can meaningfully identify the spin-$2$ Weyl tensor with that of the AdS-Schwarzschild black hole with mass proportional to $M$. 
However, near $r=0$ the Weyl tensors are large and the strong coupling between infinitely many fields of all spins may give rise to important deviations from the standard results in gravity. 
This raises the very interesting question whether the non-localities associated to the unbroken higher-spin symmetry suppress the short-distance singularities. 

A remarkable feature of the Didenko-Vasiliev solution, shared with $4D$ gravity black holes, is that it linearizes the full equations of motion. 
Technically, this property is encoded into the fact that $4D$ black hole metrics as well as the higher-spin gauge fields of the solution can be written in Kerr-Schild form \cite{Didenko:2009td} in a certain gauge. 
In the Didenko-Vasiliev solution, this is achieved by factorizing a certain operator appearing in the Vasiliev equations, known as the inner Kleinian operator, into a product of two delta functions on ${\cal T}$ \cite{berezin,Didenko:2009td}, and by expressing the fluctuation part of the master fields via a spacetime-dependent vacuum projector related to the above-mentioned $AdS_4$ Killing vector, providing an Ansatz that simultaneously solves the linearized equations and trivializes all nonlinear corrections.

\scss{Summary of our main results}

In this paper, we find six families of exact solutions to four-dimensional bosonic higher-spin gravities by combining the gauge-function method on ${\cal X}$ with the aforementioned factorization property of the inner Kleinians on ${\cal T}$. 
The latter facilitates the separation of variables in the twistor-space ${\cal T}\stackrel{\rm loc}{\cong}{\cal Y}\times {\cal Z}\setminus {\cal D}$, where ${\cal Y}\cong \Comp^2$ is a non-commutative fibre space on which the master fields admit expansions in terms of symbols belonging to associative algebras with well-defined traces, ${\cal Z}\cong \Comp^2$ is a non-commutative base space, and ${\cal D}$ stands for some submanifold of ${\cal Y}\times{\cal Z}$ on which the master fields may develop suitable singularities --- as it happens on some of the solutions, as we shall see.
The corresponding factorized Ansatz amounts to expanding the master fields in terms of projector algebras on ${\cal Y}$ times coefficient matrices on ${\cal Z}$ that solve the deformed oscillator problem (modulo subtleties having to do with potential non-trivial ${\cal D}$ developing over certain points in spacetime).

All our solutions possess the Kerr-Schild property of the Didenko-Vasiliev solution, \emph{i.e.} the full Weyl tensors coincide with the linearized ones. 
This implies that the full Weyl zero-forms belong to linear spaces, while gauge fields and internal connections on ${\cal Z}$ contain interference terms (that do not cancel out in the gauge we use, though they do in other gauge choices \cite{Didenko:2009td}). 

The projectors are functions of pairs of generators in the complexified Cartan subalgebra of $\mso(3,2)$, that can be any inequivalent combination of rotations $J$, boosts $iB$ or spatial and time translations $iP$ and $E$, respectively, namely $(E,J)$, $(J,iB)$ and $(iB,iP)$. 
This yields three classes of $\mso(2)\times\mso(2)$-invariant\footnote{
More precisely, the solutions are left invariant by the intersection of the enveloping algebra of $\mso(2)\times\mso(2)$ with the underlying higher-spin symmetry algebra. } solutions, that we shall refer to as being biaxially symmetric (or simply axisymmetric), consisting of master fields that are diagonalized over bases of eigenstates $\ket{\bf n}$ of the above pairs of generators.
These solution spaces are coordinatized by massive deformation parameters $\nu_{\bf n}$ representing the eigenvalues of the Weyl zero-form master field in the aforementioned bases.

The two Cartan generators can be identified as the linear combinations of the number operators acting in two Fock spaces, say ${\cal F}_i^+$ ($i=1,2$), and the corresponding anti-Fock spaces ${\cal F}_i^-$. 
The total state space $({\cal F}_1^+\oplus {\cal F}_1^-)\otimes ({\cal F}_2^+\oplus {\cal F}_2^-)$ decomposes under $\mso(3,2)$ into four sub-sectors $(\pm,\pm)\equiv {\cal F}^\pm_1\otimes {\cal F}^\pm_2$.
In each sub-sector, one of the two Cartan generators is either positive or negative definite,
and is to be referred to as the principal Cartan generator.
The full equations admit discrete global symmetries (the $\tau$-map) that relate $(+,+)$ to $(-,-)$ and $(+,-)$ to $(-,+)$. Hence $(+,+)\oplus (-,-)$ and $(+,-)\oplus(-,+)$ form two independent families of solutions, which can be labelled by their principal Cartan generators.
In other words, denoting each family by ${\cal M}_{K_{(\pm)}}(\mh_\Real)$ with distinct symmetry sub-algebras $\mh_\Real=\mso(2)_{(+)}\oplus\mso(2)_{(-)}\subset \mso(3,2)\cong \msp(4;\Real)$ and principal Cartan generator $K_{(\pm)}\in\msp(4;\Comp)$ (formed as either the sum $(+)$ or difference $(-)$ of the aforementioned number operators), the six families of solutions can be organized into the following three pairs:
\be {\cal M}_{E}(E,J)\ ,\quad {\cal M}_{J}(E,J)\ ;\qquad {\cal M}_{J}(J,B)\ ,\quad {\cal M}_{iB}(J,B)\ ;\qquad {\cal M}_{iB}(B,P)\ ,\quad {\cal M}_{iP}(B,P)\ .\ee
In case the principal Cartan generator is imaginary, the reality condition implies that the corresponding master fields must contain both Fock-space and anti-Fock-space projectors.
The required projector algebra can be realized by presenting the dependence on the Cartan-subalgebra generators via inverse Laplace-like transforms introducing auxiliary closed-contour integrals, which we refer to as regular presentations.
The co-existence of Fock-space and anti-Fock-space projectors is also required for the minimal-model projection of all our solutions. 

In this paper we shall mainly focus on the two solution spaces ${\cal M}_{E}(E,J)$ and ${\cal M}_{J}(E,J)$. Drawing on the results obtained in \cite{Iazeolla:2008ix}, the projectors used in the $(+,+)$ and $(-,-)$ sub-sectors of ${\cal M}_{E}(E,J)$ are related to supersingleton  and anti-supersingletons states, respectively, while the $(+,-)$ and $(-,+)$ sub-sectors of ${\cal M}_{J}(E,J)$ are related to analogous ultra-short albeit non-unitary $\mso(3,2)$-irreps.
Specific combinations of such axisymmetric solutions give rise to solutions with enhanced spherical $\mso(3)\oplus\mso(2)$-symmetry or cylindrical $\mso(2,1)\oplus\mso(2)$-symmetry (and their higher-spin extensions), arising from enhancing either $\mso(2)_J$ to $\mso(3)$ or $\mso(2)_E$ to $\mso(2,1)$, respectively (see Table \ref{Table1}). 
In the general-relativistic terminology, which is valid in the asymptotic weak-curvature regions, this amounts to that the non-enhanced Killing vector becomes hypersurface-orthogonal; if the latter is time-like, the corresponding stationary solution is, in fact, static. 
This is the case for all the solutions belonging to the rotationally-invariant family, one member of which is the Didenko-Vasiliev solution that we find here being based on the singleton ground-state projector ${\cal P}_1(E):=4e^{-4E}$. 

All the solutions found in this paper are algebraically special. In particular, all the Weyl tensors of the symmetry-enhanced solutions are always of generalized Petrov-type D. This means that the Weyl tensors have two principal spinors, \emph{i.e.}, that at every spacetime point there exists a (normalized) tangent-space twistor basis $(u^+_\a(x),u^-_\a(x))$ (a \emph{spin-frame}, in the terminology of \cite{PenroseRindler}) on which the self-dual part of the spin-$s$ Weyl tensor takes the form\footnote{We use the shorthand notations $T_{a(n)}$ to denote a tensor with $n$ totally symmetrized indices $T_{a_1\ldots a_n}=T_{(a_1\ldots a_n)}$. Repeated non-contracted indices are also to be understood as totally symmetrized, $S_{aa} T_{aa}:=S_{(a_1a_2}T_{a_3 a_4)}$.} $C_{\a(2s)}\sim  f(x) (u^+_\a u^-_\a)^s $, which we shall also refer to as type-$\{s,s\}$ (analogously for the anti-selfdual part with the complex conjugate twistors $(\bar u^+_{\ad}(x),\bar u^-_{\ad}(x))$). Furthermore, the principal spinors of the Weyl tensors of the above solutions are those of the Killing two-form $\vark_{\m\n}=\nabla_\m v_\n$ where $v_\m$ is a specific $AdS_4$ Killing vector, \emph{i.e.}, the Weyl tensors can be rewritten as $C_{\a(2s)}\sim  F(x) (\vark_{\a\a})^s $. In four-dimensional Einsten gravity, if the corresponding Killing vector is asymptotically time-like, this is a local hallmark of black-hole solutions \cite{Mars,Didenko:2008va,Didenko:2009tc}. The Weyl tensors of the axisymmetric solutions are less special, as we shall see:  they are algebraically general for spin $s\leq k$ and type-$\{s-k,s-k,\underbrace{1,\dots,1}_{2k}\}$ for $s > k$ (the integer $k$ depending on the projector the solution is built on), which we shall refer to as \emph{almost type-D}.

For our spherically-symmetric solutions, which are based on energy-dependent supersingleton state projectors ${\cal P}_n(E)$, the aforementioned specific $AdS_4$ Killing vector is asymptotically time-like and given by the time-translation $\partial/\partial t$. These solutions contain an infinite tower of Weyl tensors, one for every spin, of the form
\bea C^{(n)}_{\a(2s)}\ \sim \ \frac{i^{n-1}\m_n}{r^{s+1}}\,(u^+_\a u^-_\a)^s \ ,\eea
where $\m_n=i^{-n}\nu_n$ are real deformation parameters, as explained in Sections \ref{Sec:internal} and \ref{Sec:Weyl}. As first observed in \cite{Didenko:2009td} for the case $n=1$, the $s=2$ sector coincides with the AdS-Schwarzschild Weyl tensor. Interestingly, the curvatures are real for $n$ odd and imaginary for $n$ even, which suggests that solutions built on projectors  ${\cal P}_n(E)$ over combinations of states belonging to the scalar ($n$ odd) or spinor ($n$ even) singleton representation, related to the Type A or Type B models\footnote{As explained in \cite{Iazeolla:2008ix}, the projectors ${\cal P}_n(E)$ belong to $\mD(1/2,0)\otimes\mD^\ast(1/2,0)$ for $n$ odd and to $\mD(1,1/2)\otimes\mD^\ast(1,1/2)$ for $n$ even, where $\mD(1/2,0)$ is the scalar-singleton representation and $\mD(1,1/2)$ the spinor singleton representation. From the point of view of a two-sided, twisted-adjoint action they are an enveloping-algebra realization of states belonging to the tensor product of two scalar and two spinor singletons, respectively, which are in their turn related to the spectrum of the Type A and Type B minimal bosonic models defined in \cite{Sezgin:2003pt}.} \cite{Sezgin:2003pt},  are connected via a generalized electric/magnetic duality. 

Starting with the asymptotically space-like Killing vector $\partial/\partial\varphi$, on the other hand, leads to cylindrically-symmetric solutions, based on the rotation-dependent projectors ${\cal P}_n(J)$, the Weyl curvatures of which exhibit a non-singular behaviour,
\bea C^{(n)}_{\a(2s)} \ \sim \ \frac{i^{n+s+1}\m_n}{(1+r^2\sin^2\theta)^{\ft{s+1}2}}\,(u^+_\a u^-_\a)^s \ ,\eea
where $\m_n$ are real deformation parameters.

Finally, in the case of strictly axisymmetric solutions in ${\cal M}_{E}(E,J)$, the Weyl tensors diverge at the origin with a power-law that is in general different from that of the spherically-symmetric case (see for instance Eq. \eq{axisexc}).
On the other hand, in ${\cal M}_{J}(E,J)$, the Weyl tensors inherit the regularity of the cylindrically-symmetric solutions. 

Therefore, for a number of reasons, it is tempting to identify the singular family ${\cal M}_{E}(E,J)$ as higher-spin generalizations of black holes. A more detailed study of whether or not, for instance, the singularity is physical and not a gauge artifact, and whether or not these solutions possess an event horizon, can be performed by analyzing the propagation of small fluctuations over them. 
The deviations from Einstein gravity in the strong-curvature region, as discussed above, may be radical, essentially due to the non-locality of interactions induced by the unbroken higher-spin symmetry. 
To probe this region, it may be necessary to extend the usual tools of differential geometry to the higher-spin context, since standard concepts such as the relativistic interval are not higher-spin invariant. 

However, we already have at our disposal some useful instruments for distinguishing gauge-inequivalent solutions and for characterizing them physically even in strong-field regions, namely zero-form charges \cite{Sezgin:2005pv,Iazeolla:2007wt,Colombo:2010fu,Sezgin:2011hq}, that we shall use in this paper. They are a set of functionals of the zero-form master-fields, defined via the trace of the $\star$-product algebra, that are conserved on the field equations and provide classical observables that do not break the higher-spin gauge symmetries. As the non-locality on ${\cal T}$ of the star-product is mapped via the field equations to spacetime non-locality, the zero-form charges hide their higher-derivative nature into the $\star$-products between master-fields, and this facilitates their evaluation. We find that certain zero-form charges involving the spacetime curvatures are well-defined on our solutions, and amount to linear combinations of powers of the squared deformation parameters $\m^2_{n}$, that therefore characterize the various field configurations in a gauge-independent way. Interestingly enough, all these invariants are finite everywhere -- unless the solution under consideration is based on infinitely many projectors and the eigenvalues $\m_{n}$ are not too small. Whether or not this is the signal of a true singularity in higher-spin gravity is a question that needs further study. We speculate on this and related issues in the final Section of this paper, where we also mention a possible way of turning on an angular momentum, the details of which we leave for future work. 

\scss{Plan of the paper}

The organization of this paper is as follows. In Section \ref{sec:in} we recall some relevant aspects of the Vasiliev's equations, in a general fashion that also encompasses details that are relevant for the off-shell extensions and globally defined formulations \cite{Boulanger:2011dd,Sezgin:2011hq}, including some discussion of observables in higher-spin gravity. Moreover, in Section \ref{gaugefmethod} we recall the gauge-function solution method and give the reduced, twistor-space equations that we shall solve in the next Section. In Section \ref{Sec:internal} we spell out our solution strategy and Ans\"atze: in particular, we give the details of the gauge functions we work with, introduce the main objects the solutions will be made up of, discuss their spacetime meaning, and solve the deformed oscillator problem. Section \ref{sec3summary} contains a summary of the obtained internal, twistor-space solutions, where we also show that they form a subalgebra under $\star$-product, and analyze their singular points in ${\cal Y}\times{\cal Z}$ --- \emph{i.e.}, the appearance of a non-trivial submanifold ${\cal D}$ of singular points in ${\cal Y}\times{\cal Z}$. In Section \ref{Sec:sptfields} we obtain the spacetime master-fields by evaluating the relevant $\star$-products with the gauge function. We show that all the master-fields and the generating functions of the gauge fields are regular, except possibly at spacetime points where the curvatures are singular. Thus, introducing the spacetime coordinates softens the behaviour of the singular solutions on ${\cal T}$: for example, in the spherically-symmetric case the radial coordinate $r$ enters the Weyl master zero-form as the parameter of a limit representation of a delta function in twistor space. Section \ref{Sec:Weyl} is devoted to the study of the individual spin-$s$ Weyl tensors (to which one can assign a physical meaning only in asymptotic regions, where the curvatures are weak and the different spins effectively decouple) focusing on solutions depending on $E$ and $J$. We then turn, in Section \ref{Sec:0f}, to characterizing the solutions also in strong-field regions by evaluating the above-mentioned zero-form observables. Finally, in Section \ref{Sec:concl} we draw our conclusions and mention a number of directions for future study and open problems. The paper is completed by seven appendices, where we spell out our conventions (Appendix \ref{App:conv}), give some general discussion of the $\star$-product algebras for the Vasiliev system and orderings (Appendix \ref{App:A}), recall some background material (Appendices \ref{terminology} and \ref{weakfields}), and collect some results that are used in the main body of the paper (Appendices \ref{App:L-rot}, \ref{AppProj} and \ref{App:D}).


\scs{Vasiliev's Equations for Four-Dimensional Bosonic Models}\label{sec:in}

In this section we describe various aspects of Vasiliev's four-dimensional higher-spin gravities, with focus on bosonic models. The direct requisites for the construction of the exact solutions are found mainly in Sections \ref{discretesymm}, \ref{innerkleinians}, \ref{sec:defosc}, \ref{manifestlor}, \ref{gaugefmethod} and \ref{0formcharges}, and in Appendix \ref{App:chiraldomain}\,; in particular, the original form of the Vasiliev equations, which we shall solve using the gauge-function method, is given in Eqs. \eq{MC}--\eq{INT3}. For the general picture and some terminology we refer to Appendix \ref{terminology}, and for the weak-field expansion we refer to Appendix \ref{weakfields}. 

\scss{Kinematics}

\scsss{Correspondence space}

The basic variables of Vasiliev's formulation of higher-spin gravity are differential forms on $\mathfrak{C}$, a non-commutative symplectic manifold with symplectic structure $\C$, that we shall refer to as the correspondence space.
Locally, $\mathfrak{C}$ is the product of a phase-spacetime, containing the ordinary (commutative) spacetime, and internal directions.
Like in ordinary gravity, the moduli space of the theory, say ${\cal M}$, consists of super-selection sectors ${\cal M}^{(\Sigma)}$, here labelled by $\Sigma$, related to various classes of boundary condition on $(\mathfrak{C})$, as we shall detail in Section \ref{gaugefmethod}.
In higher-spin gravity, each sector, which consists of field configurations on-shell, arises inside a larger, graded-associative differential algebra defined off-shell, here denoted by $\Omega^{(\Sigma)}(\mathfrak{C})$, that is endowed with a differential $\widehat d$ and a binary composition rule $\star$, such that if $\widehat f,\widehat g\in\O^{(\Sigma)}(\mathfrak{C})$ then 
\be \widehat d\left(\widehat f\star\widehat g\right)~=~\left(\widehat d\,\widehat f\right)\star\widehat g+(-1)^{{\rm deg}(\widehat f)}\widehat f\star\left(\widehat d\,\widehat g\right)\ ,\qquad \widehat d^{\,\,2}~=~0\ ,\ee
and a hermitian conjugation operation obeying 
\be \left(\widehat f\star\widehat g\right)^\dagger~=~(-1)^{{\rm deg}(\widehat f) {\rm deg}(\widehat g)}\big(\widehat g\big)^\dagger\star \big(\widehat f\,\big)^\dagger\ ,\qquad \left(\widehat d\,\widehat f\right)^\dagger~=~\widehat d\left(\big(\widehat f\,\big)^\dagger\right)\ .\ee
The graded bracket $\left[\cdot,\cdot\right]_\star$ is defined by
\be \left[\widehat f\,,\, \widehat g\right]_\star~=~\widehat f \star \widehat g-(-1)^{{\rm deg}(\widehat f) {\rm deg}(\widehat g)}\widehat g\star\widehat f\ .\ee
In the atlas approach, the manifold $\mathfrak{C}$ consists of charts $\mathfrak{C}_I$ covered by real canonical coordinates $\Xi^{\underline M}_I$ obeying 
\be [\Xi^{\underline M}_I,\Xi^{\underline N}_I]_\star ~=~2i\C^{\underline{MN}}\ ,\qquad [\Xi^{\underline M}_I,d\Xi^{\underline N}_I]_\star ~=~0\ ,\qquad [d\Xi^{\underline M}_I,d\Xi^{\underline N}_I]_\star ~=~0\ ,\ee
where $\C^{\underline{MP}}\C_{\underline{MN}}=\delta^P_N$ and we shall write $d\,\Xi^{\underline M}_I d\,\Xi^{\underline N}_I:= d\,\Xi^{\underline M}_I \star d\,\Xi^{\underline N}_I$. 
The charts are glued together via canonical transformations 
\be \Xi^{\underline M}_{I'}~=~(\widehat T_{I'}^{I})^{-1}\star \Xi^{\underline M}_{I}\star \widehat T_{I'}^{I}\ ,\label{tf}\ee
such that $\C|_{\mathfrak{C}_I}=\frac12 d\,\Xi^{\underline M}_I \star d\,\Xi^{\underline N}_I \C_{\underline{MN}}$ where $d\,\Xi^{\underline M}_I\equiv \widehat d (\Xi^{\underline M}_I)$ and $\C_{\underline{MN}}$ is a constant symplectic matrix. 
As we shall discuss below, the globally-defined elements $\widehat f\in \O^{(\Sigma)}(\mathfrak{C})$ are represented by sets $\left\{\widehat f_I(\Xi^{\underline M}_{I},d\Xi^{\underline M}_{I})\right\}$ of locally-defined composite operators. 
The $\Xi^{\underline M}_{I}$-dependence of the latter is expanded in a basis presented using a suitable prescription, or regular presentation, with the following two key properties: i) it is adapted to the boundary conditions related to $\Sigma$; and ii) it provides $\O^{(\Sigma)}(\mathfrak{C}_I)$ with two algebraic structures, namely that of iia) an associative $\star$-product algebra; and iib) a separate left- and right-module for the $\star$-product algebra consisting of arbitrary polynomials in $(\Xi^{\underline M}_{I},d\Xi^{\underline M}_{I})$.  
In other words, the regularly-presented basis elements must have $\star$-product compositions among themselves as well as with arbitrary polynomials that are finite as well as compatible with associativity, and their symbols must be functions  that facilitate the imposition of the boundary conditions in question.

Turning to Vasiliev's formulation of four-dimensional higher-spin gravity,  the local splitting of the correspondence space into a phase-spacetime and internal directions is of the form:
\be \mathfrak{C}_I~\cong~ T^\ast {\cal X}_I\times {\cal Y}\times {\cal Z}\ ,\qquad \C|_{\mathfrak{C}_I}~=~\C_{T^\ast{\cal X}_I}+\C_{\cal Y}+\C_{\cal Z}\ ,\label{calCI}\ee
where $T^\ast {\cal X}_I$ is a chart of a phase space $T^\ast{\cal X}$ and ${\cal Y}\times {\cal Z}$ is a twistor space; the corresponding canonical coordinates ($\underline\a=(\a,\ad)$; $\a,\ad=1,2$)
\be \Xi^{\underline M}~=~(X^M,P_M; Y^{\underline\a};Z^{\underline\a})\ ,\qquad (Y^{\underline\a};Z^{\underline\a})~=~(y^\a,\yb^{\ad};z^\a,-\zb^{\ad})\ ,\ee
are defined such that 
\be \C_{T^\ast{\cal X}}~=~dP_M dX^M\ ,\quad \C_{\cal Y}~=~\frac12( dy^\a dy_\a+  d\yb^{\ad}d\yb_{\ad})\ ,\quad \C_{\cal Z}~=~-\frac12 (dz^\a dz_\a+d\zb^{\ad} d\zb_{\ad})\ ,\label{gammay}\ee
\be (X^M)^\dagger~=~ X^M\ ,\qquad (P_M)^\dagger~=~P_M\ ,\qquad (y^\a)^\dagger~=~\yb^{\ad}\ ,\qquad (z^\a)^\dagger~=~\zb^{\ad}\ ,\ee
which implies
\be [X^M,P_N]_\star~=~i\delta^M_N\ ,\qquad [y^\a,y^\b]_\star~=~2i\e^{\a\b}\ ,\qquad [z^\a,z^\b]_\star~=~-2i\e^{\a\b}\ ,\label{oscillators}\ee
and hermitian conjugates, and our conventions for spinors are collected in Appendix \ref{App:conv}.
In what follows, we shall construct exact solutions on submanifolds of ${\cal X}$ by first projecting to the reduced correspondence space
\be \check{\mathfrak{C}}~\stackrel{\rm loc}{\cong}~ {\cal X}\times {\cal Y}\times {\cal Z}\ ,\ee
and then further down to four-dimensional submanifolds of ${\cal X}$.

\scsss{Bosonic master fields}\label{discretesymm}

The fundamental fields are a locally-defined zero-form $\widehat \Phi_I$; a locally-defined one-form $\widehat A_I$; and a globally-defined complex two form $(\widehat J,\,\,\widehat{\!\!\bar J})$. These master fields obey the reality conditions 
\be (\widehat \Phi,\widehat A,\widehat J,\widehat{\bar J})^\dagger~=~(\pi(\widehat \Phi),-\widehat A,-\widehat{\bar J},-\widehat J)\ .\label{reality}\ee
In bosonic models, they also obey the projections
\be \pi\bar\pi(\widehat \Phi,\widehat A)~=~(\widehat \Phi,\widehat A)\ ,\qquad \pi(\widehat J,\,\,\widehat{\!\!\bar J})~=~\bar\pi(\widehat J,\,\,\widehat{\!\!\bar J})~=~(\widehat J,\,\,\widehat{\!\!\bar J})\ ,\label{piofj}\ee
where $\pi$ and $\bar\pi$ are the involutive automorphisms defined by $\widehat d\,\pi=\pi\,\widehat d$, $\widehat d\,\bar\pi=\bar\pi\,\widehat d$ and
\be \pi(X^M,P_M;y^\a,\yb^{\ad};z^\a,\zb^{\ad})~=~(X^M,P_M;-y^\a,\yb^{\ad};-z^\a,\zb^{\ad})\ ,\qquad \pi(\widehat f\star\widehat g)~=~\pi(\widehat f)\star \pi(\widehat g)\ ,\ee
\be \bar\pi(X^M,P_M;y^\a,\yb^{\ad};z^\a,\zb^{\ad})~=~(X^M,P_M;y^\a,-\yb^{\ad};z^\a,-\zb^{\ad})\ ,\qquad \bar\pi(\widehat f\star\widehat g)~=~\bar\pi(\widehat f)\star \bar\pi(\widehat g)\ .\ee
In minimal bosonic models, the master fields obey the stronger projections
\be \tau(\widehat \Phi,\widehat A,\widehat J,\widehat{\bar J})~=~(\pi(\widehat \Phi),-\widehat A,-\widehat J,-\widehat{\bar J})\ ,\label{min}\ee
where $\tau$ is the graded anti-automorphism defined by $\widehat d\,\tau ~=~\tau\, \widehat d$ and
\be \tau(X^M,P_M;Y^{\underline\a};Z^{\underline\a})~=~(X^M,-P_M;iY^{\underline\a};-iZ^{\underline\a})\ ,\qquad \tau(\widehat f\star\widehat g)~=~(-1)^{\widehat f\widehat g} \tau(\widehat g)\star \tau(\widehat f)\ ,\ee
and obeying $\tau^2=\pi\bar\pi$.
The perturbative spectra of the bosonic and minimal bosonic models consist of real Fronsdal fields of integer and even-integer spins, respectively, with each spin occurring in the spectrum with multiplicity one. 

\scsss{Inner Kleinians}\label{innerkleinians}

The automorphisms $\pi$ and $\bar\pi$ are inner and generated via the adjoint action of inner Kleinians as follows:
\bea \pi(\widehat f)&=&\widehat \kappa\star\widehat f\star\widehat \kappa\ ,\qquad \widehat \k~=~\cos_\star (\pi \widehat w)\ ,\qquad \widehat\k\star\widehat\k~=~1\ ,\\[5pt]
\bar\pi(\widehat f)&=&\widehat{\bar\kappa}\star\widehat f\star\widehat{\bar\kappa}\ ,\qquad \widehat{\bar\kappa}~=~(\widehat \k)^\dagger~=~\cos_\star (\pi \widehat {\bar w})\ ,\qquad \widehat{\bar\k}\star\widehat{\bar\k}\ =\ 1\ ,\eea
where the holomorphic and anti-holomorphic (shifted) number operators, respectively, are realized as 
$\widehat w=\frac12\left\{ \widehat a^-_\a,\widehat a^{+\a}\right\}_\star$, with $(\widehat a^+_\a,\widehat a^-_\a)=\frac12( y_\a+z_\a,-i y_\a+i z_\a)$ obeying $\left[\widehat a^-_\a,\widehat a^{+\b}\right]_\star=\delta_\a^\b$, and $\widehat{\bar w}=(\widehat w)^\dagger$, 
such that 
\be \widehat w~=~\frac{i}2 y^\a\star z_\a\ ,\qquad \widehat{\bar w}~=~-\frac{i}2 \bar y^{\ad}\star \bar z_{\ad}\ .\ee
The Kleinians can be expressed in various ordering schemes; for details, see Appendix \ref{C4}. For the weak-field expansion, it is convenient to normal-order with respect to the complexified Heisenberg algebra $(\widehat a^+_\a,\widehat a^-_\a)$ \cite{more}, which we denote by $\widehat N_+$-order, where the induced $\star$-product among symbols is given by \eq{E28}, that is
\be [\widehat f_1]^{\widehat N_+}\star\, [\widehat f_2]^{\widehat N_+}~=~ \int_{{\cal R}_C} {d^4U d^4V\over (2\pi)^4} e^{i( v^\a u_\a+\bar v^{\ad} \bar u_{\ad}) }[\widehat f_1]^{\widehat N_+}(y+u,\bar y+\bar u;z+u,\bar z-\bar u)[\widehat f_2]^{\widehat N_+}(y+v,\bar y+\bar v;z-v,\bar z+\bar v)\ ,\ee
where $[\cdot]_B$ and $[\cdot]^B$ denote the Wigner map to the basis $B$ and its inverse, respectively, that is, $[\cdot]_B$ maps totally-symmetric operators to $B$-ordered operators and $[\cdot]^B$ maps operators to totally-symmetric symbols (for further details, see Appendix \ref{App:A}). In the $\widehat N_+$-order, one has \cite{vasiliev,more}
\be \widehat\kappa~=~\left[\exp(iy^\a z_\a)\right]_{\widehat{\rm N}_+}\ ,\qquad\widehat{\bar\k}~=~\left[\exp(-i\yb^{\ad}\zb_{\ad})\right]_{\widehat{\rm N}_+}\ .
\ee
while in the Weyl order  
\be
\widehat\kappa~=~\left[(2\pi)^2 \delta^2(y)\delta^2(z)\right]_{\rm Weyl}\ ,\qquad \widehat{\bar\k}~=~\left[(2\pi)^2\delta^2(\yb)\delta^2(\zb)\right]_{\rm Weyl}\ ,\label{Weyl}
\ee
which implies the factorization property \cite{Didenko:2009td}
\be \widehat\kappa~=~\kappa_y\star \kappa_z\ ,\qquad \kappa_y~=~[2\pi\delta^2(y)]_{\rm Weyl}\ ,\qquad \kappa_z~=~[2\pi\d^2(z)]_{\rm Weyl}\ ,\label{factorizekappa}\ee
where $\k_y$ and $\k_z$ are the inner Kleinians for the chiral oscillator algebras generated by $y_\a$ and $z_\a$, respectively (for further details, see Appendix \ref{C4}). This factorization property, which holds in all orders, is crucial for the separation of twistor-space variables that we shall use below. 
 
\scss{Unfolded equations of motion}\label{fieldequations}

\scsss{Quasi-free differential algebra}\label{qfda}

The unfolded equations of motion of the four-dimensional bosonic higher-spin gravities that we shall study can be written as\footnote{In the topological open-string C-model proposed in \cite{Engquist:2005yt} as a microscopic origin for Vasiliev's equations, the $\star$-multiplication by $\C_{\cal Y}^{\star2}$ has a natural interpretation as the insertion into the path integral of delta-functions for fermionic zero-modes. }
\be \C_{\cal Y}^{\star2}\star \widehat D\widehat \Phi~=~0\ ,\qquad  \C_{\cal Y}^{\star2}\star \left(\widehat F+{\cal F}(\widehat \Phi)\star \widehat J+\,\overline{\!\cal F}(\widehat \Phi)\star \,\,\widehat{\!\!\bar J}\,\right)~=~0\ ,\label{e1}\ee
\be \C_{\cal Y}^{\star2} \star \widehat d\,\widehat J~=~0\ ,\qquad \C_{\cal Y}^{\star 2} \star \widehat d \,\,\widehat{\!\!\bar J}~=~0\ ,\label{e3}\ee
with Yang-Mills-like curvatures $\widehat F:=\widehat d\,\widehat A+\widehat A\star\widehat A$ and $D\widehat \Phi:=\widehat d\,\,\widehat \Phi+[\widehat A,\widehat\Phi]_\pi$, where
$[\widehat f,\widehat g]_\pi:=\widehat f\star\widehat g-(-1)^{{\rm deg}(\widehat f){\rm deg}(\widehat g)}\widehat g\star\pi(\widehat f)$ for $\widehat f,\widehat g\in\O(\mathfrak{C})$.
The interaction ambiguities ${\cal F}$ and $\overline{\!\cal F}=({\cal F})^\dagger$ are given by
\be {\cal F}(\widehat \Phi)~=~\sum_{n=0}^\infty f_{2n+1}(\widehat \Phi\star\pi(\widehat\Phi)) \left(\widehat\Phi\star\pi(\widehat\Phi)\right)^{\star n}\star \widehat \Phi\ ,\ee
where $f_{2n+1}$ are complex-valued zero-form charges obeying 
\be \widehat d f_{2n+1}~=~0\ ,\label{f2n+1constraints}\ee
as we shall describe in more detail below.
Integrability requires the algebraic constraints
\be \widehat J\star \pi(\widehat \Phi,\widehat A)~=~(\widehat\Phi,\widehat A)\star \widehat J\ ,\qquad
\,\,\widehat{\!\! \bar J}\star \pi(\widehat \Phi,\widehat A)~=~(\widehat\Phi,\widehat A)\star \,\,\widehat{\!\! \bar J}\ ,\label{Jconstraints}\ee
modulo terms that are annihilated by $\C_{\cal Y}^{\star2}\star$. In other words, Eqs. \eq{e1}--\eq{e3} and Eq. \eq{f2n+1constraints} are compatible with $\widehat d^2\equiv 0$ modulo Eq. \eq{Jconstraints}, hence defining a universal (\emph{i.e.} valid on any ${\cal X}$) quasi-free associative differential algebra. Factoring out perturbative redefinitions of $\widehat \Phi$, the ambiguity residing in ${\cal F}$ reduces down to \cite{more,Vasiliev:1999ba,Sezgin:2011hq} 
\be {\cal F}~=~{\cal B}\star\widehat\Phi\ ,\qquad {\cal B}~=~\exp_\star\left(i\theta[\widehat\Phi\star\pi(\widehat\Phi)]\right)\ ,\label{calB}\ee\be \theta[\widehat\Phi\star\pi(\widehat\Phi)]~=~ \sum_{n=0}^\infty \theta_{2n}[\widehat\Phi\star\pi(\widehat\Phi)]\,\left(\widehat\Phi\star\pi(\widehat\Phi)\right)^{\star n}\ ,\label{Theta}\ee
which breaks parity except in the following two cases \cite{Sezgin:2003pt}:
\bea \mbox{Type A model (scalar)}&:& \theta~=~0\ ,\qquad P(\widehat \Phi,\widehat A,\widehat J)~=~(\widehat \Phi,\widehat A,\widehat J)\ ,\\[5pt]
\mbox{Type B model (pseudo-scalar)}&:& \theta~=~\frac{\pi}2\ ,\qquad P(\widehat \Phi,\widehat A,\widehat J)~=~(-\widehat \Phi,\widehat A,-\widehat J)\ ,\eea
where the parity operation is the automorphism of $\O(\mathfrak{C}_I)$ defined by
\be P(X^M,P_M,y^\a,\yb^{\ad},z^\a,\zb^{\ad})~=~(X^M,P_M,\yb^{\ad},y^\a,-\zb^{\ad},-z^\a)\ ,\qquad \widehat  d P~=~P \widehat d\ .\ee
The gauge transformations read
\be \delta_{\widehat\e}\widehat \Phi~=~-[\widehat \e,\widehat\Phi]_\pi\ ,\qquad \delta_{\widehat\e}\widehat A~=~\widehat D\widehat\e\ ,\quad \delta_{\widehat\e}\widehat J~=~0\ ,\ee
with $\widehat D\widehat\e:=\widehat d\widehat\e+[\widehat A,\widehat\e]_\star$, and where $\widehat \e$ is subject to the same kinematic conditions as $\widehat A$. In globally-defined formulations, the transition functions $T_I^{I'}$ defined in \eq{tf} glue together the locally-defined configurations $(\widehat \Phi_I,\widehat A_I,\widehat J_I)$ as follows:
\be \widehat \Phi_I~=~(\widehat T_I^{I'})^{-1}\star\widehat \Phi_{I'}\star \pi(\widehat T_I^{I'})\ ,\qquad 
\widehat A_I~=~(\widehat T_I^{I'})^{-1}\star(\widehat A_{I'}+\widehat d)\star \widehat T_I^{I'}\ ,\qquad \widehat J_I~=~\widehat J_{I'}\ .\label{glue}\ee

\scsss{Free differential algebra and deformed oscillators}\label{sec:defosc}

The projection implied by the $\star$-multiplication by $\C_{\cal Y}^{\star2}$ can be solved locally on $\mathfrak{C}_I$ by taking the master fields to be forms on $T^\ast{\cal X}_I\times {\cal Z}$ valued in the algebra $\O^{[0]}({\cal Y})$ of zero-forms on ${\cal Y}$. Thus 
\be \widehat A~=~\widehat U+\widehat V\ ,\ee
where
\be \widehat U~=~ dX^M \widehat U_M(X,P;Z;Y)+ dP_M \widehat U^M(X,P;Z;Y)\ ,\ee\be  \widehat V~=~ dZ^{\underline\a}\widehat V_{\underline\a}(X,P;Z;Y)~=~dz^\a \widehat V_\a(X,P;Z;Y)+dz^{\ad} \widehat V_{\ad}(X,P;Z;Y)\ ,\ee
and the algebraic constraints \eq{Jconstraints} admit the solution 
\be \widehat J~=~ -\frac{i}4 dz^\a \wedge dz_\a\,\widehat \k\ ,\qquad \,\,\widehat{\!\!\bar J}~=~-\frac{i}4 d\zb^{\ad}\wedge d\zb_{\ad}\,\widehat{\bar \kappa}\ .\label{J}\ee 
In order to find exact solutions, it is convenient to cast\footnote{As an intermediate step, the twistor-space components of the master field equations can be rewritten as  
\be \partial_\a\widehat \Phi+\widehat V_\a\star\widehat\Phi+\widehat\Phi\star\pi(\widehat V_\a)\,=\,0\ ,\quad \partial_{\ad}\widehat \Phi+\widehat {\bar V}_{\ad}\star\widehat\Phi+\widehat\Phi\star\bar\pi(\widehat{\bar V}_{\ad})\,=\,0\ ,\nn\ee\be 
 d\widehat V_{\underline\a}+\partial_{\underline\a} \widehat U+\left[\widehat U,\widehat V_{\underline\a}\right]_\star\,=\,0\ ,\quad \widehat F_{\a\b}\,=\, -\frac{ib}2 \e_{\a\b}{\cal B}\star\widehat \Phi\star\widehat\k\ ,\quad \widehat F_{\a\bd}\,=\,0\ ,\quad \widehat  F_{\ad\bd}\,=\,  -\frac{i\bar b}2 \e_{\ad\bd}\overline {\cal B}\star\widehat \Phi\star\widehat{\bar\k}\ ,\nn\ee
where $\widehat F_{\underline{\a\b}}=2\partial_{[\underline\a}\widehat V_{\underline\b]}+[\widehat V_{\underline\a},\widehat V_{\underline\b}]_\star$ and $\partial_{\underline\a}\equiv \partial/\partial Z^{\underline\a}$. } the remaining differential constraints into Vasiliev's original deformed-oscillator format\footnote{This format exhibits two global symmetries: Firstly, the $\mathbb{Z}_2$-transformations $(\widehat \Phi,\widehat W,\widehat S_\a,\widehat{\bar S}_{\dot\a})\rightarrow(\widehat \Phi,\widehat W,-\widehat S_\a,-\widehat{\bar S}_{\dot\a})$. Secondly, the transformations 
\bea \left(\ba{c} y_\a \\ z_\a\ea \right)&\rightarrow & \left(\ba{cc} A_\a{}^\b& B_\a{}^\b \\ B_\a{}^\b &A_\a{}^\b
\ea \right)\left(\ba{c} y_\a \\ z_\a\ea \right)\ ,\nn\eea
that preserve (i) the $\star$-product algebra, which requires $(A+B)(A^t-B^t)=-1$ ; (ii) the inner Kleinian operators, which requires $A^t A-B^t B=-1$ ; and (iii) the bosonic projection conditions, finally fixing non-minimal $GL(2;\Comp)$-transformations with $B\neq0$, or minimal $SL(2;\Comp)_{\rm diag}$-transformations with $B=0$. In the non-minimal case, the $GL(2,\Comp)$-action is the closure of $SL(2;\Comp)_{\rm diag}$ and the discrete transformation $(y_\a,z_\a)\rightarrow i(z_\a, y_\a)$, that is broken by the $\tau$-condition in the minimal models.
}:
\be d \widehat U+\widehat U\star \widehat U~=~0\ ,\qquad d\widehat \Phi+\widehat U\star\widehat \Phi-\widehat \Phi\star \pi(\widehat U)~=~0\ ,\label{MC}\ee
\be d\widehat S_{\underline\a}+[\widehat U,\widehat S_{\underline\a}]_\star~=~0\ ,\label{dSa}\ee
\be \widehat S_\a\star\widehat\Phi+\widehat\Phi\star\pi(\widehat S_\a)~=~0\ ,\quad
\widehat {\bar S}_{\ad}\star\widehat\Phi+\widehat\Phi\star\bar\pi(\widehat {\bar S}_{\ad})~=~ 0  \ ,\label{INT1}\ee\be
[\widehat S_\a,\widehat S_\b]_\star~=~ -2i\e_{\a\b}(1-{\cal B}\star\widehat \Phi\star\widehat \kappa)\ ,\quad [\widehat {\bar S}_{\ad},\widehat {\bar S}_{\bd}]_\star~=~ -2i\e_{\ad\bd}(1- \overline{\cal B}\star  \widehat\Phi\star\widehat {\bar \kappa}) \ ,\label{INT2}\ee\be
[\widehat S_\a,\widehat{\bar S}_{\ad}]_\star~=~ 0\ ,\label{INT3}\ee
where we have defined $d=dX^M \partial_M+dP_M\partial^M$ and 
\be \widehat S_{\underline\a}~=~Z_{\underline\a} -2i\widehat V_{\underline\a}~=~( \widehat S_{\a}, -\widehat{\bar S}_{\ad})~=~(z_\a-2i \widehat V_\a,-\zb_{\ad}+2i \widehat{\bar V}_{\ad})\ .\ee
The integrability of the system implies the gauge transformations 
\be \delta_{\widehat \e} \,\widehat \Phi~=~-[\widehat\e,\widehat\Phi]_\pi\ ,\quad \delta_{\widehat \e}\,\widehat S_{\underline\a}\ =\ -[\widehat \e,\widehat V_{\underline\a}]_\star\ ,\quad \delta_{\widehat \e}\, \widehat U~=~ d\widehat \e+[\widehat U,\widehat \e\,]_\star\ .\ee

\scsss{Manifest Lorentz invariance}\label{manifestlor}

Manifest local Lorentz covariance can be achieved by means of the field redefinition \cite{Vasiliev:1999ba,Sezgin:2002ru,Sezgin:2011hq}
\be \widehat W~:=~ \widehat U-\widehat K\ ,\qquad \widehat K~:=~ \frac1{4i} \left(\o^{\a\b} \widehat M_{\a\b}+\bar \o^{\ad\bd} \widehat {\overline M}_{\ad\bd}\right)\ ,\label{fieldredef}\ee
where $(\o^{\a\b},{\bar \o}^{\ad\bd})$ is the canonical Lorentz connection, and 
\bea \widehat M_{\a\b}&:=& \widehat M^{(0)}_{\a\b}+\widehat M^{(S)}_{\a\b}\ ,\qquad
\widehat {\overline M}_{\ad\bd}\: =\ \widehat{\overline M}^{(0)}_{\ad\bd}+\widehat{\overline{M}}^{(\bar S)}_{\ad\bd}\ ,\label{fullM}\eea
are the full Lorentz generators, consisting of the internal part
\bea \widehat M^{(0)}_{\a\b}&:=& y_{(\a} \star y_{\b)}-z_{(\a} \star z_{\b)}\ ,\qquad
\widehat{\overline M}^{(0)}_{\ad\bd}\ :=\ \yb_{(\ad}\star \yb_{\bd)}- \zb_{(\ad} \star \zb_{\bd)}\ ,\label{M(0)}\eea
rotating the $Y$ and $Z$ oscillators, and the external part
\bea \widehat M^{(S)}_{\a\b}&:=& \widehat S_{(\a}\star \widehat S_{\b)}\ ,\qquad
\widehat{\overline M}^{(\bar S)}_{\ad\bd}\ :=\  \widehat {\bar S}_{(\ad}\star \widehat {\bar S }_{\bd)}\ ,\label{M(S)}\eea
rotating the spinor indices carried by $(\widehat S_\a,\widehat{\bar S}_{\ad})$. As a result, the master equations read
\bea &\nabla\widehat W+\widehat W\star \widehat W + \frac1{4i} \left(r^{\a\b} \widehat M_{\a\b}+\bar r^{\ad\bd} \widehat {\overline M}_{\ad\bd}\right)\ =\ 0\ ,\quad \nabla\widehat \Phi+\widehat W\star\widehat \Phi-\widehat \Phi\star\pi(\widehat W)\ =\ 0\ ,\label{2.55}&\\[5pt]
&\nabla\widehat S_\a+\widehat W\star\widehat S_\a-\widehat S_\a\star \widehat W\ =\ 0\ ,\quad \nabla\widehat {\bar S}_{\ad}+\widehat W\star\widehat {\bar S}_{\ad}-\widehat {\bar S}_{\ad}\star \widehat W\ =\ 0&\\[5pt]
&
\widehat S_\a\star\widehat\Phi+\widehat\Phi\star\pi(\widehat S_\a)\ =\ 0\ ,\quad
\widehat {\bar S}_{\ad}\star\widehat\Phi+\widehat\Phi\star\bar\pi(\widehat {\bar S}_{\ad})\ =\ 0 &\\[5pt]
&
[\widehat S_\a,\widehat S_\b]_\star\ =\ -2i\e_{\a\b}(1-{\cal B}\star\widehat \Phi\star\widehat\kappa)\ ,\quad [\widehat {\bar S}_{\ad},\widehat {\bar S}_{\bd}]_\star\ =\ -2i\e_{\ad\bd}(1- \overline{\cal B}\star \widehat\Phi\star\widehat{\bar \kappa})&\\[5pt]
&
[\widehat S_\a,\widehat{\bar S}_{\ad}]_\star\ =\ 0\ ,&\label{2.56}\eea
where $r^{\a\b}:=d\o^{\a\b}+\o^{\a\c}\o^{\b}{}_\c$ and $\bar r^{\ad\bd}:=d\bar \o^{\ad\bd}+\o^{\ad\cd}\o^{\bd}{}_{\cd}$,
and
\bea \nabla \widehat W&:=& d\widehat W+\frac1{4i} \left[\o^{\a\b} \widehat M^{(0)}_{\a\b}+\bar \o^{\ad\bd} \widehat {\overline M}^{(0)}_{\ad\bd}~,~\widehat W\right]_\star\ ,\label{lorcovfirst}\\[5pt]
\nabla \widehat \Phi&:=& d\widehat \Phi+\frac1{4i} \left[\o^{\a\b} \widehat M^{(0)}_{\a\b}+\bar \o^{\ad\bd} \widehat {\overline M}^{(0)}_{\ad\bd}~,~\widehat \Phi\right]_\star\ ,\\[5pt]
\nabla \widehat S_\a&:=& d\widehat S_\a+\o_\a{}^\b\widehat S_\b+\frac1{4i} \left[\o^{\b\c} \widehat M^{(0)}_{\b\c}+\bar \o^{\bd\cd} \widehat {\overline M}^{(0)}_{\bd\cd}~,~\widehat S_\a\right]_\star\ ,\\[5pt]
\nabla \widehat S_{\ad}&:=& d\widehat S_{\ad}+\bar \o_{\ad}{}^{\bd}\widehat {\bar S}_{\bd}+\frac1{4i} \left[\o^{\b\c} \widehat M^{(0)}_{\b\c}+\bar \o^{\bd\cd} \widehat {\overline M}^{(0)}_{\bd\cd}~,~\widehat {\bar S}_{\ad}\right]_\star\ .\label{lorcovlast}\eea
Besides their manifest local Lorentz symmetry, these equations are by construction also left invariant under the local shift-symmetry with parameter $(\varsigma^{\a\b},{\bar \varsigma}^{\ad\bd})=dX^M(\varsigma_M{}^{\a\b},{\bar \varsigma}_M{}^{\ad\bd})+dP_M(\varsigma^M{}^{\a\b},{\bar \varsigma}^M{}^{\ad\bd})$ acting such that 
\be \delta_{\varsigma} (\widehat U,\widehat \Phi,\widehat S_\a,\widehat{\bar S}_{\ad})~=~0\ ,\qquad  \delta_{\varsigma} (\o^{\a\b},{\bar \o}^{\ad\bd})~=~ (\varsigma^{\a\b},{\bar \varsigma}^{\ad\bd})\quad\Rightarrow\quad \delta_{\varsigma}\widehat W~=~ -\frac1{4i} \left(\varsigma^{\a\b} \widehat M_{\a\b}+\bar \varsigma^{\ad\bd} \widehat {\bar M}_{\ad\bd}\right)\ .\ee
The canonical Lorentz connection can be embedded into the full theory by using the aforementioned shift-symmetry to impose 
\be\left.{\partial^2\over \partial y^\a \partial y^\b} \widehat W\right|_{Y=Z=0}~=~0\ ,\qquad \left.{\partial^2\over \partial \bar y^{\ad} \partial {\bar y}^{\bd}} \widehat W\right|_{Y=Z=0}~=~0\ .\label{shiftgauge}\ee

\scsss{Spacetime projection and component fields}

For the projection of Eqs. \eq{2.55}--\eq{2.56} to manifestly generally-covariant equations of motion for dynamical component fields in four-dimensional spacetime ${\cal X}_4$, see Appendix \ref{weakfields}. In essence, after choosing a manifestly $Sp(4;\Real)_{\rm diag}$-invariant ordering scheme, eliminating the auxiliary fields related to the unfolded description on ${\cal X}_4$ and ${\cal Z}$, and fixing suitable physical gauges (such as the twistor gauge condition \eq{twistorgauge} on $\widehat{V}_\a$  and generalized holonomic gauges on $W_\m$), there remains a set of dynamical fields consisting of a physical scalar field 
\be \phi~\equiv ~C~:=~\widehat \Phi|_{Y=Z=0}\ ,\ee
which together with the self-dual Weyl tensors $C_{\a(2s)}$ ($s\geqslant 1$) make up the generating function ($s\geqslant 0$)
\be {\cal C}~:=~\widehat \Phi|_{Z=0,\yb =0}\ ,\qquad C_{\a(2s)}~:=~\left.\frac{\partial^{2s}}{\partial^{\a_1}\cdots\partial^{\a_{2s}}}{\cal C}\right|_{y=0}\ ,\label{2.60}\ee
and a tower of manifestly Lorentz-covariant, symmetric and doubly-traceless tensor gauge fields, or Fronsdal tensors, given by ($s\geqslant 1$)
\be \phi_{\mu(s)}~:=~\left.2i e^{\a_1\ad_1}_{(\mu_1} \cdots e^{\a_{s-1}\ad_{s-1}}_{\mu_{s-1}} \frac{\partial^{2s-2}}{\partial^{\a_1}\cdots\partial^{\a_{s-1}}\bar\partial^{\ad_1}\cdots \bar\partial^{\ad_{s-1}}} W_{\mu_s)}\right|_{Y=0}\ ,\ee
where $x^\mu$ are local coordinates on ${\cal X}_4$ and 
\be W~:=~ \widehat W|_{Z=0}~=~\left.\left(\widehat U-\frac1{4i} \left(\o^{\a\b} (y_\a\star y_\b+\widehat S_\a\star\widehat S_\b)+\bar \o^{\ad\bd} (\yb_{\ad}\star\yb_{\bd}+\widehat {\bar S}_{\ad}\star\widehat{\bar S}_{\bd})\right)\right)\right|_{Z=0}\ .\label{2.61}\ee 
As a result, the regular presentation of the master fields, by its very definition, provides a regularization scheme for the strongly-coupled derivative expansions of the interaction vertices in the component-field formulation that is compatible with higher-spin gauge symmetry.
In this sense, the naive spacetime picture, based on a tower of interacting Fronsdal fields without any regular presentation attached to it, does not contain the same amount of information as the full  formulation in terms of master fields in correspondence space, as we shall comment on in the Conclusions. 

\scss{Gauge functions and moduli}\label{gaugefmethod}

Equations \eq{MC} and \eq{dSa} can be solved (on a chart $\mathfrak{C}_I$) by 
\be \widehat U_I~=~\widehat L^{-1}_I\star d\widehat L_I\ ,\quad \widehat \Phi_I~=~\widehat L^{-1}_I\star \widehat \Phi'\star \pi(\widehat L_I)\ ,\quad
\widehat S_{I;\underline\a}~=~ \widehat L^{-1}_I\star \widehat S'_{\underline\a}\star \widehat L_I\ ,\label{Lrot}\ee
where $\widehat L_I(X,P,Y,Z)$ is a gauge function, assumed to obey
\be \widehat L_I|_{X=P=Y=Z=0}~=~1\ ,\ee
and $(\widehat\Phi',\widehat S'_{\underline\a})$ are integration constants for the zero-forms on $T^\ast{\cal X}$ given by 
\bea (\widehat\Phi',\widehat S'_{\underline\a})&=& (\widehat\Phi,\widehat S_{\underline\a})|_{X=P=0}\eea
and obeying the remaining twistor-space equations
\bea
&\widehat S'_\a\star\widehat\Phi'+\widehat\Phi'\star\pi(\widehat S'_\a)\ =\ 0\ ,\quad
\widehat {\bar S}{}'_{\ad}\star\widehat\Phi'+\widehat\Phi'\star\bar\pi(\widehat {\bar S}{}'_{\ad})\ =\ 0 &\label{prime1}\\[5pt]
&
[\widehat S'_\a,\widehat S'_\b]_\star\ =\ -2i\e_{\a\b}(1-{\cal B}\star\widehat \Phi'\star\kappa)\ ,\quad [\widehat {\bar S}{}'_{\ad},\widehat {\bar S}{}'_{\bd}]_\star\ =\ -2i\e_{\ad\bd}(1- \overline{\cal B}\star \widehat\Phi'\star\bar \kappa)&\label{prime2}\\[5pt]
&
[\widehat S'_\a,\widehat{\bar S}{}'_{\ad}]_\star\ =\ 0\ .&\label{prime3}\eea
Given a solution to these equations, the generating functions \eq{2.60} and \eq{2.61} take the form ${\cal C}_I=\left(\widehat L^{-1}_I\star \widehat \Phi'\star \pi(\widehat L_I))\right|_{Z=0,\yb=0}$ and 
\be W_I~=~ \left.\widehat L^{-1}_I\star \left[d- \frac1{4i}\left(\o^{\a\b} \left( y_\a \star y_\b+ \widehat S'_\a\star\widehat S'_\b\right) + \bar\o^{\ad\bd} \left( \yb_{\ad} \star\yb_{\bd}+\widehat {\bar S}'_{\ad}\star\widehat {\bar S}'_{\bd}\right)\right)\right]\star \widehat L_I\right|_{Z=0}\ ,\label{Wmu}\ee
subject to \eq{shiftgauge}, which serves to determine $(\o_M^{\a\b},\bar \o_M^{\ad\bd})$.

%
%
%
%
A particular class of solutions, containing the exact solutions listed in Section \ref{Sec1.2}, admits perturbative expansions 
\be\widehat \Phi'~=~\sum_{n=1}^\infty \widehat \Phi^{\prime(n)}\ ,\qquad  \widehat S'_{\underline\a}~=~ \sum_{n=0}^\infty \widehat S^{\prime(n)}_{\underline\a}~\equiv~Z_{\underline\a}-2i\sum_{n=0}^\infty \widehat V^{\prime(n)}_{\underline\a} \ ,\label{solclass}\ee
where $(\widehat S^{\prime(n)}_{\underline\a},\widehat \Phi^{\prime(n)})$ are of the $n$th order in the integration constant 
\be \Phi'(Y)~=~\widehat \Phi'(Y,Z)|_{Z=0}\ ,\label{ic}\ee
and $\widehat S^{\prime(0)}_{\underline\a}$ is a flat connection in twistor space obeying 
\bea [\widehat S^{\prime(0)}_{\underline\a},\widehat S^{\prime(0)}_{\underline\b}]_\star ~=~ -2iC_{\underline{\a\b}}  \ := \ -2i\left(\ba{cc}\e_{\a\b} & 0 \\ 0& \e_{\ad\bd}\ea\right) \ .\label{fc}\eea
Depending on the boundary conditions on $\widehat S^{\prime(0)}_{\underline\a}$ in twistor space there are various natural approaches to solving these equations: If the boundary conditions are chosen such that there exists a gauge where $\widehat V^{\prime(0)}_{\underline\a}=0$, one may adapt the perturbative scheme based on \eq{rho1} and \eq{rho2} to the case at hand. 
In this paper, we shall instead obtain solutions of the form \eq{solclass} by solving the deformed oscillator problem \eq{prime1}--\eq{prime3} using separation of variables and the non-perturbative method of \cite{Prokushkin:1998bq,Sezgin:2005pv} spelled out in Appendix \ref{App:D}. 
This method also encompasses non-trivial flat connections $\widehat V^{\prime(0)}_{\underline\a}$, essentially by activating Fock-space projectors in the space of functions on ${\cal Y}\times {\cal Z}$. The resulting solutions appear naturally in gauges that differ radically from the aforementioned radial twistor gauge in the sense that the space of residual symmetries is not isomorphic to $\mathfrak{hs}(4)$ or its non-minimal extension, as we shall discuss below. 

In constructing exact solutions, we are thus led to the following moduli (for a more detailed discussion on (iii) and (iv),  see \cite{Boulanger:2011dd,Sezgin:2011hq}):
\begin{itemize}
\item[(i)] local degrees of freedom contained $\Phi'(Y)$\,;
\item[(ii)] boundary degrees of freedom contained in $\widehat L_I|_{\partial{\mathfrak C}}$ where $\partial{\mathfrak C}$ in particular contains the boundary of its four-dimensional spacetime sub-manifold\,; 
\item[(iii)] monodromies and projectors contained in flat connections $\widehat V^{\prime(0)}$ on ${\cal Z}\times {\cal Y}$ and $\widehat U^{(0)}$ on $T^\ast{\cal X}$ \,;
\item[(iv)] windings contained in the transition functions $\widehat T_I^{I'}$ defined in \eq{tf} and \eq{glue}\,;
\end{itemize}
In what follows, we shall mainly activate (i), (ii) and to some extent (iii). The Weyl zero-form moduli have so far been examined mainly in the following sectors:
\begin{itemize}
\item[(ia)] the non-unitarizable twisted-adjoint sectors consisting of arbitrary twistor-space polynomials \cite{Vasiliev:1990vu,Prokushkin:1998bq,Bekaert:2005vh} and plane waves \cite{Bolotin:1999fa,Colombo:2010fu};
\item[(ib)] the unitarizable sector consisting of states with compact $\mso(3,2)$-weights belonging to the massless
lowest-weight spaces $\mathfrak{D}^{(-)}\cong \bigoplus_{s=0,2,4,\dots}\mathfrak{D}(s+1;(s))\cong
[\mathfrak{D}(\frac12;(0))]^{\otimes 2}_{\rm symm}$ and $\mathfrak{D}^{(+)}\cong  \bigoplus_{s=2,4,\dots}\mathfrak{D}(s+1;(s,1))\oplus \mathfrak{D}(2;(0))\cong [\mathfrak{D}(1;(\frac12))]^{\otimes 2}_{\rm 
anti-symm}$ with oscillator realization in terms of operators represented by finite-dimensional matrices in the scalar and spinor singleton weight-spaces $\mathfrak{D}(\frac12;(0))$ and $\mathfrak{D}(1;(\frac12))$, respectively. These spaces are the twisted-adjoint $\mathfrak{hs}(4)$-orbits of the scalar ground states $T^{(0)}_{1;(0)}=\exp(-4E)$ and $T^{(0)}_{2;(0)}=\exp(-4E)(1-8E)$, respectively, which are proportional to the projectors ${\cal P}_{1}(E)\equiv {\cal P}_{1;(0)}=\ket{\frac12;(0)}\bra{\frac12;(0)}$ and ${\cal P}_{2}(E)\equiv {\cal P}_{2;(0)}=\ket{1;(\frac12)}^i\,{}_i\bra{1;(\frac12)}$; for further details, see Appendix \ref{AppProj} and \cite{Iazeolla:2008ix};
\item[(ic)] the unitarizable sector of states with compact $\mso(3,2)$-weights belonging to the generalized Verma modules ${\cal W}^{(+)}$ and ${\cal W}^{(-)}$ given by the twisted-adjoint $\mathfrak{hs}(4)$-orbits of the reference states $T^{(0)}_{0;(0)}=\ft{\sinh 4E}{4E}$ and $[T^{(0)}_{0;(1)}]_r=\frac{3}{64E^3} P_r (4E \cosh 4E-\sinh 4E)$, respectively, modulo the ideal subspaces $\mathfrak{D}^{(\pm)}$  \cite{Iazeolla:2008ix};
\item[(id)] the sector of states with compact  $\mso(3,2)$-weights belonging to the spaces ${\cal S}^{(\pm)}=\mD^{(\pm)}\star \kappa_y$.
\end{itemize}
The sector (ia) contains the initial data for the exact $\mso(3,1)$-invariant solutions given together with twistor-space moduli of type (ii) in \cite{Sezgin:2005pv,Iazeolla:2007wt}. The sector (ib) contains Anti-de Sitter analogs of flat-space plane waves and the sector (ic) contains runaway solutions \cite{Iazeolla:2008ix}. The completion of the latter two sectors into exact solutions, which includes providing regular presentations along the lines discussed in Section \ref{regpres}, remains an open problem at this stage. Finally, the sector (id) consists of initial data for solutions with at least two Killing symmetries corresponding to the energy operator and one compact spin of $\mso(3,2)$. Below, we provide this sector with a regular presentation that together with other tools facilitates the construction of corresponding exact solutions for both minimal and non-minimal models, including a non-minimal solution that appears to be gauge-equivalent to the extremal solution of \cite{Didenko:2009td}. We also treat related sectors singled out by other choices of commuting pairs of Killing vectors. Finally, we dress the resulting solution spaces with additional twistor-space moduli, corresponding to insertions of additional $Z$-dependent projectors into the connection $\widehat V_{\underline\a}$.

\scss{Classical observables}\label{Sec:0charges}

In order to provide a gauge-invariant characterization of exact solutions that remains valid in strong-coupling regions where the weak-field expansion (see Appendix \ref{weakfields}) breaks down, it is useful to develop a formalism for classical observables.
These are functionals of the locally-defined master fields and transition functions, defined in \eq{tf}, that are defined globally in generalized spacetimes carrying various higher-spin geometric structures \cite{Sezgin:2011hq}. 
There are several globally-defined formulations, or phases, of the theory, based on different unbroken gauge groups, or structure groups.
In what follows, we shall mainly focus on zero-form charges, which are observables in the unbroken phase.
We then present certain $p$-form charges that may play an important r\^ole in the characterization of solutions, as we shall comment on in the Outlook.

\scsss{Zero-form charges}\label{0formcharges}

In the unbroken phase, the classical observables do no break any gauge symmetries. 
The basic such observables are Wilson loops in commuting sub-manifolds of $T^\ast{\cal X}$.
These loops can be decorated with insertions of zero-form composites that transform as adjoint elements \cite{Sezgin:2011hq}. 
In the case of trivial monodromy, these can be contracted down to a single point $T^\ast{\cal X}$ resulting in zero-form charges given by the generating function
\be {\cal I}(\s,k,\bar k;\l,\bar \l)~=~ {\widehat{\rm  Tr}}_{\Real}\left[(\widehat\k\widehat{\bar \k})^{\star\s}\star \exp_\star(\l^\a\widehat S_\a+\bar\l^{\ad}\widehat{\bar S}_{\ad})\star (\widehat \Phi\star\widehat\k)^{\star k}\star (\widehat \Phi\star\widehat{\bar \k})^{\star\bar k}\right]\ ,\ee
where $\widehat{\rm  Tr}_{\Real}$ is the chiral trace defined by the integral in \eq{chiraltrace} with integration domain \eq{chiraldomain}\,; $(\s,k,\bar k)$ are natural numbers defined modulo $(\s,k,\bar k)\sim (\s\pm 2,k,\bar k)\sim (\s,k\pm 2,\bar k\mp2)\sim (\s\pm1,k\pm1,\bar k\mp1)$\,; and $(\l^\a,\bar\l^{\ad})$ are commuting spinors. 
The zero-form charges are manifestly higher-spin gauge invariant and hence defined globally on any base manifold; it follows that 
\be d\,{\cal I}(\s,k,\bar k;\l,\bar\l)~=~0\ ,\ee
modulo the equations of motion. The derivatives of ${\cal I}(\s,k,\bar k;\l,\bar\l)$ with respect to $(\l,\bar \l)$ can be re-written as traces of $\star$-commutators involving the internal Lorentz generators $(\widehat M^{(S)}_{\a\b},\widehat{\bar M}^{(S)}_{\ad\bd})$ defined in \eq{M(S)}, 
whose evaluation requires a careful examination of boundary terms in twistor space. 
In what follows, we shall mainly be concerned with\footnote{These classical observables can be identified as the on-shell values of certain deformations of the topological action of \cite{Boulanger:2011dd}, which can be interpreted as generators of semi-classical amplitudes. These amplitudes were in their turn proposed in \cite{Engquist:2005yt} to correspond to correlation functions for a topological open-string C-model.}
${\cal I}(\s,k,\bar k):={\cal I}(\s,k,\bar k;0,0)$, 
and in particular the supertraces 
\be {\cal I}_{2N}~:=~{\cal I}(1,2N,0)~=~\widehat{\rm Tr}_{\Real}[\widehat\kappa\widehat{\bar\kappa}\star (\widehat \Phi\star\pi(\widehat\Phi))^{\star N}]\ . \label{fullamp}\ee
In the Weyl order, $[\widehat\kappa\widehat{\bar\kappa}]_{\rm Weyl}=(2\pi)^4\delta^4(Y)\delta^4(Z)$, and the zero-form charges assume the localized form
\bea {\cal I}_{2N}&=&\left.\left[(\widehat \Phi\star\pi(\widehat\Phi))^{\star N}\right]^{\rm Weyl} \right|_{Y=Z=0}\ . \label{fullamp00}\eea
On the other hand, in the normal order where $[\widehat\kappa\widehat{\bar\kappa}]^{\widehat{\rm N}^+}=\exp i( y^\a z_\a-\yb^{\ad}\zb_{\ad})$,  the zero-form charges assume the non-local form
\bea {\cal I}_{2N}&=&\int_{{\cal R}_{\Real}} {d^4Y d^4Z\over (2\pi)^4} \exp i( y^\a z_\a-\yb^{\ad}\zb_{\ad}) \left[(\widehat \Phi\star\pi(\widehat\Phi))^{\star N}\right]^{\widehat{\rm N}^+}\ . \label{fullamp00b}\eea
In particular, if $\widehat \Phi$ does not depend on $Z$, that is $\Phi:=\left.\widehat\Phi\right|_{Z=0}=\widehat \Phi$, then the equality of \eq{fullamp00} and \eq{fullamp00b} follows immediately from $[\Phi]^{\widehat{\rm N}^+}=[\Phi]^{\rm Weyl}$.
In what follows we shall evaluate the zero-form charges ${\cal I}_{2N}$ on exact type-D and almost-type-D solutions. Moreover, in \cite{Colombo:2010fu} their perturbative $\Phi'$-expansion has been shown to be well-defined for $\Phi'$ in the twistor-space plane-wave sector\footnote{In this sector, the first sub-leading term of ${\cal I}_{2N}$ vanishes, \emph{i.e.}
\be \left.{\cal I}_{2N}\right|_{\rm tw. plane-waves}~=~ \left.(\Phi'\star\pi(\Phi'))^{\star N}\right|_{Y=0}+{\cal O}((\Phi')^{2N+2})\ .\nn\ee
In general, the arguments involved as well as the localization procedure require a careful study since the trace may reduce to a boundary term in ${\cal Z}$\,; for example, from $(q\widehat V+\widehat V\star\widehat V)^{\star 2}=-\frac18 dz^2d\zb^2 (\widehat \Phi\star\pi(\widehat \Phi)) \star \widehat\kappa\widehat{\bar\kappa}$, with $q=dZ^{\underline\a}\partial_{\underline\a}$,
it follows that ${\cal I}_2= - \frac1{2\pi^4}\int_{{\cal R}_{\Real}} {d^2y\,d^2\yb} \,q (\widehat V\star q\widehat V+\frac23 \widehat V\star\widehat V\star \widehat V)$ that one can argue leads to protection of ${\cal I}_2$ to all orders in the $\Phi'$-expansion.}. These two sectors thus remain well-defined within bosonic models with 
\be \theta_{2n}~=~\theta_{2n}\left({\cal I}_{2N}\right) \label{theta}\ee
in the phase factor ${\cal B}$ defined in \eq{calB} and \eq{Theta}.

\scsss{Comments on observables in broken phases}\label{Sec:soldering}

The characteristic observables of broken phases break some of the higher-spin gauge symmetries off-shell\,; these broken symmetries re-surface on-shell albeit with restricted gauge parameters forming sections belonging to bundles associated to the principal bundle of the (unbroken) structure group \cite{Boulanger:2011dd,Sezgin:2011hq}. One can show that the characteristic observables, also referred to as the order parameters, are actually diffeomorphic invariant.

One such broken phase, proposed in \cite{Sezgin:2011hq}, is the soldered phase with soldering one-form 
$\widehat E:=\frac12(1-\pi)\widehat W$, 
with $\widehat W$ defined in \eq{fieldredef}. Its order parameters are functionals ${\cal O}\left[\widehat E_I,\widehat \Phi_I,\widehat S_{\underline\a,I};\widehat T_I^{I'}\right]$ that depend explicitly on $\widehat E$ and that are manifestly invariant under the unbroken gauge transformations with locally-defined parameters $\widehat \L_I=\pi(\widehat \L_I)$, that is $\delta_{{\widehat\L}}{\cal O}\equiv0$ without using the equations of motion or any boundary conditions. Moreover, they are invariant on-shell under gauge transformations with broken gauge parameters $\widehat \xi_I=-\pi(\widehat \xi_I)$ that belong to sections, that is $\delta_{\widehat\xi} {\cal O}=0$ modulo the equations of motion and the transitions \eq{glue} and $\xi_I=(\widehat T_I^{I'})^{-1}\star  \xi_{I'}\star \widehat T_I^{I'}$.  
An example of such order parameters are the charges of complex abelian $p$-forms given by ($p=2,4,\dots$)\footnote{These classical observables can be identified as the on-shell values of certain deformations of the topological action of \cite{Boulanger:2011dd}, which can be interpreted as generators of semi-classical amplitudes associated to the boundaries of $\Sigma$. Another set of order parameters are minimal $(p+1)$-volumes constructed from Cayley-like determinants of generalized metrics such as ($s=2,4,\dots$)
$G_{M_1\dots M_{s}}= \widehat{Tr}_{\Real}\left[\widehat\kappa\widehat{\bar\kappa}\star \widehat E_{(M_1}\star \cdots \star \widehat E_{M_s)}\right]$ and generalizations thereof obtained by inserting adjoint impurities. One may ask whether there exist preferred metrics that can be used to compute not just minimal areas but also $p$-brane partition functions. 
}
\be Q_{\Sigma}~=~\oint_{\Sigma} \widehat{Tr}_{\Real}\left[\left(\widehat E\star \widehat E+\frac12(1+\pi)\widehat r^{(S)}\right)^{\star (p/2)}\star \widehat\kappa\right]\ ,\label{abelianpform}\ee
where $\widehat r^{(S)}=\frac1{4i}(r^{\a\b}\widehat M^{(S)}_{\a\b}+\bar r^{\ad\bd}\widehat{\overline{M}}^{(\bar S)}_{\ad\bd})$ and $\Sigma$ is a non-trivial cycle in a Lagrangian sub-manifold of $T^\ast{\cal X}$. 


\scss{Super-selection sectors and regular presentations}\label{regpres}

Let us end this Section by commenting on super-selection sectors, regular presentations and the interrelations between these two notions. 
Drawing on the general structure of the theory, it is natural to adopt a working hypothesis, yet to be fully explored, that, like in the case of ordinary gravity, there exists super-selections rules that partition the classical moduli space into super-selection sectors ${\cal M}^{(\Sigma)}(\mathfrak{C})$ that by their very definition are to be coordinatized by preferred sets of classical observables, here denoted by $\Sigma$.
In other words, by this hypothesis, super-selection is tantamount to identifying super-selection sectors as charts of moduli space, which is a good definition in the sense that it amounts to insisting on the finiteness of a set of observable quantities (a property that cannot change through local fluctuations or deformations).
In practice, to implement these rules, one separately constructs families of classical observables and classical solutions  and evaluates the former on the latter. 
As solutions of the equations of motion are characterized by boundary conditions or asymptotic behaviors in the correspondence space, it is natural to expect that different classes of boundary conditions are paired up with corresponding sets of observables.
This form of duality allows one to identify super-selection sectors as classes of boundary conditions, which is a slightly less abstract way of realizing the former than that of charts in moduli space.

Moreover, from our discussion in Section \ref{gaugefmethod}, we see that the boundary conditions form various representations of the underlying higher-spin Lie algebra. 
In particular, the local degrees of freedom of the theory fall into the twisted-adjoint module.
The latter has an indecomposable structure \cite{Iazeolla:2008ix} consisting of sub-representations generated from ground states (in finite-dimensional representations of a sub-algebra of the higher-spin algebra);
for example, it contains particles, run-away solutions, instantons and solitons, of which the latter are to become the main topic of the remainder of this paper.
As in standard classical perturbative field theory, one may start by extending a class of boundary conditions into linearized bulk fields on-shell, assuming some bulk vacuum such as anti-de Sitter spacetime,
that can then be dressed into interacting fields on-shell by various techniques, such as the gauge-function method. 

In a given sector of boundary conditions, or super-selection sector by the above reasoning, the on-shell master fields are composite operators whose functional nature in twistor space may require the usage of a suitable regular presentation, leading to well-defined $\star$-product compositions (being finite as well as compatible with associativity).
These presentations involve not only a choice of ordering scheme (as discussed in Appendix \ref{App:A}) but possibly also the usage of auxiliary integration variables entering via Laplace transforms and open or closed-contour generalizations thereof; it would be interesting to develop further the already quite far-going analogies to Schwinger's  first-quantized proper-time presentation of the Feynman propagators in spacetime.

Although the study of regular presentations is a crucial and non-trivial physical problem in higher-spin gravity, it has so far been addressed only in limited number of  contexts.
The original works \cite{Vasiliev:1990vu,Prokushkin:1998bq,Bekaert:2005vh} concerned the determination of the regular presentation of the dependence on internal canonical coordinates of formal perturbative expansions around maximally-symmetric spacetime backgrounds, later refined to generally-covariant weak-field expansions in \cite{Sezgin:2002ru} (see also Appendix \ref{weakfields}).
As these perturbative expansions do not refer to any specific spacetime boundary conditions, their regular presentations concern the class of arbitrary polynomials in twistor space, or equivalently, the class of twistor-space plane waves, dressed by a certain type of non-polynomial functions stemming from the internal Kleinians.
The resulting presentations involve homotopy-contracting integrals, which can be taken along either an open or a closed contour \cite{Iazeolla:2008ix,Iazeolla:2008bp}, entering only at sub-leading orders in the perturbative approach; for a recent analysis of the r\^ole of closed-contour homotopy integrals and the duality between zero-form charges and twistor-space plane waves, see \cite{Colombo:2010fu}.
Other recent works \cite{Iazeolla:2008ix} (for a discussion, see also \cite{Iazeolla:2008bp}) have initiated a systematic study of the interplay between more non-trivial spacetime boundary conditions and the need for non-trivial closed-contour regular presentations of the linearized Weyl zero-form.
As we shall see next, the latter are required in order to define minimal models with generalized Type-D solutions, and in particular with spherically symmetric solutions.
%



\scs{Gauge Function Ansatz for Generalized Type-D Solutions}\label{Sec:internal}


This section describes Ans\"atze for families of exact type-D and almost-type-D solutions to Vasiliev's equations based on gauge functions in an $AdS_4$ background spacetime and separation of variables in twistor space: the latter is achieved by expanding the master-fields in Fock-space projectors realized as functions on $\Omega^{[0]}({\cal Y})$, leading to a tractable deformed-oscillator problem. For the notations and conventions used in $AdS_4$ spacetime and the terminology used in Petrov classification, see Appendix \ref{App:conv}; for the explicit realizations of spin-frames and projectors, see Appendices \ref{App:L-rot} and \ref{AppProj}; and for details concerning the deformed oscillators, see Appendix \ref{App:D}.

\scss{Solution strategy and basic notation}

We divide the presentation of our solutions into the following steps\footnote{In order to sooner recognize the higher-spin generalization of the familiar type-D Weyl tensors characteristic of isolated massive objects in general relativity, the reader may take steps (V) and (VI) directly after step (II). }:
\begin{enumerate}
\item[I)] gauge functions and spin-frames (in Section \ref{gaugefansatz}); 
\item[II)] separation of twistor-space variables by expansion in projectors (in Section \ref{Sec:intansatz}, with details on the projectors in Sections \ref{Sec:3emb}, \ref{Sec:kinematic} and their regular presentation in Appendix \ref{AppProj}); 
\item[III)] solution of the deformed-oscillator problem (in Section \ref{Sec:defosc} and Appendix \ref{App:D});
\item[IV)] reconstruction of full gauge fields (in Sections \ref{Sec:intconn} and \ref{Sec:gfields});
\item[V)] reconstruction of the full Weyl zero-form (in Section \ref{Sec:fullWeyl});
\item[VI)] weak-field analysis: disentangling individual spin-$s$ Weyl tensors in regions where all zero-forms approach their vacuum values, \emph{i.e.} vanish (in Section \ref{Sec:Weyl});
\item[VII)] strong-field analysis: calculation of $p$-form charges that remain well-defined in regions where the individual zero-forms blow up (in Section \ref{Sec:0f}).
\end{enumerate} 
The method spelled out above yields six families of solutions organized into three pairs that we denote by ${\cal M}_{K_{(\pm)}}(\mh_\Real)$ with distinct symmetry sub-algebras $\mh_\Real=\mso(2)_{(+)}\oplus\mso(2)_{(-)}\subset \mso(3,2)\cong \msp(4;\Real)$ and ``principal'' Cartan generator $K_{(\pm)}\in\msp(4;\Comp)$ defined modulo $Sp(4;\Real)$ rotations, as will be clarified in more detail below. There are four possible such generators, namely
\be K~=~E\,,\ J\,,\ iB\,,\ iP \ ,\label{the4K}\ee
where $E=P_0=M_{0'0}$ is the AdS energy, $J:=M_{12}$ is a spin, $B:= M_{03}$ is a boost and $P:=P_1=M_{0'1}$ is a translation, leading to the three pairs
\be {\cal M}_{E}(E,J)\ ,\quad {\cal M}_{J}(E,J)\ ;\qquad {\cal M}_{J}(J,B)\ ,\quad {\cal M}_{iB}(J,B)\ ;\qquad {\cal M}_{iB}(B,P)\ ,\quad {\cal M}_{iP}(B,P)\ .\ee
Each family is a space of biaxially symmetric solutions coordinatized by a set of deformation parameters as follows ($\ve=\pm$, $\mathbb N_{\pm \ft12}:=\left\{\pm\ft12,\pm\ft32,\dots\right\}$):
\be {\cal M}_{K_{(\ve)}}(\mh)~=~\left\{ \nu_{n_1,n_2}; \sigma_{n_1,n_2}; \theta_{n_1,n_2}\right\}_{(n_1,n_2)\in (\mathbb N_{+\ft12}\times \mathbb N_{\ft{\ve\phantom{1\!\!\!}} 2})\,\cup \,(\mathbb N_{-\ft12}\times \mathbb N_{-\ft{\ve\phantom{1\!\!\!}} 2})}\ ,\ee
where $\nu\in\Comp$ are ``$\Phi$-moduli'' whose real and imaginary parts are related to generalized masses and TAUB-NUT charges, or generalized electric and magnetic charges, depending on which terminology one prefers to use, while $\sigma, \theta\in \{\pm 1\}$ are ``$S$-moduli'' related to boundary conditions on the twistor-space connection. 

The family ${\cal M}_{E}(E,J)$ contains the spherically-symmetric solutions, \emph{i.e.} solutions with enhanced $\mso(3)\oplus\mso(2)$-symmetry, while the remaining five families contain solutions with enhanced $\mso(2,1)\oplus\mso(2)$-symmetry. The symmetry-enhanced spin-$s$ Weyl tensors are of generalized Petrov-type D, \emph{i.e.} type-$\{s,s\}$, while those of the generic $\mso(2)_{(+)}\oplus\mso(2)_{(-)}$-symmetric solutions are less special: for a given fixed projector they are algebraically general for spin $s\leq k$ and of type $\{s-k,s-k,1,\dots,1\}\equiv \{(s-k)^2,1^{2k}\}$ for $s>k$, where the integer $k$ depends on the projector, which we shall refer to as almost type-D (see Appendix \ref{App:conv} for more details on the generalization of the Petrov classification to higher spins).

\scss{Spacetime gauge function}\label{gaugefansatz}

We equip the four-dimensional sub-manifold ${\cal X}_4\subset T^\ast{\cal X}$ with coordinates $x^\mu\in {\cal R}_4\subset \Real^4$, and define Gaussian gauge functions
\be \widehat L_{(K)}(x|Y,Z)~=~L(x|Y)\star \tilde L_{(K)} (x|Z)\ ,\label{gaugef}\ee
realized as $\star$-exponentials of bilinears in $Y^{\underline\a}$ and $Z^{\underline\a}$, respectively, and where $L:~{\cal R}_{4}\rightarrow Sp(4;\Real)/SL(2;\Comp)$ reconstructs spacetime and $\tilde L_{(K)}:~{\cal R}_{4}\rightarrow SL(2;\Comp)/C_{SL(2;\Comp)}(K^{L})$ aligns the spin-frame of ${\cal Z}$ with a $K$-adapted spin-frame of ${\cal Y}$, as we shall describe below. 
In the above, $C_{SL(2;\Comp)}(M)$ denotes the subgroup of $SL(2;\Comp)$ that commutes with 
\be M~:=~\ft18 M_{\underline{\a\b}} Y^{\underline\a}\star Y^{\underline{\b}}~\in~ \msp(4;\Comp)\ ;\label{MComp}\ee
$K\in \msp(4;\Comp)$ is the aforementioned principal Cartan generator; and we use the notation 
\be f^L(Y)~:=~ L^{-1}(x|Y)\star f(Y)\star L(x|Y)\ ,\qquad f^{\tilde L_K}(Z)~:=~ (\tilde L_K)^{-1}(x|Z)\star f(Z)\star \tilde L_{(K)}(x|Z)\ .\label{fL}\ee
It follows that
\be K^{L}_{\underline{\a\b}}~=~L_{\underline\a}{}^{\underline\a'}(x) L_{\underline\b}{}^{\underline\b'}(x) K_{\underline{\a'\b'}}\ ,\label{killingmatrix}\ee
where the matrix representation $L_{\underline{\a\b}}(x)$ of $L$, \emph{idem} $\tilde L_{{(K)}}$, are defined via
\be Y^{L}_{\underline\a}~:=~L^{-1}\star Y_{\underline\a}\star L~=~ L_{\underline\a}{}^{\underline\b} Y_{\underline\b}\ ,\qquad Z^{\tilde L_{{(K)}}}_{\underline\a}~:=~\tilde L_{{(K)}}^{-1}\star Z_{\underline\a}\star \tilde L_{{(K)}}~=~ (\tilde L_{{(K)}})_{\underline\a}{}^{\underline\b} Z_{\underline\b}\ .\label{Lmatrix}\ee
More generally, one has that 
\be \left[f^L(Y)\right]^{\rm Weyl}~=~\left[f(Y^L)\right]^{\rm Weyl}\ ,\qquad \left[f^{\tilde L_{(K)}}(Z)\right]^{\rm Weyl}~=~\left[f(Z^{\tilde L_{(K)}})\right]^{\rm Weyl}\ .\ee
The Gaussian gauge functions $\widehat L_{(K)}$ are related to the non-Gaussian dittos $\widehat L_{(v)}$ of the twistor gauge \eq{twistorgauge} via a higher-spin gauge transformation 
\be \widehat L_{(v)}(x|Y,Z)~=~ \widehat G^{(K)}_{(v)} \star \widehat L_{(K)}\ ,\label{Gnew}\ee
whose construction we defer to future studies.

The vacuum configuration $(\widehat \Phi,\widehat S_{\underline\a},\widehat U)=(\widehat \Phi,\widehat S_{\underline\a},\widehat U)^{(0)}_{(K)}$ given by 
\be \widehat\Phi^{(0)}_{(K)}~=~ 0\ ,\qquad \widehat S^{(0)}_{{(K)}\underline \a}~=~ Z^{\tilde L_{(K)}}_{\underline \a}\ ,\ee\be \widehat U^{(0)}_{(K)}~=~ \widehat L_{(K)}^{-1}\star \,d\widehat L_{(K)}~=~ \O^{(0)}+ \tilde L_{(K)}^{-1}\star \,d\tilde L_{(K)}\ ,\qquad \O^{(0)}~:=~L^{-1}\star \,dL\ ,\label{vac0}\ee
solves the full Vasiliev equations. Imposing \eq{shiftgauge} and using $\tilde L_{(K)}^{-1}\star d\tilde L_{(K)}|_{Z=0}=0$, it follows that
\be e^{(0)}_{\a\ad}~=~ \left.2i\l {\partial^2\over \partial y^\a \partial \yb^{\ad}} \O^{(0)}\right|_{Y=0}\ ,\qquad (\omega^{(0)}_{\a\b},\bar\omega^{(0)}_{\ad\bd})~=~2i \left.\left({\partial^2\over \partial y^\a \partial y^{\b}},{\partial^2\over \partial \yb^{\ad} \partial \yb^{\bd}}\right) \O^{(0)}\right|_{Y=0}\label{vac02}\ ,
\ee
describing an $AdS_4$ background. The above configuration can be brought via
\be \widehat G^{(0)}{}^{(K)}_{(v)}~=~\tilde L^{-1}_{(K)}\ ,\ee
to $(\widehat \Phi,\widehat S_{\underline\a},\widehat U)=(\widehat \Phi,\widehat S_{\underline\a},\widehat U)^{(0)}_{(v)}=(0,Z_{\underline\a},\O^{(0)})$ in the twistor-gauge \eq{twistorgauge} with $v^{\underline\a}=Z^{\underline\a}$. 
The vacuum is invariant under higher-spin gauge transformations obeying $\delta_{\widehat\e^{(0)}_{(K)}}(\widehat \Phi,\widehat S_{\underline\a},\widehat U)^{(0)}_{(K)}=0=\delta_{\widehat\e^{(0)}_{(v)}}(\widehat \Phi,\widehat S_{\underline\a},\widehat U)^{(0)}_{(v)}$, \emph{viz.}
\be \widehat \e^{(0)}_{(K)}~=~\widehat L^{-1}_{(K)}\star \e'(Y)\star \widehat L_{(K)}~=~\e^{\prime L}\ ,\qquad \widehat \e^{(0)}_{(v)}~=~(\widehat G^{(0)}{}^{(K)}_{(v)})^{-1}\star \widehat \e^{(0)}_{(K)}\star \widehat G^{(0)}{}^{(K)}_{(v)}~=~\e^{\prime L}\ ,\ee
where $\e'$ belongs to either $\mhs(4)$ or $\mhs_1(4)$ as defined under \eq{residualhs(4)}. 
In particular, the bilinear sector consists of Killing matrices $\e^{\prime L}_{\underline{\a\b}}\in\msp(4;\Real)$ whose complexifications $M^L_{\underline{\a\b}}\in \msp(4;\Comp)$ obey
\bea D^{(0)}  M^{L}_{\underline{\a\b}} \ = \ 0 \ ,\eea
where $D^{(0)}=d+\O^{(0)}$ is the AdS-covariant derivative. The decomposition \cite{Didenko:2008va,Didenko:2009tc}
\bea M^L_{\underline{\a\b}} \ = \ \left(\ba{cc} \vark^L_{\a\b} & v^L_{\a\bd} \\[5pt] \bar{v}^L_{\ad\b} & \bar{\vark}^L_{\ad\bd} \ea \right) \ ,\qquad v^L_{\a\bd}~=~\bar v^L_{\bd\a}\ , \label{kv}\eea
\emph{idem} $M_{\underline{\a\b}}$ yields a complexified $AdS_4$ Killing vector $v^L_{\a\bd}(x)=\bar{v}^L_{\bd\a}(x)$ and the self-dual and anti-self-dual components $\vark^L_{\a\b}(x)$ and  $\bar{\vark}^L_{\ad\bd}(x)$, respectively, of the corresponding Killing two-form $\vark^L_{\m\n} : = \nabla^{(0)}_{\m} v^L_\n$. 

The exact solutions that we shall construct are generalized type-D and almost type-D deformations of the $AdS_4$ vacuum whose symmetry algebra is essentially given by the higher-spin (enveloping) extension of the centralizer $\mathfrak{c}_{\msp(4;\Real)}(K)$ of principal Cartan generators $K\in\msp(4;\Comp)$ that obey
\be K_{\underline{\a}}{}^{\underline\b}K_{\underline{\b\c}}~=~-C_{\underline{\a\b}}\ .\label{K2=1a}\ee
As we shall see in Section \ref{Sec:3emb}, \eq{K2=1a} selects the four cases listed in \eq{the4K}, \emph{i.e.} $K=E$, $J$, $iB$, $iP$, modulo $Sp(4;\Real)$-rotations. To the $\pi$-odd principal generators $E$ and $iP$ correspond block off-diagonal $K_{\underline{\a\b}}$ matrices ($K_{\underline{\a\b}}=-(\C_0)_{\underline{\a\b}}$ and $K_{\underline{\a\b}}=-i(\C_1)_{\underline{\a\b}}$, respectively, for which $\vark_{\a\b}=0$ and $v_{\a\bd}\neq 0$) and to the $\pi$-even generators $J$ and $iB$ block diagonal ones ($K_{\underline{\a\b}}=-(\C_{12})_{\underline{\a\b}}$ and $K_{\underline{\a\b}}=-i(\C_{03})_{\underline{\a\b}}$, respectively, for which $\vark_{\a\b}\neq 0$ and $v_{\a\bd}=0$). Using the decomposition \eq{kv}, the condition \eq{K2=1a} reads\footnote{The latter equation in \eq{K2=1cpbb} is equivalent to the statement that the corresponding $AdS_4$ Killing vector is hypersurface-orthogonal, \emph{i.e.} $v^{L}_{[\m}\nabla^{(0)}_{\n} v^L_{\r]}=0$ (see, for instance, \cite{wald}). In gravity, the hypersurface-orthogonality of a time-like Killing vector $v^\mu$ of a metric $g_{\m\n}$, \emph{i.e.} $v_{[\m}\nabla_{\n} v_{\r]}=0$, means that the otherwise stationary metric is actually static. 
As found in \cite{Didenko:2009tc}, the (consistent) truncation of unfolded vacuum gravity to the type-D sub-system, described by a free differential algebra consisting of $(e^{\a\ad};\o^{\a\b},\bar\o^{\ad\bd};K_{\underline{\a\b}})$, admits an integrating flow that deforms $v_\mu:=e_\mu{}^{\a\ad} v_{\a\ad}$ while preserving the two invariants $K^{\underline{\a\b}}K_{\underline{\a\b}}$ and $\textrm{Tr}(K^4)$ along the flow, and hence the aforementioned property of $v$ if it holds to begin with in the vacuum. }:
\bea \vark^2+v^2~=~\bar\vark^2 + \bar v^2 ~ = ~ 1 \ , \qquad
\vark_\a{}^\b v_{\b\cd} + v_\a{}^{\bd}\bar{\vark}_{\bd\cd} ~=~ 0 \ ,\label{K2=1cpbb}\eea
where 
\be v^2~:=~\ft12 v^{\a\bd}v_{\a\bd}\ ,\qquad \vark^2~:=~\ft12 \vark^{\a\b}\vark_{\a\b}\ ,\ee
\emph{idem} $\bar{\vark}$. 
Note that the condition \eq{K2=1a} (equivalently \eq{K2=1cpbb}) also holds for the $L$-rotated elements $K^L$ ($\vark^L$ and $v^L$). For a given $K$ and at points where both $(v^L)^2$ and $(\vark^L)^2$ are non-vanishing, the eigenspinors of $(\vark^L)_\a{}^\b$ define a one-function family of $K$-adapted (normalized) spin-frames $\tilde U\equiv \tilde U_{(K)}$ for ${\cal Y}$ defined by
\be (\tilde u^\pm_{\a}(x),\tilde {\bar u}^\pm_{\ad}(x))~\sim~ (e^{\pm \chi(x)} \tilde u^\pm_{\a}(x), e^{\pm (\chi(x))^\ast} \tilde {\bar u}^\pm_{\ad}(x))\ ,\ee\be (\tilde u^\pm_{\a})^\dagger~=~\tilde {\bar u}^\pm_{\ad}\ ,\quad \tilde u^{-}_{\a} \tilde u^+_{\b}-\tilde u^+_{\a} \tilde u^-_{\b}~=~\e_{\a\b}\ ,\label{spinframe2}\ee
and by 
\be \vark^L_{\a\b}~=~ 2\Theta(x) \tilde u^+_{(\a}(x) \tilde u^-_{\b)}(x)\ ,\qquad v^L_{\a\bd}~=~ \breve\Theta(x) \left(\tilde u^+_{\a}(x) \tilde {\bar u}^{-\s_\dagger\s_\pi} _{\bd}(x)+ \tilde u^-_{\a}(x) \tilde {\bar u}^{\s_\dagger\s_\pi} _{\bd}(x)\right)\ ,\label{spinframe1}\ee
\be (\Theta,\breve \Theta)^\ast~=~(-\s_\pi\Theta,\s_\dagger \breve \Theta)\ ,\qquad \Theta^2-\breve\Theta^2~=~\s_\pi\ ,\label{spinframe3}\ee
\be (\chi)^\ast~=~\s_\dagger\s_\pi \chi\ ,\ee
where $\s_\dagger$ and $\s_\pi$ are signs related to properties of $K$, $K^\dagger:=\s_\dagger K$, $\pi(K):=\s_\pi K$. The r\^ole of $\tilde L_{(K)}$ is to align the spin-frame of ${\cal Z}$ with $\tilde U$, that is (see Appendix \ref{App:L-rot} for an explicit example)
\be u^{\pm\b} (\tilde L_{(K)})_\b{}^\a~:=~\tilde u^{\pm\a}\ ,\label{alignment}\ee
where $u^\pm_\a$ is a fixed common spin-frame of ${\cal Y}$ and ${\cal Z}$.
While $(\vark^L_{\a\b},v^L_{\a\ad})$ are well-defined at any point where $L$ is well-defined, the $K$-adapted spin-frame and hence $\tilde L_K$ are well-defined only at points where both $\Theta$ and $\breve\Theta$ are non-vanishing, that is
\be {\cal R}_4~=~\left\{ x^\mu~:~ \mbox{$L$ is well-defined and $\Theta,\,\breve\Theta\neq 0$}~\right\}\ . \ee
%
%
Working in units where $\l=1$ and defining  
\be x^a~=~-\frac12 (\s^a)^{\a\ad}x_{\a\ad}\ ,\qquad x^{\a\ad}~=~(\s_a)^{\a\ad} x^a\ ,\ee
one may choose $L$ to be manifestly Lorentz-covariant leading to \cite{Bolotin:1999fa,Sezgin:2005pv,Iazeolla:2008ix}
\be  L~=~\exp_\star (4i\xi  x^a  P_a) \ =\ {2h\over 1+h} \left[\exp {4ix^a P_a\over 1+h}\right]_{\rm Weyl}\ ,\qquad x^2  ~<~  1\ ,\qquad x^2~:=~ x^a x_a\ ,\label{L}\ee
\be\xi~:=~(1-h^2)^{-\ft12}\tanh^{-1}\sqrt{\ft{1-h}{1+h}}\ ,\qquad h~:=~\sqrt{1-x^2}\ , \ee
corresponding to the following matrix representation: 
\be 
L_{\underline{\a}}{}^{\underline{\b}} ~=~ \left(\ba{cc}\cosh(2\xi\,x)\,\d_\a{}^\b & \sinh(2\xi\,x)\frac{x_\a{}^{\bd}}{x} \\[5pt] \sinh(2\xi\,x)\frac{\bar x_{\ad}{}^{\b}}{x} & \cosh(2\xi\,x)\,\d_{\ad}{}^{\bd}\ea\right)\ .\label{3.20}\ee
In the notation of Appendix \ref{App:conv}, the vacuum connection $\O^{(0)}$ consists of the $AdS_4$ vierbein $e_{(0)}{}^{\a\ad}= - h^{-2}(\s^a)^{\a\ad}dx_a$ and Lorentz connection $\o_{(0)}{}^{\a\b}=- h^{-2} (\s^{ab})^{\a\b} dx_a x_b$ corresponding to presenting the metric in stereographic coordinates as\footnote{The metric remains well-defined for $x^2>1$ such that the regions $x^2<1$ and $x^2>1$ together yield a single cover of $AdS_4$. } $ds^2_{(0)}=4 (1-x^2)^{-2}dx^2$. For relations to global embedding coordinates and global spherically-symmetric coordinates, see Appendix \ref{App:conv}. 
The resulting decompositions \eq{kv} of $K^L_{\underline{\a\b}}$ take the following forms:
\be \s_\pi=-1~: \qquad v^L_{\a\bd} ~=~ \frac{1}{h^2}\left(v_{\a\bd}-x_\a{}^{\cd}\bar v_{\cd}{}^{\c}x_{\c\bd}\right) \ , \qquad \vark^L_{\a\b} ~=~ \frac{1}{h^2}\left(x_\a{}^{\ad}\bar v_{\ad\b}-v_{\a}{}^{\ad}\bar x_{\ad\b}\right) \ ,\label{kLv}\ee
\be \s_\pi=+1~: \qquad v^L_{\a\bd} ~=~ \frac{1}{h^2}\left(x_\a{}^{\ad}\bar \vark_{\ad\bd}-\vark_{\a}{}^{\c}x_{\c\bd}\right) \ , \qquad \varkappa^L_{\a\b} ~=~ \frac{1}{h^2}\left(\vark_{\a\b}-x_\a{}^{\bd}\bar \vark_{\bd}{}^{\cd}x_{\cd\b}\right)  \ ,\label{kLk}\ee
where $h$ is defined in \eq{L}, $\s_\pi=-1$ for $K=E,iP$ and $\s_\pi=+1$ for $K=J,iB$, and $v_{\a\bd}$ and $\vark_{\a\b}$ are the $2\times 2$ blocks of the corresponding $K_{\underline{\a\b}}$ matrices given above.
Consequently, using \eq{convert}, the pseudo-norm of the $AdS_4$ Killing vector $v^L_\mu:=e_\mu^{(0)\,\a\ad} v^L_{\a\ad}$ is given by 
\be g^{\mu\nu}_{(0)}v^L_{\mu} v^L_{\nu} ~=~ - (v^L)^2~=~ \left\{\ba{c} -\frac{4}{(1-x^2)^2} \left[x^2+(x^a v_a)^2\right] -1 \ , \qquad \mbox{ $\s_\pi=-1$} \ ,\\[5pt] -\frac{1}{(1-x^2)^2} x^a\,\vark_a{}^b\,\left(\vark_{bc}+i\widetilde{\vark}_{bc}\right)x^c \ , \qquad \mbox{$\s_\pi=1$}\ ,\ea\right.\ee
where $\widetilde\vark_{ab}:=\ft12\e_{abcd}\vark^{cd}$, and $(\vark^L)^2=1-(v^L)^2$. The corresponding expressions for $\Theta$ are listed in Table \ref{Table1}. We note that for $K=E$ and $K=J$ the corresponding $AdS_4$ Killing vectors $\ft{\partial}{\partial t}$ and $\ft{\partial}{\partial\varphi}$ are globally time-like and space-like, respectively, \emph{viz.}
\bea (\ft{\partial}{\partial t})^\m (\ft{\partial}{\partial t})_\m & = & -(1+r^2) \ , \qquad (\vark^L)^2=-r^2 \\[5pt]  (\ft{\partial}{\partial\varphi})^\m (\ft{\partial}{\partial\varphi})_\m & = & r^2\sin^2\theta \ , \qquad (\vark^L)^2 \ = \ 1+r^2\sin^2\theta \ ,
\eea
here expressed in global spherically-symmetric coordinates. On the other hand, the imaginary Killing vectors corresponding to $iB$ and $iP$ have indefinite pseudo-norm, though they are time-like and space-like, respectively, at spatial infinity (of the anti-de Sitter background).
The importance of the principal Cartan generators $K$ stems from the fact that the corresponding $\vark^L_{\a\b}$ determines the spacetime behaviour of the generalized (almost) type-D Weyl tensors, as we shall see in Sections \ref{Sec:fullWeyl} and \ref{Sec:Weyl}.


\scss{Separation of twistor-space variables}\label{Sec:intansatz}


In order to separate the dependence of the internal master fields on the twistor variables $Y$ and $Z$, one may take them to be elements of the algebra
\be \Omega^{(\Sigma_{\mathfrak P})}({\cal Y}\times {\cal Z}) ~:=~\left\{ \widehat {\cal O}(Y,Z)~=~\sum_{\mathbf n,\mathbf n'}\sum_{k=0,1 }  P_{\mathbf n|\mathbf n'}(Y) \star \kappa_y^{\star k} \star \check {\cal O}_{k;\mathbf n|\mathbf n'}(Z)\right\}\ ,\label{OmegamathfrakP}\ee
where $\mathfrak P$ refers to a set of generalized projectors $P_{\mathbf n|\mathbf n'}=\pi\bar\pi(P_{\mathbf n|\mathbf n'})$ assumed to obey ($i = 1,2$)
\be P_{\mathbf n|\mathbf n'} \star P_{\mathbf m|\mathbf m'} \ = \ \delta_{\mathbf n',\mathbf m} P_{\mathbf n|\mathbf m'} \ , \label{3.31}\ee
with $(\bf n, \bf n')$ being discrete indices, and to form a set that is invariant under the operations 
$\pi$, $\dagger$ and $\tau$ and $\star$-multiplication by $ \kappa_y\bar \kappa_{\yb}$, such that 
\be \pi(P_{\mathbf n|\mathbf n'})~=:~P_{\pi(\mathbf n)|\pi(\mathbf n')}\ ,\qquad
(P_{\mathbf n|\mathbf n'})^\dagger~=:~P_{I(\mathbf n')|I(\mathbf n)}\ ,\qquad 
\tau(P_{\mathbf n|\mathbf n'})~=:~P_{\t(\mathbf n')|\t(\mathbf n)}\ ,\label{indexmaps}\ee 
\be P_{\mathbf n|\mathbf n'}\star \kappa_y\bar \kappa_{\yb}~=:~ \kappa_{\mathbf{n}'} P_{\mathbf n|\mathbf n'}\ ,\label{phasephi}\ee
with $\pi^2({\bf n})=I^2({\bf n})=\t^2({\bf n})=n$ and $(\kappa_{\mathbf{n}})^2=1$. 
For explicit realizations of generalized-projector algebras, see Appendix \ref{AppProj}; in particular, for the proof of associativity, see the discussion below Eq. \eq{recort}. The binary product rule in $\Omega^{(\Sigma_{\mathfrak P})} $ takes the form
\be \widehat {\cal O}\star \widehat {\cal O}'~=~ \sum_{\mathbf n,\mathbf n'}
P_{\mathbf n|\mathbf n'} \star\left( \left(\check {\cal O}_0\star \check {\cal O}'_0+\check {\cal O}_1\star \pi_y(\check {\cal O}'_1)\right)_{{\bf n}|{\bf n}'}+\kappa_y\star\left(\check {\cal O}_0\star \check {\cal O}'_1+\check {\cal O}_1\star \pi_y(\check {\cal O}'_0)\right)_{{\bf n}|{\bf n}'}\right) \ ,\ee
using the matrix notation 
\be (\check F(Z)\star \check F'(Z))_{{\bf n}|{\bf n}'}~:=~ \sum_{\bf m}(\check F(Z))_{{\bf n}|{\bf m}}\star (\check F'(Z))_{{\bf m}|{\bf n}'}\ ,\qquad ({\mathbf 1})_{{\bf n}|{\bf n}'}~:=~\delta_{{\bf n},{\bf n}'}\ ,\ee\be   (\pi_y(\check F(Z)))_{{\bf n}|{\bf n}'}~:=~(\check F(Z))_{\pi({\bf n})|\pi({\bf n}')}\ ,\qquad (\pi_z(\check F(Z)))_{{\bf n}|{\bf n}'}~:=~\pi((\check F(Z))_{{\bf n}|{\bf n}'})\ .\ee
This composition rule is associative provided that $\check {\cal O}_{k;\mathbf n|\mathbf n'}(Z)$ belong to an associative $\star$-product algebra.
In what follows, the latter algebra shall in addition be assumed to remain closed under $\star$-multiplication by $\kappa_z$ and $\bar \kappa_{\bar z}$. 

To construct a shell within $\Omega^{(\Sigma_{\mathfrak P})}({\cal Y}\times {\cal Z})$, one first expands
\bea \widehat\Phi'&=& \sum_{\mathbf n,\mathbf n' }  P_{\mathbf n|\mathbf n'}(Y)  \star \kappa_y \star \check \Phi_{\mathbf n|\mathbf n'}(Z)\ ,\qquad \pi\bar\pi(\check \Phi_{\mathbf n|\mathbf n'})~=~\check \Phi_{\mathbf n|\mathbf n'}\ ,\label{3.6}\\[5pt]
\widehat S_{\underline \a}'&=& Z_{\underline \a}-2i \sum_{\mathbf n,\mathbf n' }  P_{\mathbf n|\mathbf n'} (Y) \star (\check V_{\underline\a})_{\mathbf n|\mathbf n'}(Z)\ ,\qquad \pi\bar\pi((\check V_{\underline\a})_{\mathbf n|\mathbf n'})~=~-(\check V_{\underline\a})_{\mathbf n|\mathbf n'}\ ,\eea
The reality condition \eq{reality} requires 
\be\sum_{\mathbf n,\mathbf n'} (P_{\mathbf n|\mathbf n'})^\dagger\,\star\,(\check \Phi_{\mathbf n|\mathbf n'})^\dagger ~=~  \sum_{\mathbf n,\mathbf n'} P_{\mathbf n|\mathbf n'}\,\star\kappa_y\bar\k_{\bar y} \,\star\,\pi(\check \Phi_{\mathbf n|\mathbf n'})\ ,\ee
\be \sum_{{\bf n},{\bf n}'} (P_{{\bf n}|{\bf n}'})^\dagger\star ((\check V_\a)_{{\bf n}|{\bf n}'})^\dagger~=~- \sum_{{\bf n},{\bf n}'} P_{{\bf n}|{\bf n}'}\star (\check{\bar V}_{\ad})_{{\bf n}|{\bf n}'}\ ,\ee
that is, 
\be (\check \Phi_{\mathbf n|\mathbf n'})^\dagger \ = \   \kappa_{I(\mathbf n)}\,\pi(\check \Phi_{I(\mathbf n')|I(\mathbf n)})\ ,\qquad ((\check V_\a)_{{\bf n}|{\bf n}'})^\dagger~=~- (\check{\bar V}_{\ad})_{I({\bf n}')|I({\bf n})}\ .\label{herm1}\ee
In the minimal-bosonic case, it follows from $\tau(\k_y)= -\k_y$ that \eq{min} requires 
\bea \tau(\check \Phi_{\mathbf n|\mathbf n'}) \ = \ -\pi(\check \Phi_{\t(\mathbf n')|\t(\mathbf n}))\ , \qquad  \tau((\check V_{\underline\a})_{\mathbf n|\mathbf n'}) \ = \ -i(\check V_{\underline\a})_{\t(\mathbf n')|\t(\mathbf n)} \ ,\label{tau1}\eea
implying that the range of $(\bf n,\bf n')$ be symmetric under overall sign inversion. 
Defining
\be (\check \Sigma_{\underline\a})_{{\bf n}|{\bf n}'}~:=~ \delta_{{\bf n},{\bf n}'}Z_{\underline\a}-2i (\check V_{\underline\a})_{\mathbf n|\mathbf n'}\ ,\qquad (\check \kappa_z)_{\mathbf{n}|\mathbf{n'}}~:=~ \delta_{\mathbf{n},\mathbf{n'}}\kappa_z \ , \qquad (\check{\bar\kappa}_{\bar{z}})_{\mathbf{n}|\mathbf{n'}}  ~:=~ \delta_{\mathbf{n},\mathbf{n'}}\kappa_{\mathbf{n}}{\bar\kappa}_{\bar{z}}\ee
\be {\cal B}~\equiv~\exp_\star i\theta[\widehat \Phi\star\pi(\widehat \Phi)]~=:~\sum_{{\bf n},{\bf n}'} P_{{\bf n}|{\bf n}'}\star \check {\cal B}_{{\bf n}|{\bf n}'}\ ,\ee
the factorization property \eq{factorizekappa} implies that the Ansatz must obey the matrix equations
\bea &\check \S_\a \star \check \Phi +\check \Phi \star \pi_z(\check \S_\a)\ =\ 0\ ,\qquad \check{\bar \S}_{\ad}\star \check \Phi +\check \Phi\star \bar\pi_{\zb}(\check {\bar \S}_{\ad})\ =\ 0\ ,\\[5pt]&[\check \S_\a,\check \S_\b]_\star\ =\ -2i\e_{\a\b}(\mathbf{1}-\check {\cal B}\star \check \Phi \star \check\kappa_z)\ ,\qquad [\check {\bar \S}_{\ad},\check{\bar \S}_{\bd}]_\star\ =\ -2i\e_{\ad\bd}({\mathbf 1}-{\check {{\bar {\cal B}}}}\star 
\check\Phi \star \check{\bar\kappa}_{\zb})\ ,&\\[5pt]
& [\check \S_\a,\check {\bar \S}_{\bd}]_\star\ =\ 0\ .&\eea
Expanding also $\widehat \e\,'(Y,Z)=\sum_{{\bf n},{\bf n}'} P_{{\bf n}|{\bf n}'}(Y)\star \check\e_{{\bf n}|{\bf n}'}(Z)$, the gauge transformations take the matrix form
\be \delta_{\check \e} \check\Phi~=~-[\check\e,\check \Phi]_{\pi_z}\ ,\qquad \delta_{\check\e}\check \S_{\underline\a}~=~-[\check\e,\check\S_{\underline\a}]_\star\ .\ee
The space of solutions covered by the Ansatz contains a subspace\footnote{Whether or not its complement is non-trivial remains to be investigated.} consisting of the gauge orbits reached from diagonal solutions obeying
\be \check \Phi_{\mathbf n|\mathbf n'}(Z)~=~\delta_{\mathbf n,\mathbf n'} \,\Phi_{\bf n}(Z)\ .\label{3.51}\ee
In the perturbative sector, this implies that 
\be (\check V_{\underline\a})_{\mathbf n|\mathbf n'}(Z)~=~\delta_{\mathbf n,\mathbf n'}V_{\underline\a}^{\mathbf n}(Z)\ ,\qquad  (\check \Sigma_{\underline\a})_{\mathbf n|\mathbf n'}(Z)~=~\delta_{\mathbf n,\mathbf n'}\Sigma_{\underline\a}^{\mathbf n}(Z)\ ,\label{3.24}\ee
modulo gauge artifacts. For diagonal solutions we shall use the notation 
\be P_{\bf n}~:=~P_{{\bf n}|{\bf n}}\ ,\qquad \check {\cal B}_{{\mathbf n}|{\mathbf n}'}~=:~\delta_{\mathbf n,\mathbf n'} \,{\cal B}_{\bf n}\ .\label{diagansatz2}\ee
The resulting partially gauge-fixed equations of motion read
\bea &\S^{\mathbf{n}}_\a\star \Phi_{\mathbf{n}} +\Phi_{\mathbf{n}}\star \pi_z(\S^{\mathbf{n}}_\a)\ =\ 0\ ,\quad \bar \S^{\mathbf{n}}_{\ad}\star \Phi_{\mathbf{n}} +\Phi_{\mathbf{n}}\star \bar\pi_{\zb}(\bar \S^{\mathbf{n}}_{\ad})\ =\ 0\ ,\\[5pt]&[\S^{\mathbf{n}}_\a,\S^{\mathbf{n}}_\b]_\star\ =\ -2i\e_{\a\b}(1-{\cal B}_{\bf n}\star \Phi_{\mathbf{n}}\star \kappa_z)\ ,\\[5pt]& [\bar \S^{\mathbf{n}}_{\ad},\bar \S^{\mathbf{n}}_{\bd}]_\star\ =\ -2i\e_{\ad\bd}(1-\kappa_{\bf n}\bar{\cal B}_{\bf n}\star \Phi_{\mathbf{n}}\star \bar\kappa_{\zb})\ ,&\\[5pt]
& [\S^{\mathbf{n}}_\a,\bar \S^{\mathbf{n}}_{\bd}]_\star\ =\ 0\ .&\eea
Perturbative expansion in the initial datum 
\be \nu_{\mathbf n}~:=~\Phi_{\mathbf{n}}|_{Z=0}\ ,\ee
and taking all gauge artifacts that are either $\pi$-odd or non-holomorphic to vanish, leads to a holomorphic Ansatz obeying 
\be \Phi_{\mathbf{n}}~=~ \nu_{\mathbf{n}}\ ,\ee\be \bar\partial_{\ad}\S^{\mathbf{n}}_\a~=~ 0\ ,\qquad \S^{\mathbf{n}}_\a~=~-\pi_z(\S^{\mathbf{n}}_\a)\ ,\qquad \partial_\a \bar \S^{\mathbf{n}}_{\ad}~=~ 0\ ,\qquad \bar \S^{\mathbf{n}}_{\ad}~=~- \bar\pi_{\zb}(\bar \S^{\mathbf{n}}_{\ad})\ ,\label{defosc1}\ee\be [\S^{\mathbf{n}}_\a,\S^{\mathbf{n}}_\b]_\star~=~ -2i\e_{\a\b}(1-{\cal B}_{\bf n} \nu_{\mathbf{n}} \kappa_z)\ ,\qquad  [\bar \S^{\mathbf{n}}_{\ad},\bar \S^{\mathbf{n}}_{\bd}]_\star~=~ -2i\e_{\ad\bd}(1-\kappa_{\bf n}\bar {\cal B}_{\bf n} \bar \nu_{\mathbf{n}} \bar\kappa_{\zb})\ ,\label{defosc2}\ee
which are defined modulo the residual holomorphic gauge transformations
\be \delta_{\e_{\mathbf{n}}}\Sigma^{\mathbf{n}}_\a~=~[\Sigma^{\mathbf{n}}_\a,\e^{\mathbf{n}}]_\star\ ,\qquad  \delta_{\bar\e_{\mathbf{n}}}\bar\Sigma^{\mathbf{n}}_{\ad}~=~[\bar\Sigma^{\mathbf{n}}_{\ad},\bar\e^{\mathbf{n}}]_\star\ ,\label{hologauge}\ee
\be \bar\partial_{\ad}\e^{\mathbf{n}}~=~0\ ,\qquad \partial_{\a}\bar \e^{\mathbf{n}}~=~0\ .\ee


\scss{Deformed oscillators}\label{Sec:defosc}


The deformed oscillators $(\Sigma^{\mathbf{n}}_\a(z),\bar{\Sigma}^{\mathbf{n}}_{\ad}(\bar z))$ defined by \eq{defosc1} and \eq{defosc2} can be obtained explicitly by adapting the $\circ$-product method of \cite{Prokushkin:1998bq}, later refined in \cite{Sezgin:2005pv} (see also \cite{Iazeolla:2007wt}), resulting in the following two steps: 

\begin{itemize} 
\item[i)] using a spin-frame $u^\pm_\a$ to split ($u^{\a+}u^-_\a=1$)
\be \Sigma^{\mathbf{n}}_\a(z)~=~ u^-_\a \Sigma^{\mathbf{n}+}(z)-u^+_\a \Sigma^{\mathbf{n}-}(z)\ ,\qquad [\Sigma^{\mathbf{n}-},\Sigma^{\mathbf{n}+}] _\star~=-2i(1-{\cal B}_{\mathbf{n}}\nu_{\mathbf{n}} \k_z)\ ,\label{defoscpm}\ee
and representing the Weyl-ordered\footnote{For the (anti-)normal-ordered forms of the deformed oscillators, see Appendix \ref{App:D}.} symbols $(\left[\Sigma^{\mathbf{n}}_\a(z)\right]^{\rm Weyl},\left[\bar{\Sigma}^{\mathbf{n}}_{\ad}(\bar z)\right]^{\rm Weyl})$ by the generalized Laplace transforms ($z^\pm:=u^{\pm\a}z_\a$, $w_z:=z^+z^-$, $[z^-,z^+]_\star = -2i $)
\be \left[\Sigma^{\mathbf{n}\pm}\right]^{\rm Weyl} ~\equiv~ u^{\pm\,\a}\Sigma_\a^{\mathbf{n}\pm} ~=~ 4z^{\pm}\int_{-1}^1 \frac{dt}{(t+1)^2}\,f^{\mathbf{n}\pm}_{\s_{\mathbf{n}}}(t)  \,e^{i\s_{\mathbf{n}}\ft{t-1}{t+1} w_z}\ ,\label{WSigmaansatz}\ee
where $(\s_{\mathbf{n}})^2=1$ can be chosen independently for each $\mathbf{n}$, and 
\be f^{\mathbf{n}\pm}_{\s_\mathbf{n}}(t) ~:=~ \delta(t-1)+ j^{\mathbf{n}\pm}_{\s_\mathbf{n}}(t)\ee
obey the integral equations ($\left[\kappa_z\right]^{\rm Weyl}=2\pi \delta^2(z)$) 
\bea 4\int_{-1}^1 dt\int_{-1}^1 dt'\,\frac{f^{\mathbf{n}-}_\s(t)f^{\mathbf{n}+}_\s(t')}{(tt'+1)^2}\,\left[1+i\s\frac{tt'-1}{tt'+1}w_z \right]\,e^{i\s\ft{tt'-1}{tt'+1}w_z} ~=~1-{\cal B}_{\mathbf{n}}\nu_{\mathbf{n}} \left[\k_z\right]^{\rm Weyl}\ ;\label{stepi}\eea
\item[ii)] inserting $1=\int_{-1}^1 du\,\d(tt'-u)$ into the left-hand side of \eq{stepi} and changing order of integration, using  
\bea  (h_1 \circ h_2)(u) ~:=~ \int_{-1}^1 dt\int_{-1}^1 dt'\,h_1(t)\,h_2(t')\,\d(tt'-u)\ ,\label{ringo}\eea
which defines a commutative and associative product on the space of functions on the unit interval, and the representation 
\be \lim_{\varepsilon\rightarrow 0}\frac{1}{\varepsilon}e^{-i\s\ft1\varepsilon w_z}~=~\s \left[\k_z\right]^{\rm Weyl}\ , \label{deltalimit}\ee
yields the integral equations
\be  (f^{\mathbf{n}-}_\s\circ f^{\mathbf{n}+}_\s)(t) ~=~ \d(t-1)-\frac{\s_{\mathbf{n}} {\cal B}_{\mathbf{n}}\n_{\mathbf{n}}}{2} \ ,\label{ringeq}\ee
with the following solution space for each value of $\mathbf n$ (for details, see Appendix \ref{App:D}):
\be f^\pm_\s(t)~=~ g^{\circ(\pm1)}_\s \circ f_\s\ ,\qquad f_\s~=~\delta(t-1)+j_\s(t)\ ,\label{j2}\ee
\be j_\s(t)~=~q_\s(t)+\sum_{k=0}^\infty \l_{\s,k} p_k(t)  \ ,\qquad q_\s(t)~=~-\frac{\s{\cal B}\nu}{4}{}_1F_1\left[\frac{1}{2};2;-\frac{\s{\cal B}\nu}{2}\log t^{2}\right]\ ,\label{j}\ee
where $g_\sigma$ is a gauge artifact and we use the notation $g^{\circ(+1)}=g$ and $g^{\circ(-1)}\circ g=1$; $p_k(t):=\ft{(-1)^k}{k!}\,\d^{(k)}(t)$ act as projectors in the $\circ$-product algebra; and $\l_k$ are given by \eq{lambdak} and \eq{Lkq}.
\end{itemize}
The first step relies on the fact that if sources $\r^\a$ are used to write (in what follows we suppress $\mathbf n$)
\bea \left[\Sigma^{\pm}_{\s}\right]^{\rm Weyl} \ = \ -4i\frac{\partial}{\partial\r^{\pm}} \int_{-1}^1 \frac{dt}{t+1} \,f^{\pm}_\s(t)\, \left. e^{\ft{i}{t+1} \left(\s(t-1)w_z +\r^+z^+ +\r^-z^-\right)}\right|_{\r^\pm=0} \ , \label{WSigma} \eea
then the space of generalized Laplace transforms over $[-1,1]$ with sources is closed under the $\star$-product, as can be seen from the following $\star$-product formula in the Weyl-order:
\bea & \frac{1}{t+1}\,e^{\ft{i}{t+1} \left(\s (t-1) w_z +\r^+z^+ +\r^-z^-\right)}\,\star \,\frac{1}{t'+1}\, e^{\ft{i}{t'+1} \left(\s (t'-1)w_z +\r^{\prime+}z^+ +\r^{\prime-}z^-\right)}  &\nn\\[5pt]  &  \ = \  \frac{1}{2(\tilde{t}+1)} \,e^{\ft{i}{\tilde{t}+1}\left(\s(\tilde{t}-1)w_z +\tilde{\r}^+z^+ +\tilde{\r}^-z^--\ft12\r^+\r^{\prime-}+\ft12\r^-\r^{\prime+}-\s \frac{t'-1}{t+1}\r^+\r^--\s\frac{t-1}{t'+1}\r^{\prime +}\r^{\prime -}\right)} \ ,&  \label{selfrep} \eea
\be \tilde{t}~:=~tt'\ ,\qquad \tilde{\r}^\pm~:=~\ft{(t'-1)(1\mp\s)+2}2\r^\pm+\ft{(t-1)(1\pm\s)+2}2\r^{\prime \pm}\ ,\label{tildet}\ee
where thus the induced map $(t,t')\in[-1,1]\times[-1,1]\rightarrow \tilde{t}\in[-1,1]$. 
In particular,
 \bea \left[\frac1{(t+1)^{2}} z^-\,e^{i\s \ft{t-1}{t+1}w_z}\, ,\, \frac1{(t'+1)^{2}}z^+\,e^{i\s \ft{t'-1}{t'+1}  w_z}\right]_\star  \ = \  -\frac{i}{2(\tilde{t}+1)^2}\,\left(1+i\s \frac{\tilde{t}-1}{\tilde{t}+1}w_z\right)\, e^{i\s \ft{\tilde{t}-1}{\tilde{t}+1}w_z}  \ . \eea
In the second step, letting $h(u) : = (f^-_\s\circ f^+_\s)(u)$, Eq. \eq{stepi} implies 
\bea 4\int_{-1}^1\frac{du}{(u+1)^2}\,h(u)\left[1+i\s\,\frac{u-1}{u+1}w_z\right]\,e^{i\s\ft{u-1}{u+1}w_z} ~=~1-2\pi {\cal B}\nu \delta^2(z)\ ,\label{penult}\eea
that in view of \eq{deltalimit} admits the unique solution 
\be h(u)~=~\d(u-1)-\ft{\s {\cal B}\n}2\ ,\ee
such that Eq. \eq{stepi} is equivalent to the $\circ$-product equation \eq{ringeq}, which is solvable essentially due to the commutative and associative nature of $\circ$. 
We also note that the presentation \eq{deltalimit} is compatible with $\kappa_z\star f(z)=f(-z)\star \kappa_z$, $\kappa_z\star\kappa_z=1$, $\tau(\kappa_z)=-\kappa_z$ and $\kappa_y\star \kappa_z=\widehat \kappa$, and that the fact that $g_\s$ contains gauge artifacts follows by using holomorphic gauge parameters in \eq{hologauge} of the form  
\be \e_\s(z)~=~\int_{-1}^1 \frac{dt}{1-t^2}\breve{\e}_\s(t) e^{i\s\ft{t-1}{t+1} w_z}\ ,\ee
which induce 
\be \delta_{\e_\s} f^\pm_\sigma(t)~=~\pm \ft{\s}{2} (\breve{\e}_\s\circ f^\pm_\sigma)(t)\ .\ee

\scss{Three inequivalent embeddings of complexified Heisenberg algebras}\label{Sec:3emb}

As we shall see below, generalized projectors obeying \eq{3.31} with $\mathbf n = (n_1,n_2)\in (\mathbb Z+\frac12)\times (\mathbb Z+\frac12)$ can be obtained by taking 
\be (w_i-n_i)\star P_{\mathbf n|\mathbf n'} \ = \ 0 \ , \qquad  P_{\mathbf n|\mathbf n'}\star (w_{i}-n'_i) \ = \ 0 \ , \label{3.8}\ee
where $w_i$ are the shifted number operators of the mutually-commuting complexified Heisenberg algebras ($i,j=1,2$, $\ve,\ve'=\pm$, $\e^{-+}=1$)
\be [y^{\ve}_i,y^{\ve'}_j]_\star~=~\e^{\ve\ve'}\delta_{ij}\ .\label{HA}\ee
The latter can be realized as ($\s,\s'=\pm$)
\be y^\ve_{1}~=~(2i)^{-\ft12}U^{\underline\a}_1\, Y_{\underline\a}^{\ve,\ve}\ ,\qquad y^\ve_{2}~=~(2i)^{-\ft12}U^{\underline\a}_2\, Y^{\ve,-\ve}_{\underline\a}\ ,\qquad Y^{\s,\s'}_{\underline\a}~:=~ (\Pi^{\s,\s'})_{\underline\a}{}^{\underline\b}~Y_{\underline\b}\ , \label{3.39}\ee
using rank-one projectors $(\Pi^{\s,\s'})_{\underline{\a\b}}:=(\Pi^{\s}_{(+)})_{\underline\a}{}^{\underline\c}(\Pi^{\s'}_{(-)})_{\underline{\c\b}} = - (\Pi^{-\s,-\s'})_{\underline{\b\a}}$ given by products of commuting rank-two projectors ($q = \pm $)
\be
(\Pi^{\s}_{(q)})_{\underline{\a\b}} ~:=~ \frac{1}{2}\left(C_{\underline{\a\b}}+i\s K^{(q)}_{\underline{\a\b}}\right) ~=~-(\Pi^{-\s}_{(q)})_{\underline{\a\b}}\ , \ee
where $K^{(q)}_{\underline{\a\b}}  =  K^{(q)}_{\underline{\b\a}}\in \msp(4;\Comp)$ obey
\bea K^{(q)}_{\underline\a}{}^{\underline\c}\,K^{(q)}_{\underline\c}{}^{\underline\b}~=~-\delta_{\underline\a}^{\underline\b}\ ,\label{K2=1}\qquad [K^{(q)},K^{(q')}]_{\underline{\a\b}}~=~0\ .\label{k2=1}\eea
The commutation relations between the one-dimensional oscillators obtained above read
\bea [Y^{\s,\s'}_{\underline\a},Y^{\t,\t'}_{\underline\b}]_\star \ = \ 2i\,\d_{\s,-\t}\,\d_{\s',-\t'}\,(\Pi^{\s,\s'})_{\underline{\a\b}} \ ,\eea
whose independent components are $[Y^{-,-}_{\underline\a},Y^{+,+}_{\underline\b}]_\star=2i(\Pi^{-,-})_{\underline{\a\b}}$ and $[Y^{-,+}_{\underline\a},Y^{+,-}_{\underline\b}]_\star=2i(\Pi^{-,+})_{\underline{\a\b}}$. The corresponding shifted number operators can thus be written as
\bea N^{\s,\s'} & = & \frac{1}{2i}\,(\Pi^{\s,\s'})_{\underline{\a\b}}\,Y^{(\underline\a} \star Y^{\underline\b)} \ = \ \frac{1}{2i}\,C^{\underline{\a\b}} \,Y^{\s,\s'}_{\underline{\b}}\star Y^{-\s,-\s'}_{\underline\a} \ = \ -N^{-\s,-\s'}\ , \\[5pt]
[N^{\s,\s'},Y^{\t,\t'}_{\underline\a}]_\star & = & \d_{\s\t}\,\d_{\s'\t'}\,Y^{\s,\s'}_{\underline\a} - \d_{\s,-\t}\,\d_{\s',-\t'}\,Y^{-\s,-\s'}_{\underline\a} \ ,\eea
and one can identify related Cartan sub-algebras 
\be \mh~:=~\{K_{(+)},K_{(-)}\}\in \msp(4;\Comp)\ ,\qquad K_{(q)} ~ = ~ \frac{1}{8}\,K^{(q)}_{\underline{\a\b}}\,Y^{\underline\a} \star Y^{\underline\b}\ ,\ee
as 
\be K_{(q)}  ~:=~ \frac{1}{2}\,(w_2+ qw_1)\ ,\qquad w_1~:=~N^{+,-}\ ,\qquad w_2~:=~N^{+,+}  \ .\label{Mq}\ee
Using the basis $M_{AB}=(M_{AB})^\dagger$ defined in \eq{MAB} and $(\C_{AB})_{\underline{\a}}{}^{\underline\b}(\C_{AB})_{\underline{\b\c}}=-\eta_{AA}\eta_{BB}C_{\underline{\a\c}}$, one finds that \eq{k2=1} admits the following solutions modulo $Sp(4;\Real)$ rotations:
\be \mh~=~ \{E,J\}\ ,\qquad \mh~=~ \{J,iB\}\ ,\qquad \mh~=~\{iB,iP\}\ ,\label{csa}\ee
where $E:=P_0=M_{0'0}$, $J:=M_{12}$, $B:= M_{03}$ and $P:=M_{0'1}=P_1$.

\scss{Projectors and kinematical conditions on deformation parameters}\label{Sec:kinematic} 

For a given $\mh\cong \mso(2)_{(+)}\oplus \mso(2)_{(-)}$ with generators $K_{(\pm)}=\ft12(w_2\pm w_1)$, a set of projectors\footnote{The $\star$-product formalism refers \emph{a priori} to bi-modules rather than separate left- and right-modules; the latter types of modules can be introduced by associating the complexified Heisenberg algebras to state spaces
\be{\cal F}^{\mh}_i~=~({\cal F}^{\mh}_i)^{+}\oplus ({\cal F}^{\mh}_i)^-\ ,\qquad ({\cal F}^{\mh})_i^{\ve} \ = \ \left\{\ket{\ve n}_i~:=~\ft{(y^{\ve})^{n-\ft12}}{\sqrt{(n-\ft12)!}}\star\ket{ \ft{\ve}2}_i\,\right\}_{n\, \in\, \mathbb N+\ft12}\ ,\nn\ee
where $y^\ve_i\star\ket{-\ft{\ve}2}_i:=0$ define the (anti-)ground state of the (anti-)Fock space and $(w_i-n_i)\star\ket{n_i}_i=0$.
The resulting total state space ${\cal F}^{\mh}:={\cal F}^{\mh}_1\otimes {\cal F}^{\mh}_2$ thus decomposes under $\msp(4;\Comp)$ into 
\be {\cal F}^{\mh}~=~ \bigoplus_{\ve_1,\ve_2=\pm} ({\cal F}^{\mh})^{\ve_1,\ve_2}\ ,\qquad ({\cal F}^{\mh})^{\ve_1,\ve_2}~:=~({\cal F}^{\mh}_1)^{\ve_1}\otimes ({\cal F}^{\mh}_2)^{\ve_2}\ .\nn\ee
Introducing dual spaces $({\cal F}^{\mh}_i)^\ast$ consisting of states ${}_i\bra{n}$ obeying ${}_i\langle n|\star|m\rangle_i =\delta_{mn}$, generated from dual ground states obeying ${}_i\bra{\ft{\ve}2}\star y^\ve_i=0$, one may realize ($\ve_i:=n_i/|n_i|$)
\be P_{n_1,n_2}~=~\ket{n_1,n_2}\bra{n_1,n_2} ~=~\frac{(\ve_1)^{n_1+\ft12}(\ve_2)^{n_2+\ft12}}{(|n_1|-\ft12)! (|n_2|-\ft12)!} (y^{\ve_1}_1)^{|n_1|-\ft12}(y^{\ve_2}_2)^{|n_2|-\ft12}\star \ket{\ft{\ve_1}2,\ft{\ve_2}{2}}\bra{\ft{\ve_1}2,\ft{\ve_2}{2}}\star (y^{-\ve_1}_1)^{|n_1|-\ft12}(y^{-\ve_2}_2)^{|n_2|-\ft12}\ ,\nn\ee
which can be converted into a proper $\star$-product realization by first converting $\ket{\ft{\ve_1}2,\ft{\ve_2}{2}}\bra{\ft{\ve_1}2,\ft{\ve_2}{2}}$.}  
$P_{n_1,n_2}(w_1,w_2)$ obeying ($n_i\in \integ+\ft12$)
\be P_{n_1,n_2}\star P_{n_1',n_2'} \ = \ \d_{n_1 n_1'}\d_{n_2 n_2'}P_{n_1,n_2} \ , \qquad (w_i-n_i) \star  P_{n_1,n_2} \ = \ 0 \ , \label{ortdiag}\ee
are given by products of two (commuting) sets of projectors, \emph{viz.} $P_{n_1,n_2}=P_{n_1}(w_1)\star P_{n_2}(w_2)$, with auxiliary closed-contour integral realization \eq{intproj} subject to the prescription that $\star$-products are to be performed prior to the auxiliary integrals. As shown in Appendix \ref{AppProj}, this regular presentation ensures the orthogonality conditions in \eq{ortdiag} simply via a change of variable (see Eq. \eq{recort}) while preserving associativity.
More precisely, if $\ve_1\ve_1'=1=\ve_2\ve_2'$ then the auxiliary integrals in the quantity $P_{n_1,n_2}\star P_{n_1',n_2'}$ can be performed before the $\star$-product, and the projectors can hence be presented without the former as in \eq{intproj} and \eq{enhanced}. 
On the other hand, if $\ve_1\ve_1'=-1$ or $\ve_2\ve_2'=-1$ then the non-integral presentation leads to a divergent $\star$-product. 
The divergence can be traced back to the one arising in the $\star$-product $\left[2e^{-2w_i}\right]_{\rm Weyl}\star \left[2e^{2w_i}\right]_{\rm Weyl}$ between the non-integral presentations of the ground-state and anti-ground-state projectors, as can be seen from Eq. \eq{swtw} for $s=1=-s'$. 

The projectors $P_{n_1,n_2}$ have rank one in the sense that $Tr(P_{n_1,n_2})=1$. With the exception of the (anti-)ground-state projectors ($q=\ve_1\ve_2$)
\be P_{\ft{\ve_1}2,\ft{\ve_2}2}~=~ {\cal P}_{\ve_2}(K_{(q)}) ~=~ \left[4 \exp\left(\mp \frac{\ve_2}{2}\, Y^{\underline\a}\, K^{(q)}_{\underline{\a\b}}\,Y^{\underline\b} \right)\right]_{\rm Weyl}~=~\left[4\,e^{\mp 4\ve_2K_{(q)}}\right]_{\rm Weyl}\ ,\ee
which depend only on the principal Cartan generator, the projectors $P_{n_1,n_2}$ depend on both $K_{(+)}$ and $K_{(-)}$ and are hence $\mh$-invariant. We refer to the latter and to the solutions built on them as being biaxially symmetric (or axisymmetric) in the sense that they are invariant under two commuting rotations in the five-dimensional embedding space.
The rank-$|n|$ projectors ${\cal P}_{n}(K_{(q)})$ ($n\in\{\pm1,\pm2,\dots\}$) are given by the sum of $|n|$ rank-one projectors as in \eq{ranknproj} and have regular presentations given by the integral realization \eq{enhanced}. They depend only on the principal Cartan generator $K_{(q)}$ and are therefore invariant under the centralizer $\mathfrak{c}_{\msp(4;\Comp)}(K_{(q)})\cong \mso(2;\Comp)\,\oplus \,\mso(3;\Comp)$. We shall refer to these projectors and to the solutions built on them as being symmetry-enhanced axisymmetric solutions, or just symmetry-enhanced for simplicity.

The phase-factors $\kappa_{\bf n}$ defined in \eq{phasephi} are given by 
\be \kappa_{\bf n}~=~(-1)^{|n_1|+|n_2|-1} \kappa(K_{(\ve_1\ve_2)})\ ,\ee
where $\kappa(K_{(\ve_1\ve_2)})$ are collected in Table \ref{Table1}. 
{\footnotesize \tabcolsep=1mm \begin{table}[t]
\centering\begin{tabular}{|l|l|l|l|l|l|l|}\hline
\ $K$ \ & \ $\s_\pi(K)$ \ & \ $\s_\dagger(K)$ \ & \ $\kappa(K)$ \
& \ $\mathfrak{c}_{\msp(4;\Real)}(K)\cong\mso(2)\oplus \mg_3$ \ & $\Theta(X^A)$ & $M_2(\mg_3)$\\\hline
${E=M_{0'0}}$ & $-$ & $+$& $-1$ & \mbox{\footnotesize $
\mso(2)_{E}\,\oplus\, \mso(3)_{M_{rs}}$}& $\sqrt{(X^1)^2+(X^2)^2+(X^3)^2}$ & $S^2$
\\ ${J=M_{12}}$ & $+$&$+$
& $+1$ &  \mbox{\footnotesize $\mso(2)_{J}\,\oplus\,\mso(2,1)_{\{E,P_3,M_{03}\}} $} & $i\sqrt{(X^0)^2+(X^{0'})^2-(X^3)^2}$ & $AdS_2$\\
${iB=iM_{03}}$ & $+$ &$-$ &
$-1 $ & \mbox{\footnotesize $\mso(2)_{B}\,\oplus\,\mso(2,1)_{\{M_{12},P_1,P_2\}}$} & $i\sqrt{(X^{0'})^2-(X^1)^2-(X^2)^2}$ & $H_2, dS_2$ \\
${iP=iM_{0'1}} $&$ -$ &$-$ & $+1$ &\mbox{\footnotesize $\mso(2)_{P}\,\oplus\,\mso(2,1)_{\{M_{02},M_{03},M_{23}\}}$}&$\sqrt{(X^0)^2-(X^2)^2-(X^3)^2}$ & $H_2, dS_2$ \\
\hline
\end{tabular}
\caption{{\footnotesize Properties of ground-state projectors.
The signs $\s_\pi$ and $\sigma_\dagger$ are defined above Eq. \eq{spinframe1} and the signs $\kappa(K)$ are defined by $4\, e^{\mp 4K}\star \kappa_y\bar\kappa_{\yb}=\kappa(K) e^{\mp 4K}$ and evaluated in Appendix \ref{AppProj} using Gaussian integration. The centralizers leave $\Theta$ invariant, as becomes manifest in the global embedding coordinates $X^A$ obeying $\eta_{AB} X^A X^B=-1$. In global spherical coordinates, one has $\sqrt{(X^1)^2+(X^2)^2+(X^3)^2}=r$ and $\sqrt{(X^0)^2+(X^{0'})^2-(X^3)^2}=\sqrt{1+r^2\sin^2 \th}$ (see Appendix \ref{App:conv}). The manifolds $M_2(\mg_3)$ are two-dimensional maximally-symmetric foliates with rank-$3$ isometry algebras $\mg_3$. While the foliations are unique for the solutions with principal Cartan generators $E$ and $J$ (with corresponding Killing vectors having definite pseudo-norm everywhere), the solutions based on $iB$ and $iP$ generators (with Killing vector fields having indefinite pseudo-norm, see Section \ref{gaugefansatz}) have different local foliates, \emph{i.e.}, the hyperbolic spacetime $H_2$ and the two-dimensional de Sitter spacetime $dS_2$ in the regions where $(X^3)^2-(X^{0})^2+1>0$ and $(X^3)^2-(X^{0})^2+1<0$  or $(X^1)^2-(X^{0'})^2+1>0$ and $(X^1)^2-(X^{0'})^2+1<0$ , respectively.
}}
\label{Table1}
\end{table}}
As for the various discrete maps acting on $P_{\bf n}$ we refer to Table \ref{Table2}, where we have used $\pi(E,J,iB,iP)=(-E,J,iB,-iP)$, $(E,J,iB,iP)^\dagger=(E,J,-iB,-iP)$ and $\tau(E,J,iB,iP)=(-E,-J,-iB,-iP)$ which implies
\bea \mh~=~\{E,J\}&:& \pi(w_1,w_2)~=~(-w_2,-w_1)\ ,\qquad (w_1,w_2)^\dagger~=~(w_1,w_2)\ ,\\[5pt]
\mh~=~\{J,iB\}&:& \pi(w_1,w_2)~=~(w_1,w_2)\ ,\qquad (w_1,w_2)^\dagger~=~(w_2,w_1)\ ,\\[5pt]
\mh~=~\{iB,iP\}& :&\pi(w_1,w_2)~=~(w_2,w_1)\ ,\qquad (w_1,w_2)^\dagger~=~(-w_1,-w_2)\ ,\eea
{\footnotesize \tabcolsep=1mm \begin{table}[t]
\centering\begin{tabular}{|l|l|l|l|l|}\hline
\ \ $\mh=\{K_{(+)},K_{(-)}\}$ \ & \ \  $\pi(P_{n_1,n_2})$ \ & \ \ $(P_{n_1,n_2})^\dagger$ \ &\ \ $\tau(P_{n_1,n_2})$\ \ &\ \ $\kappa_{n_1,n_2}$\ \\ \hline
\ \  $\{E,J\}$ &\ \  $P_{-n_2,-n_1}$ & \ \  $P_{n_1,n_2}$ &\ \  $P_{-n_1,-n_2}$ &\ \  $(-1)^{n_1+n_2}$ \\ 
\ \  $\{J,iB\}$ &\ \  $P_{n_1,n_2}$&\ \  $P_{n_2,n_1}$&\ \  $P_{-n_1,-n_2}$&\ \  $(-1)^{n_1+n_2+1}$ \\ 
\ \  $\{iB,iP\}$\ \  &\ \  $P_{n_2,n_1}$&\ \  $P_{-n_1, -n_2}$ &\ \  $P_{-n_1,-n_2}$&\ \  $(-1)^{n_1+n_2}$ \\ 
\hline
\end{tabular}
\caption{{\footnotesize Properties of rank-one projectors. The Cartan generators $K_{(\pm)}:=\ft12(w_2\pm w_1)$, such that $w_1=K_{(+)}-K_{(-)}$ and $w_2=K_{(+)}+K_{(-)}$. The phase factors $\kappa_{n_1,n_2}$ are defined by $P_{n_1,n_2}\star \kappa_y\bar\kappa_{\yb}=\kappa_{n_1,n_2}P_{n_1,n_2}$.}}
\label{Table2}
\end{table}}
For the diagonal ansatz, the reality condition \eq{herm1} simplifies to 
\be (\nu_{n_1,n_2})^\dagger~=~\kappa_{I(n_1),I(n_2)} \nu_{I(n_1),I(n_2)}~=~\kappa_{n_1,n_2} \nu_{I(n_1),I(n_2)}\ ,\ee
which implies 
\bea \nu_{n_1,n_2} \ = \ \left\{\ba{ll} i^{n_1+n_2}\,\m_{n_1,n_2}\ ,  &\mh=\{E,J\} \\[5pt] (-1)^{n_1+n_2+1}\,\n^\ast_{n_2,n_1}\ , & \mh=\{J,iB\} \\[5pt] (-1)^{n_1+n_2}\,\n^\ast_{-n_1 , - n_2}\ ,  &\mh=\{iB,iP\}  \ea\right. \label{Ms}\eea
where $\m_{n_1,n_2}$ are real constants. We note that the ranges of $n_1$ and $n_2$ are identical for $\mh=\{J,iB\} $ and separately symmetric around zero for $\mh=\{iB,iP\}$. Moreover, from Eq. \eq{Ms} it follows that if the principal Cartan generator is imaginary then the master fields must contain Fock-space as well as anti-Fock-space projectors which requires the regular presentation based on the auxiliary closed-contour integrals, as discussed above.

In the minimal-bosonic models, the $\tau$-projection \eq{tau1}, that is, $\nu_{n_1,n_2}=-\nu_{-n_1,-n_2}$, implies that 
\bea \mh~=~\{E,J\}&:& \nu_{n_1,n_2}~=~ i^{n_1+n_2}\,\m_{n_1,n_2}\ ,\qquad  \m_{n_1,n_2}~=~(-1)^{n_1+n_2+1} \mu_{-n_1,-n_2}\ ,\\[5pt]
\mh~=~\{J,iB\}&:& \nu_{n_1,n_2}~=~(-1)^{n_1+n_2+1} \n^\ast_{n_2,n_1}~=~(-1)^{n_1+n_2} \n^\ast_{-n_2,-n_1}\ ,\\[5pt]
\mh~=~\{iB,iP\}&:& \nu_{n_1,n_2}~=~i^{n_1+n_2+1}\mu_{n_1,n_2}\ ,\qquad \m_{n_1,n_2}=(-1)^{n_1+n_2+1} \mu_{-n_1,-n_2}\ ,\eea
requiring the auxiliary closed-contour presentation of the projectors in all cases.

In the symmetry-enhanced case, it follows from
\be ({\cal P}_n(E))^\dagger~ = ~ {\cal P}_n(E)  \ , \quad ({\cal P}_n(J))^\dagger ~ = ~ {\cal P}_n(J) \ , \quad
({\cal P}_n(iB))^\dagger ~ = ~ {\cal P}_{-n}(iB)  \ , \quad ({\cal P}_n(iP))^\dagger ~ = ~ {\cal P}_{-n}(iP) \ ,\ee
that the deformation parameters $\nu_n$ in non-minimal models must obey 
{\footnotesize \tabcolsep=1mm \begin{table}[t]
\centering\begin{tabular}{|l|l|l|}\hline
${\cal P}_n(K)$ &  Non-minimal models &  Minimal models \\ 
\hline
$K=E$& $\nu_n~=~i^{n}\,\m_{n}$& $\mu_n~=~(-1)^{n+1}\mu_{-n}$ \\
$K=J$& $\nu_n~=~i^{n+1}\,\m_{n}$& $\mu_n~=~(-1)^{n+1}\mu_{-n}$ \\
$K=iB$& $(\nu_n)^\ast~=~(-1)^{n} \nu_{-n}$& $\nu_n~=~i^{n+1}\mu_n$\,,\quad$\mu_n~=~(-1)^{n+1}\mu_{-n}$ \\
$K=iP$& $(\nu_n)^\ast~=~(-1)^{n+1} \nu_{-n}$& $\nu_n~=~i^{n}\mu_n$\,,\quad $\mu_n~=~(-1)^{n+1}\mu_{-n}$ \\
\hline
\end{tabular}
\caption{{\footnotesize Reality properties of the deformation parameters $\nu_n$ for different types of rank-$n$, symmetry-enhanced projectors ${\cal P}_n(K_{(q)})$. $\m_n$ are real parameters.}}
\label{Table3}
\end{table}}
\bea  \nu_n  \ = \ \left\{\ba{ll}  i^n\mu_n \ , & \mbox{for ${\cal P}_n(E)$} \\[5pt]  i^{n+1}\,\m_{n} \ , & \mbox{for ${\cal P}_n(J)$}    \\[5pt] (-1)^{n}\,\n^\ast_{-n} \ , & \mbox{for ${\cal P}_n(iB)$}    \\[5pt] (-1)^{n+1}\,\n^\ast_{-n} \ , & \mbox{for ${\cal P}_n(iP)$}  \ea\right. \label{nuenhanced}\eea
where $\m_n$ are real constants. In the minimal-bosonic cases, it follows from $\nu_n=-\nu_{-n}$ that $\n_n = i^{n+1}\mu_n $ for ${\cal P}_n(iB)$, and that $\n_n = i^{n}\mu_n $ for ${\cal P}_n(iP)$, and that $\m_n=(-1)^{n+1}\m_{-n}$ in all cases; see Table \ref{Table3} for a summary of these results.
Concerning the need for closed-contour presentations of the projectors, the same considerations hold as in the biaxially symmetric cases.

As for the internal connection, the reality conditions \eq{herm1} implies
\bea (V^{n_1 ,n_2}_\a)^\dagger \ = \ \left\{\ba{ll} -{\bar V}_{\ad}^{n_1, n_2}  \ , &\mh=\{E,J\} \\[5pt] -{\bar V}_{\ad}^{n_2 ,n_1}\ ,&\mh=\{J,iB\}   \\[5pt] -{\bar V}_{\ad}^{-n_1, -n_2}\ ,& \mh=\{iB,iP\}  \ea\right.   \eea
and, in the symmetry-enhanced cases, 
\bea  (V^n_\a)^\dagger~=~ \left\{\ba{ll} -\bar V_{\ad}^n \ , & \mbox{for ${\cal P}_n(E)$ and ${\cal P}_n(J)$} \\[5pt]  -\bar V_{\ad}^{-n}\ , & \mbox{for ${\cal P}_n(iB)$ and ${\cal P}_n(iP)$}   \ea\right.  \label{nuenhanced2}\eea
In the minimal-bosonic models, the $\tau$-condition implies 
\be \tau(V_\a^{\mathbf{n}}) = -i V_\a^{-\mathbf{n}}\ .\ee

\scss{Summary of internal solution and minimal-bosonic projection}\label{sec3summary}

In summary, the diagonal internal solution is given explicitly by
\be \widehat \Phi' ~=~\sum_{\mathbf{n}\in(\integ+\ft12)^2} \nu_{\mathbf{n}} P_{\mathbf{n}}\star \kappa_y\ ,\qquad \widehat S'_{\underline\a}~=~Z_{\underline\a}- 2i \sum_{\mathbf{n}\in(\integ+\ft12)^2}P_{\mathbf{n}}\star V_{\underline\a}^{\mathbf{n}}\ ,\label{factansatz}\ee
where $\nu_{\mathbf{n}}$ are complex coefficients; the projectors are given in Weyl-order (see \cite{Iazeolla:2008ix} and Appendix \ref{AppProj}) by ($\mathbf{n}=(n_1,n_2)$; $\ve_i=n_i/|n_i|$)
\be \left[P_{\mathbf{n}}\right]^{\rm Weyl} ~=~ 4(-1)^{|n_1|+|n_2|-1}\oint_{C(\ve_1)} \frac{ds_1}{2\pi i}\frac{(s_1+1)^{n_1-\ft12}}{(s_1-1)^{n_1+\ft12}}\oint_{C(\ve_2)}\frac{ds_2}{2\pi i}\frac{(s_2+1)^{n_2-\ft12}}{(s_2-1)^{n_2+\ft12}}
e^{-4\breve K(s_1,s_2)}\ ,\label{PnWeyl}\ee
\bea \breve K&=&\ft12 (s_1w_1+s_2w_2)~=~\ft12(s_1+s_2) K_{(+)}+\ft12(s_2-s_1) K_{(-)}  \\[5pt]&=:& \ft18 (y^\a y^\b\breve \vark_{\a\b}+\bar y^{\ad} \bar y^{\bd}\breve {\bar\vark}_{\ad\bd}+2y^a\bar y^{\bd}\breve v_{\a\bd})\ ;\label{Kbreve}\eea
and $V_{\underline\a}^{\mathbf{n}}=(V_{\a}^{\mathbf{n}}(z),\bar V_{\ad}^{\mathbf{n}}(\zb))$ with holomorphic part given in symmetric gauge and Weyl order by
\be \left[V_{\a}^{\mathbf{n}}\right]^{\rm Weyl}~=~2i z_\a \int_{-1}^1 \frac{dt}{(t+1)^2}\, j_{\mathbf{n}}(t) e^{i\s_{\mathbf{n}}\ft{t-1}{t+1} z^+ z^-}\ , \ee
\be j_{\mathbf{n}}(t)~=~q_{\mathbf{n}}(t)-2\sum_k \th_{\mathbf{n},k} \left[1-{1+(-1)^k\over
2}\left(1-\sqrt{1-{\s_{\mathbf{n}}{\cal B}_{\mathbf{n}}\nu_{\mathbf{n}}\over 1+k}}\right)\right] p_k(t)\ ,\ee
\be  q_{\mathbf{n}}(t)~=~-\frac{\s_{\mathbf{n}}{\cal B}_{\mathbf{n}}\nu_{\mathbf{n}}}{4}\,{}_1F_1\left[\frac{1}{2};2;\frac{\s_{\mathbf{n}}{\cal B}_{\mathbf{n}}\nu_{\mathbf{n}}}{2}\log\frac{1}{t^2}\right]\ ,\qquad p_k(t)~=~ {(-1)^k\over k!} \d^{(k)}(t)\ ,\ee
where $\s_{\mathbf{n}}\in\{\pm1\}$ and $\th_{\mathbf{n},k}\in\{0,1\}$.
In minimal-bosonic models, the internal connection 
\be\widehat S^{\prime({\rm min})}_{\underline\a} ~=~Z_{\underline\a} -2i\sum_{\mathbf{n}\in  (\mathbb N+\ft12)\times (\integ +\ft12)}(1+i\tau)(P_{\mathbf{n}}\star V^{\mathbf{n}}_{\underline\a})\ , \label{Smin}\ee
which can be written equivalently as
\bea \widehat S^{\prime({\rm min})}_{\underline\a} ~=~Z_{\underline\a} -2i\sum_{\mathbf{n}\in (\integ +\ft12)^2} P_{\mathbf{n}}\star V^{\mathbf{n}}_{\underline\a} \ , \qquad \n_{-\mathbf{n}}~=~-\nu_{\mathbf{n}}\ ,\qquad \s_{-\mathbf{n}} ~= ~-\s_{\mathbf{n}}\ ,\label{Smin2}\eea
as can be seen from ($V_{\underline\a}(\nu_{\mathbf{n}},\s_{\mathbf{n}}):=V^{\mathbf{n}}_{\underline\a}$)
\be i\tau\left(P_{\mathbf{n}}\star V_{\underline\a}(\nu_{\mathbf{n}},\s_{\mathbf{n}})\right) ~=~
 P_{-\mathbf{n}}\star V_{\underline\a}(-\nu_{\mathbf{n}},-\s_{\mathbf{n}})~=~P_{-\mathbf{n}}\star V_{\underline\a}(\nu_{-\mathbf{n}},\s_{-\mathbf{n}})\ .\ee
The $\star$-product compositions ($\sigma,k,\bar k,m\in \mathbb N$)
\be (\widehat\kappa\star\widehat{\bar\kappa})^{\star \sigma}\star (\widehat \Phi'\star\widehat \kappa)^{\star k} \star (\widehat \Phi'\star\widehat {\bar\kappa})^{\star \bar k}\star \widehat S'_{\underline \a_1}\star\cdots \star\widehat S'_{\underline \a_m}~\in~\widehat {\cal A}'_{\rm diag}\ ,\ee\be \widehat {\cal A}'_{\rm diag}~:=~ \left\{\sum_{\mathbf n} P_{\mathbf n}(Y)\star\sum_{i,\bar i,j,\bar j=0,1} V^{\mathbf n}_{i,\bar i,j,\bar j}(z)\star \overline V^{\mathbf n}_{i,\bar i,j,\bar j}(\bar z)\star(\kappa_y)^{\star i}\star (\bar\kappa_{\bar y})^{\star \bar i}\star(\kappa_z)^{\star j}\star (\bar\kappa_{\bar z})^{\star \bar j}\right\}\ ,\ee
\be V^{\mathbf n}_{i,\bar i,j,\bar j}~:=~ \int_{-1}^1 \frac{dt}{t+1} \Delta^{\mathbf{n}}_{i,\bar i,j,\bar j}(t;\partial^{(\r)})\left[ e^{\ft{i}{t+1} \left(\s^{\mathbf n}(t-1) z^+ z^-
+\rho^+ z^+ +\rho^- z^-\right)}\right]_{\rm Weyl}\ ,\ee
where for each fixed $\mathbf n$, the operators 
\be \Delta^{\mathbf{n}}_{i,\bar i,j,\bar j}(t;\partial^{(\r)})[\cdot]~:=~\sum_{p=0}^\infty f^{\mathbf{n},\a_1\dots \a_p}_{i,\bar i,j,\bar j;p}(t) \left.\partial^{(\r)}_{\a_1}\cdots \partial^{(\r)}_{\a_p}(\cdot)\right|_{\r_i=0}\ee
belong to a space with the commutative and associative composition rule 
\be  \left(\Delta(\partial^{(\r)}) \circ \Delta'(\partial^{(\r)})\right)(t)\left[\cdot\right]~=~ \sum_{p,p'=0}^\infty  (f^{\a_1\dots \a_p}_{p}\circ f^{\prime\,\a_{p+1}\dots \a_{p+p'}}_{p'})(t) \left.\partial^{(\r)}_{\a_1}\cdots \partial^{(\r)}_{\a_{p+p'}}(\cdot)\right|_{\r=0}\ .\label{Deltaring}\ee
The $\star$-product compositions of elements in $\widehat {\cal A}_{\rm diag}$ thus involve
\be V(z;\s) \,\star\, V'(z;\s)\hspace{15cm}\nn\ee\be=~ \int_{-1}^1 dt \int_{-1}^1  dt' \frac{1}{2(\tilde t+1)}  \Delta(t;\partial^{(\r)}) \Delta'(t';\partial^{(\r')})\left[  e^{\ft{i}{\tilde t+1}\left(\s(\tilde t-1)z^+z^- +\tilde{\r}^+ z^ ++\tilde{\r}^-z^--\ft12 \r^+\r^{\prime -}+\ft{1}2\r^{\prime +}\r^-\right)} \right]_{\rm Weyl}\ ,\label{Apm^2} \ee
where $\tilde t=tt'$ and $\tilde \r^\pm$ are defined in \eq{tildet}. Rearranging 
\be \ft12 \Delta(t;\partial^{(\r)}) \Delta'(t';\partial^{(\r')}) \left[e^{\ft{i}{\tilde t+1}\left(\tilde{\r}^+z^+ +\tilde{\r}^-z^--\frac12 \r^+\r^{\prime -}+\ft{1}2\r^{\prime +}\r^-\right)}\right]_{\rm Weyl} \hspace{9cm}\nn\ee\be ~=:~  \sum_{I} \Delta^I(t;\partial^{(\r)}) \Delta^{\prime I}(t';\partial^{(\r)}) \left[e^{\ft{i}{\tilde t+1} (\r^{+}z^++\r^{-}z^- )}\right]_{\rm Weyl}\ ,\ee
where $f^{I,\a_1\dots\a_p}_p(t)$ and $f^{\prime I,\a_1\dots\a_p}_p(t')$ are linear combinations of $f^{\a_1\dots\a_p}_p(t)$ and $f^{\prime \a_1\dots\a_p}_p(t')$, respectively, with coefficients given by finite polynomials in $t$ and $t'$, yields 
\be V(z;\s) \star V'(z;\s)~=~ \sum_I \int_{-1}^1 \frac{ dt}{t+1}  (\Delta^I(\partial^{(\r)})\circ \Delta^{I\prime}(\partial^{(\r)}))(t) \left[e^{\ft{i}{ t+1}\left(\s(t-1)z^+z^- +\r^{+}z^++\r^{-}z^-\right)}\right]_{\rm Weyl} \ .\ee
As the the $\Delta$-operators form a well-defined $\circ$-product algebra, which is associative by construction, and as the projector algebra spanned by $P_{\mathbf n}$ is associative as well, using the prescription based on regular presentations spelled out in Appendix \ref{AppProj} (for details, see analysis below Eq. \eq{recort}), it follows that $\widehat {\cal A}_{\rm diag}$ is an associative $\star$-product algebra.
Viewed as elements of $\O^{[0]}({\cal Y}\times {\cal Z})$, the Weyl-ordered symbols of the elements in $\widehat {\cal A}_{\rm diag}$ have singularities at hyper-planes $\subset {\cal Y}\times {\cal Z}$ in the form of delta-functions or negative integer powers of twistor coordinates. 
For example, in the case that $K_{(+)}=E$ and $K_{(-)}=J$, it follows from 
\be P_{\pm\ft12,\pm \ft12}\star \k_y~=~2\pi \left[\delta^2(y\mp i\s_0 \yb)\right]_{\rm Weyl}\ ,\qquad P_{\pm \ft12,\mp \ft12}\star \k_y~=~\pm P_{\pm \ft12,\mp \ft12} \label{deltaint}\ee
that if $E$ is principal then $\big[\widehat\Phi'\big]^{\rm Weyl}$ has delta-function-like singularities on the hyper-planes $y\mp i\s_0 \yb$ while if $J$ is principal then $\big[\widehat\Phi'\big]^{\rm Weyl}$ is singularity-free. In order to exhibit the singularities in the deformed oscillators, we use\footnote{The related basic $\star$-product lemma reads ($yAy:= y^\a A_\a{}^\b y_\b$ and $uy:=u^\a y_\a$ \emph{idem} $zBz$ and $vz$)
\be \left[e^{\ft12 y A y + uy}\right]_{\rm Weyl}\,\star \, \left[e^{\ft12 z B z + vz}\right]_{\rm Weyl}~=~
\left(1+A^{\a\b} B_{\a\b}+A^2 B^2\right)^{-\ft12} e^{uy+vz+iuv+\ft12 \ft{\tilde y(A+A^2 B)\tilde y+\tilde z(B+B^2 A)\tilde z-2i \tilde y(AB-A^2 B^2)\tilde z}{1+A^{\a\b} B_{\a\b}+A^2 B^2}}\ , \ee
where $A^2=\ft 12 A^{\a\b} B_{\a\b}$ \emph{idem} $B$, and $\tilde y=y+iv$ and $\tilde z=z-iu$.} ($\bar v_{\ad\a}=v_{\a\ad}$)
\bea  &\widehat {\cal O}~:=~ \left[e^{-\ft12\left(y^\a y^\b\breve \vark_{\a\b}+\bar y^{\ad} \bar y^{\bd}\breve{\bar\vark}_{\ad\bd}+2y^a\bar y^{\bd}\breve v_{\a\bd}\right)}\right]_{\rm Weyl}\,\star \, \left[e^{\ft{i}{2(t+1)}\left(\s(t-1)z^\a z^\b {\cal D}_{\a\b}+2\r^\a z^\b{\cal I}_{\a\b}\right)} \right]_{\rm Weyl}& \nn\\[5pt]
& \ = \ \frac{1}{\sqrt{\breve \vark^2 \,G^2}}\left[e^{-\ft12\bar y^{\ad} \bar y^{\bd}\left(\breve{\bar\vark}_{\ad\bd}-\breve{\bar v}_{\ad}{}^\a\breve \vark^{-1}_\a{}^\b \breve v_{\b\bd}\right)+\ft{i}{2(t+1)}\left(\s(t-1)z^\a z^\b {\cal D}_{\a\b}+2\r^\a z^\b{\cal I}_{\a\b}\right)-\ft12 b^\a G^{-1}_\a{}^\b b_\b }\right]_{\widehat{\rm N}^+} \ ,&\label{lemmaS}\eea
where the $\widehat{\rm N}^+$-order is defined in Appendix \ref{App:A}, and  
\be {\cal D}_{\a\b}~:=~2u^-_{(\a} u^+_{\b)}\ ,\qquad {\cal I}_{\a\b}~:=~u^+_{\a} u^+_{\b}+u^-_{\a} u^-_{\b}\ ,\ee
\be G_{\a\b} ~ := ~ \breve \vark^{-1}_{\a\b}+i\s\frac{ t-1}{t+1}\,{\cal D}_{\a\b}\ , \qquad b^\a ~:=~ i\left[y^\a+\bar y^{\ad}\breve{\bar v}_{\ad}{}^\b\breve{\vark}^{-1}_\b{}^\a+\frac{1}{t+1}\left(\s (t-1)z^\b{\cal D}_\b{}^\a+\r^\b{\cal I}_\b{}^\a\right)\right]\ ,\ee
and we recall that $G^{-1}_{\a\b}=-\ft{G_{\a\b}}{G^2}$ with $G^2:=\ft12 G^{\a\b}G_{\a\b}$ and  
\be \frac{1}{\sqrt{G^2}} ~=~ \frac{(t+1)\sqrt{\breve \vark^2}}{\sqrt{(t+1)^2-i\s (t^2-1)\breve \vark^{\a\b}{\cal D}_{\a\b}+\breve\vark^2 (t-1)^2}} \ .\label{det}\ee
From the limit 
\be \left.\left[\widehat{\cal O}\right]^{\widehat{\rm N}^+}\right|_{\breve \vark=\breve{\bar\vark}=0} ~=~ e^{-y^\a \breve v_{\a\ad} \yb^{\ad}+\frac{i}{2(t+1)}\left(\s(t-1)a^\a a^\b D_{\a\b} +2 \r^\a a^\b {\cal I}_{\a\b}\right)}\ ,\qquad a_\a~:=~z_\a+i\breve v_\a{}^{\ad}\yb_{\ad}\ ,\ee
it follows that if $E$ is principal then the contribution to $\widehat S'_\a$ from the ground-state projector ${\cal P}_1(E)\equiv P_{\ft12,\ft12}$ contains a singularity of the form ($\s\equiv \s_{\ft12,\ft12}$)
\be\left[\widehat S^{\prime\pm}_{\ft12,\ft12;\s}\right]^{\widehat{\rm N}^+}~\sim~ {\cal P}_1(E) a^\pm \int_{-1}^1 {dt\over (t+1)^2}  j^\pm_{\ft12,\ft12}(t)e^{-\frac{2i\s}{t+1} a^+ a^-}~\sim~{\cal P}_1(E) \frac{1}{a^\mp}\ ,\ee
where the last step is based on performing analytical continuation on the twistor-space variables.
Similarly, taking the limit $\breve v_{\a\ad}=0=\breve{\bar v}_{\ad\a}$ one finds that if $J$ is principal ($\varkappa_{\a\b}=i {\cal D}_{\a\b}$) then the contribution to $\widehat S'_\a$ from the ground-state projector ${\cal P}_1(J)\equiv P_{-\ft12,\ft12}$ depends on the sign of $\s\equiv \s_{-\ft12,\ft12}$: for $\s=+1$ one has $G^2=4(t+1)^{-2}$ and \be\left[\widehat S^{\prime\pm}_{-\ft12,\ft12;\s=+}(y,\yb,z)\right]^{\widehat{\rm N}^+}~\sim~ {\cal P}_1(J)  a^\pm e^{-i a^+ a^-}\int_{-1}^1 dt\, j^\pm_{-\ft12,\ft12;\s=+} \left[\widehat {\cal O}'(t,y,\yb,z)\right]^{\widehat{\rm N}^+}\ ,\ee where $\left[\widehat{\cal O}'(t,y,z)\right]^{\widehat{\rm N}^+}$ is real-analytic in $(t,y,\yb,z)$, and hence $\left[\widehat S^{\prime\pm}_{-\ft12,\ft12;\s=+}(y,\yb,z)\right]^{\widehat{\rm N}^+}$ is real-analytic in $(y,\yb,z)$; for $\s=-1$ one has $G^2=4t^2(t+1)^{-2}$ and the pole at $t=0$ gives rise to an algebraic singularity.


\scs{Spacetime-dependent Master Fields}\label{Sec:sptfields}


This section contains the analysis of the spacetime-dependent master fields $(\widehat\Phi_{(K)},\widehat S_{{(K)}\underline\a},\widehat W_{(K)})$ obtained from the internal solution $(\widehat\Phi',\widehat S'_{\underline\a})$ via the gauge function $\widehat L_{(K)}$. We shall first demonstrate that if $\tilde L_{(K)}$ is chosen as to align the spin-frames in ${\cal Y}$ and ${\cal Z}$ then the internal connection $\widehat S_{{(K)}\underline\a}$ and spacetime connection $\widehat W_{{(K)}\mu}$ become singularity-free in an extended region ${\cal R}_{\rm Conn}$ of spacetime\footnote{The physical meaning of singularities in the spacetime and twistor-space connections, which may very well be gauge artifacts, can be addressed by examining observables depending on $(\widehat S_\a,\widehat{\bar S}_{\ad})$ (see Section \ref{0formcharges}); we leave this issue for future work. 
}. Finally, we show that the Weyl zero-form $\widehat \Phi_{(K)}$ is finite in a region ${\cal R}_{\rm Weyl}\supseteq {\cal R}_{\rm Conn}$. 
We also note that the combination of the Ansatz \eq{factansatz} and the $K$-adapted gauge function leads to an internal connection $\widehat S_\a$ that is not given in the twistor gauge \eq{twistorgauge} used for the perturbative analysis of Vasiliev's equations discussed in Appendix \ref{weakfields}. 


\scss{Internal connection}\label{Sec:intconn}


\scsss{Alignment of spin-frames and absence of singularities}

Using the gauge-function \eq{gaugef}, the internal connection, given by the third equation in \eq{Lrot}, takes the form
\be \widehat S_{{(K)}\a} ~=~ (\widehat L_{(K)})^{-1}\star \widehat S^{\prime}_\a\star \widehat L_{(K)}~=~(\tilde L_{(K)})_\a{}^\b z_\b-2i\sum_{\mathbf{n}}P^L_{\mathbf{n}}\star \tilde V^{\mathbf{n}}_{{(K)}\a}\ ,
\label{SNL}\ee
where the $L$-rotated projector ($\ve_i=n_i/|n_i|$, $q=\ve_1\ve_2$)
\be P_{\mathbf{n}}^L ~=~ 4(-1)^{|n_1|+|n_2|-1}\oint_{C(\ve_1)} \frac{ds_1}{2\pi i}\frac{(s_1+1)^{n_1-\ft12}}{(s_1-1)^{n_1+\ft12}}\oint_{C(\ve_2)}\frac{ds_2}{2\pi i}\frac{(s_2+1)^{n_2-\ft12}}{(s_2-1)^{n_2+\ft12}}
\left[e^{-4\breve K^L(s_1,s_2)}\right]_{\rm Weyl}\ ,  \label{PL}\ee
\be \breve K^L~=~ L^{-1}\star \breve K\star L~=~\ft12(s_1+s_2) K^L_{(+)}+\ft12(s_2-s_1) K^L_{(-)}~=~\ft18 (y^\a y^\b\breve \vark^L_{\a\b}+\bar y^{\ad} \bar y^{\bd}\breve {\bar\vark}^L_{\ad\bd}+2y^a\bar y^{\bd}\breve v^L_{\a\bd}) \ ,\label{breveKL}\ee
\be \breve \vark^L_{\a\b}~=~ \ve_2 \vark^L_{(q)\a\b}+\ft12 (s_1-\ve_1)\left(\vark^L_{(+)\a\b}-\vark^L_{(-)\a\b}\right)+\ft12 (s_2-\ve_2)\left(\vark^L_{(+)\a\b}+\vark^L_{(-)\a\b}\right) \ ,\label{brevevarkL}\ee
and the $\tilde L_{(K)}$-rotated internal connection 
\be \tilde V^{\mathbf{n}}_{{(K)}\a}~:=~ (\tilde L_{(K)})^{-1}\star V^{\mathbf{n}}_\a\star \tilde L_{(K)}\ .\ee
Defining $\widehat S^{\prime \pm}:=u^{\pm\a} \widehat S'_\a$, one has
\be \widehat S^\pm_{(K)}~:=~(\widehat L_{(K)})^{-1}\star \widehat S^{\prime \pm}\star \widehat L_{(K)}~=~ \tilde z^\pm_{(K)}-2i\sum_{\mathbf{n}} P^L_{\mathbf{n}}\star \tilde V^{\mathbf{n}\pm}_{(K)}\ ,\label{SKpm}\ee 
where $\tilde z^\pm_{(K)}:=(\tilde L_{(K)})^{-1}\star z^\pm\star \tilde L_{(K)}=\tilde u^{\pm\a}_{(K)} z_\a$, using the $K$-aligned spin-frame $\tilde U_{(K)}$ defined by \eq{alignment}, \emph{i.e.} $\tilde u^{\pm\a}_{(K)}:=u^{\pm\b} (\tilde L_{(K)})_\b{}^\a$, and
\be \left[\tilde V^{\mathbf{n}\pm}_{(K)}\right]^{\rm Weyl}~=~2i \tilde z^{\pm}_{(K)} \int_{-1}^1 \frac{dt}{(t+1)^2}\, j^\pm_{\mathbf{n}}(t) e^{i\s_{\mathbf{n}}\ft{t-1}{t+1} \tilde z^+_{(K)} \tilde z^-_{(K)}}\ .\label{intrep1}\ee
The internal connection can be represented using a source as
\be \left[\tilde V^{\mathbf{n}\pm}_{(K)}\right]^{\rm Weyl} =~2\frac{\partial}{\partial \r^\pm}\left. \int_{-1}^1 \frac{dt}{t+1}\, j^{\mathbf{n}\pm}_{\s_\mathbf{n}}(t) \, e^{\ft{i}{2(t+1)}\left(\s_\mathbf{n}(t-1)z^\a z^\b \widetilde{{\cal D}}_{(K)\a\b}+2\r^\a z^\b\widetilde{{\cal I}}_{(K)\a\b}\right)}\right|_{\r=0}\ ,\ee
\be\widetilde{{\cal D}}_{(K)\a\b}~:=~\left(\tilde u^+_{(K)}\tilde u^-_{(K)}+\tilde u^-_{(K)}\tilde u^+_{(K)}\right)_{\a\b}\ ,\qquad \widetilde{{\cal I}}_{(K)\a\b}~:=~\left(\tilde u^+_{(K)}\tilde u^+_{(K)}+\tilde u^-_{(K)}\tilde u^-_{(K)}\right)_{\a\b}\ .\label{widetildeD}\ee
From Eq. \eq{lemmaS}, which implies that
\be \left[\widehat S^{\pm}_{(K)}\right]^{\widehat N_+} -\tilde z^\pm_{(K)}~=~-16i \sum_{\mathbf{n}}(-1)^{|n_1|+|n_2|-1}\oint_{C(\ve_1)} \frac{ds_1}{2\pi i}\frac{(s_1+1)^{n_1-\ft12}}{(s_1-1)^{n_1+\ft12}}\oint_{C(\ve_2)}\frac{ds_2}{2\pi i}\frac{(s_2+1)^{n_2-\ft12}}{(s_2-1)^{n_2+\ft12}}\nn\ee\be\times  \int_{-1}^1\frac{dt}{\sqrt{(t+1)^2-i\s_{\mathbf{n}} (t^2-1)\breve\vark^{L\a\b}\widetilde {\cal D}_{(K)\a\b}+(\breve \vark^L)^2 (t-1)^2}}\,j^\pm_{\mathbf{n}}(t)\nn\ee\be \times \frac{\partial}{\partial \r^\pm}\left.e^{-\ft12\bar y^{\ad} \bar y^{\bd}\left(\breve{\bar\vark}^L_{\ad\bd}-\breve{\bar v}^L_{\ad}{}^\a(\breve\vark^{L})^{-1}_\a{}^\b \breve v^L_{\b\bd}\right)+\ft{i}{2(t+1)}\left(\s_{\mathbf{n}}(t-1) z^\a z^\b \widetilde {\cal D}_{(K)\a\b}+2\r^\a z^\b\widetilde {\cal I}_{(K)\a\b}\right)-\ft12 b^\a G^{-1}_\a{}^\b b_\b }\right|_{\r^\pm=0}\ ,\qquad\label{Spm}\ee
with $G_{\a\b} = (\breve \vark^L)^{-1}_{\a\b}+i\s_{\mathbf{n}}\frac{ t-1}{t+1}\,\widetilde {\cal D}_{(K)\a\b}$ and \be b^\a ~=~ i\left[y^\a+\bar y^{\ad}\breve{\bar v}^L_{\ad}{}^\b(\breve{\vark}^L)^{-1}_\b{}^\a+\frac{1}{t+1}\left(\s (t-1)z^\b\widetilde {\cal D}_{(K)\b}{}^\a+\r^\b\widetilde {\cal I}_{(K)\b}{}^\a\right)\right]\ ,\ee
and the prescription for the closed contours (see Appendix \ref{AppProj}), it follows that if one chooses a definite $q$-value, say $K=K_{(q)}$, then\footnote{We recall that choosing a principal Cartan generator $K=K_{(q)}$ selects a spin-frame $(\tilde u^+_{(K_{(q)})},\tilde u^-_{(K_{(q)})})$ adapted to it, \emph{i.e.} such that the matrix $K_{\underline{\a\b}}^{(q)L}$ assumes the corresponding canonical form given in Appendix \ref{App:L-rot} (see also Table \ref{Table1} and \eq{spinframe1} for the definition of $\Theta$).}
\be  \breve\vark^{L\a\b}\widetilde {\cal D}_{(K)\a\b}~=~-2\ve_2\Theta+O(s_1-\ve_1,s_2-\ve_2)\ ,\qquad (\breve \vark^L)^2 ~=~-\Theta^2+O(s_1-\ve_1,s_2-\ve_2)\ ,\ee
and the potential singularities in the integral representation of $\widehat S^\pm_K$ are shifted to the zeroes of
\be (t+1)^2-i\s_{\mathbf{n}} (t^2-1)\breve\vark^{L\a\b}\widetilde {\cal D}_{(K)\a\b}+(\breve \vark^L)^2 (t-1)^2\ee
\be =~ (t+1)^2+2i\s_{\mathbf{n}}\e_2 \Theta (t^2-1)-\Theta^2(t-1)^2+O(s_1-\ve_1,s_2-\ve_2)\ee
\be~=\Big(\big(1+i\s_{\mathbf{n}}\e_2\Theta \big)\left(t+1\right)-2i\s_{\mathbf{n}}\e_2\Theta\Big)^2+O(s_1-\ve_1,s_2-\ve_2)\ ,\ee
Moreover, the same shift of the pole in $t=-1$ takes place in the exponent of \eq{Spm}; for the spherically-symmetric case, see \eq{Spmground1} and \eq{Spmsph}. Thus the symbol $\left[\widehat S^\pm_{(K)}\right]^{\widehat N_+}$ is real-analytic in $Y$ and $Z$
if $K=E$ and $\Theta> 0$ or if $K=J$ and $\sigma_{\mathbf{n}}\e_2>0$ (in which case $-i \Theta=\sqrt{1+\Upsilon^2}\geqslant 1$), while if $K=J$ and $\sigma_{\mathbf{n}}\e_2<0$, then there remains a singularity at $t\in[0,1]$ for all values of $\Upsilon$ and $\left[\widehat S^\pm_{(K)}\right]^{\widehat N_+}$ is non-analytic in $Y$ and $Z$ for any $x$. If $K=iB$ and $K=iP$ then $\Theta$ is indefinite and there is a finite region of spacetime in which $\Theta$ is real and hence $\left[\widehat S^\pm_{(K)}\right]^{\widehat N_+}$ is real-analytic in $Y$ and $Z$.

\scsss{Spherically symmetric case}\label{Sspher}

The rotation \eq{breveKL}-\eq{brevevarkL} of the generators for solutions based on the $(E,J)$ Cartan pair proceeds as follows (see also Appendix \ref{App:L-rot}). Using the conventions in Appendix \ref{App:conv}, it follows from $K_{(+)}=E=\ft14 y^\a(\s_0)_{\a\ad}\yb^{\ad}$ that $\vark_{(E)\a\b}=0$ and $v_{(E)\a\ad}=u^+_\a\bar u^+_{\ad}+u^-_\a\bar u^-_{\ad}$. In stereographic coordinates, the $L$-rotated Killing two-form and Killing vector read
\be \vark^L_{(E)\a\b}~=~\ft{2x^i}{1-x^2}(\s_{i0})_{\a\b}\ ,\qquad v^L_{(E)\a\bd}~=~(\s_0)_{\a\bd}-\ft4{1-x^2}x_{[0}x^i(\s_{i]})_{\a\bd}\ .\ee
$E^L$ can be brought to a spin-frame $\tilde u_{(E)\a}^\pm(x)$ in which it takes the canonical form (see Appendix \ref{App:L-rot})
\be  \vark^L_{(E)\a\b} ~=~ r\,\widetilde{\cal D}_{(E)\a\b}\ ,\qquad 
 v^L_{(E)\a\bd} ~=~ \sqrt{1+r^2}\,\widetilde{\cal T}_{(E)\a\bd} \label{kLE}\ , \ee
\be \widetilde{\cal D}_{(E)\a\b}~=~\tilde u^{+}_{(E)\a} \tilde u^{-}_{(E)\b} + \tilde u^{-}_{(E)\a} \tilde u^{+}_{(E)\b} \ , \qquad \widetilde{\cal T}_{(E)\a\bd}~=~\tilde u^{+}_{(E)\a}\bar{\tilde u}^{+}_{(E)\bd}+\tilde u^{-}_{(E)\a}\bar{\tilde u}^{-}_{(E)\bd} 
\label{vLE} \ ,\ee
expressed in spherical coordinates. Similar operations can be repeated for $K_{(-)}=J$ (see Appendix \ref{App:L-rot} for the details). Note however that in general $\vark_{(+)}$ and $\vark_{(-)}$ are both type-$\{1,1\}$ but have different principal spinors.  This in particular means that on the $E$-adapted spin-frame $\tilde u_{(E)\a}^\pm$ $J$ takes a non-canonical form and, in spherical coordinates, the resulting $\breve K^L$ for the solutions based on the $(E,J)$ Cartan pair reads ($q=\ve_1\ve_2=1$; $\eta:=\ft12(s_1+s_2)\approx \ve_2$; and $\zeta=\ft12(s_2-s_1)\approx 0$ )
\bea \breve \vark^L_{\a\b} & = & (\eta r + i\z\cos\theta)\widetilde {\cal D}_{(E)\a\b}+ \zeta\sqrt{1+r^2}\sin\theta\,\widetilde {\cal I}_{(E)\a\b}\ ,\label{kE}\\[5pt]
\breve v^L_{\a\bd} & = & \eta\sqrt{1+r^2}\,\widetilde {\cal T}_{(E)\a\bd}+\zeta \,r\,\sin\th\,\widetilde {\cal S}_{(E)\a\bd} \ , \label{vE}\eea
with $\widetilde {\cal T}_{(E)\a\bd}:=(\tilde u^+_{(E)}\tilde{\bar u}^+_{(E)}+\tilde u^-_{(E)}\tilde{\bar u}^-_{(E)})_{\a\bd}$ and $\widetilde {\cal S}_{\a\bd}:=(\tilde u^+_{(E)}\tilde{\bar u}^-_{(E)}+\tilde{u}^-_{(E)}\tilde{\bar u}^+_{(E)})_{\a\bd}$, while $\widetilde {\cal D}_{(E)\a\b}$ and $\widetilde {\cal I}_{(E)\a\b}$ are defined in \eq{widetildeD}.

For instance, the solution with $n_1=n_2=\ft12$ and $\s:=\s_{\ft12,\ft12}$, corresponding to the spherically symmetric ground state, reads
\be\left[\widehat S^{\pm}_K\right]^{\widehat N_+}~=~ \tilde z^\pm+32 \left[\tilde z^\pm\pm ir\tilde y^\pm\pm i\sqrt{1+r^2}\tilde {\bar y}^\mp\right]\int_{-1}^1\frac{dt}{(t+1+i\s r(t-1))^2} \,j^\pm_{\ft12,\ft12}(t) \ \times\nn\ee\be \times \ \exp \left\{\frac{1}{t+1+i\s r(t-1)}\left[-r(t+1)\tilde y^+\tilde y^-+\tilde {\bar y}^+\tilde {\bar y}^-\big(i\s(t-1)-r(t+1)\big)-(t+1)\sqrt{1+r^2}\left(\tilde y^+\tilde {\bar y}^++\tilde y^-\tilde {\bar y}^-\right)\right.\right.\nn\ee\be
\left.\phantom{\frac{1}{t+1+i\s r(t-1)}}\hspace{-2cm}\left.+i\s (t-1)\tilde z^+\tilde z^-+\s r(t-1)\left(\tilde y^-\tilde z^+-\tilde y^+\tilde z^-\right)+\s (t-1)\sqrt{1+r^2}\left(\tilde {\bar y}^+\tilde z^+-\tilde {\bar y}^-\tilde z^-\right)\right]\right\}\ .\label{Spmground1}\ee
For higher symmetry-enhanced projectors ${\cal P}_n(E)$ it is convenient to use the integral representation \eq{enhanced} in \eq{SKpm}, and one has
\be \z=0\ ,\qquad \breve \vark^L_{\a\b}~=~\eta\vark_{(E)\a\b}^L~=~ \eta \,r \,\widetilde {\cal D}_{(E)\a\b}\ ,\qquad \breve v^L_{\a\bd}~=~\eta\, v_{(E)\a\bd}^L~=~\eta \sqrt{1+r^2}\,\widetilde {\cal T}_{(E)\a\bd}\ ,\ee
and the corresponding solutions can be conveniently cast into a more compact form using the following generalization of the modified oscillators of \cite{Didenko:2008va}:
\be \breve a_\a(\eta) \ : = \ z_\a+i(\breve \vark^L_\a{}^\b y_\b+\breve v^L_\a{}^{\bd}\bar y_{\bd}) \ , \qquad [\breve a_\a,\breve a_\b]_\star \ = \ -2i\e_{\a\b}(1+\eta^2)\ .\ee
The general spherically-symmetric internal connection can thus be written as ($\ve=n/|n|$)
\bea \left[\widehat S^{\pm}_K\right]^{\widehat N^+} & = & \tilde z^\pm+4\sum_{n=\pm1,\pm2,...}(-1)^{n-\ft{1+\ve}2}\int_{-1}^1 dt\,j^\pm_{n}(t)\oint_{C(\ve)}\frac{d\eta}{2\pi i(t+1+i\s_n\eta r(t-1))^2} \left(\frac{\eta+1}{\eta-1}\right)^n\nn\\[5pt] && \hspace{5cm} \times~~\tilde u^{\a\pm}{\breve a}_\a\,{\cal P}_1(\eta E^L)\, e^{\ft{i\s_n(t-1)}{2(t+1+i\s_n\eta r(t-1))}\tilde{\cal D}^{\a\b} {\breve a}_\a {\breve a}_\b }\ .\label{Spmsph}\eea
In particular, for $n=1$, \emph{i.e.} $n_1=n_2=\ft12$, the solution \eq{Spmground1} can be cast in the simpler form  
\bea \left[\widehat S^{\pm}_K\right]^{\widehat N^+} & = & \tilde z^\pm+8\,{\cal P}_{1}(E^L)\,\tilde a^\pm\int^1_{-1}\frac{dt}{(t+1+i\s r(t-1))^2}\,j^\pm_{1}(t)\,e^{\,\ft{i\s( t-1)}{t+1+i\s r(t-1)}\,\tilde a^+\tilde a^- } \ .\eea
where $\tilde a^\pm:=\tilde u^{\a\pm} \breve a_\a|_{\eta=1}$ coincide with the modified oscillators of \cite{Didenko:2008va}, obeying $z_\a\star P_{\ft12,\ft12}=a_\a P_{\ft12,\ft12}$. Notice that, as anticipated, the internal connection for the spherically-symmetric case may only diverge in $r=0$, as the form of the Weyl tensors \eq{o3Weylnm} suggests.


\scss{Lorentz-covariant spacetime gauge-field generating function}\label{Sec:gfields} 


We shall now show that the generating function \eq{fieldredef} of the spacetime gauge-fields, \emph{viz.}
\bea  \widehat W_{{(K)}\mu}(x|Y) \ = \ \O^{(0)}_\m +\tilde L_{{(K)}}^{-1}\star \partial_\mu \tilde L_{(K)}-\widehat K_{{(K)}\m} \ , \label{VmuwrtK}\eea
where $\O^{(0)}_\m= e_\m^{(0)}+\o^{(0)}_\m$ is the $AdS_4$ one-form connection \eq{vac0}-\eq{vac02}, has the property of being non-singular for generic spacetime points. 
From \eq{M(S)} it follows that $\widehat K_{(K)\m}$ contains terms that are linear as well as bilinear in $\widehat V_{{(K)}\a}$. The former have already been examined and shown to possess the aforementioned property, while the latter contain new structures of the form $\widehat V_{{(K)}(\a}\star\widehat  V_{{(K)}\b)}$. In the adapted spin-frame, the ``self-replication'' formula \eq{selfrep} implies
\bea \left[V_{\pm}^{\mathbf{n}}\star V_{\pm}^{\mathbf{n}}\right]^{\textrm{Weyl}} & = & 2\frac{\partial}{\partial \r^\pm}\frac{\partial}{\partial \r^{\prime\pm}} \int_{-1}^{1} dt \int_{-1}^{1} dt'\, \frac{1}{\tilde{t}+1}\,j^{\,\mathbf{n},\s_{\mathbf{n}}}(t) j^{\,\mathbf{n},\s_{\mathbf{n}}}(t')\nn\\[5pt] &\times& \quad \left. e^{\ft{i}{\tilde{t}+1}(\s_{\mathbf{n}}(\tilde{t}-1)z^+z^- +\tilde{\r}^+z^+ +\tilde{\r}^-z^--\ft12 \r^+\r^{\prime -}+\ft{1}2\r^{\prime +}\r^-)}\right|_{\r=\r'=0} \ ,  \label{Apm^2} \eea
which shows that the previous considerations for $\widehat V_{{(K)}\a}$ still apply. Thus, conjugation by $\tilde L_{(K)}$ and $\star$-multiplication by $P^L_{\mathbf{n}}$ shifts the singularity out of the homotopy integration domain; in the cylindrically-symmetric case, this restricts $\s_{\mathbf{n}}$, as found above. Thus, the spacetime gauge fields inherit the regular behaviour at generic spacetime points of the twistor-space connection.

In the case of spherically symmetric solutions, which arise for $K=E$, and recalling that $\e_{\a\b} \ = \ \tilde u^-_\a \tilde u^+_{\b}- \tilde u^+_\a \tilde u^-_{\b}$, where from now on the gauge index $(E)$ is suppressed, one has
\be \widehat K_\m (Z,Y|x) \ = \ \frac{1}{4i}\left( \o_\m^{++} \widehat M^{--} + \o_\m^{--} \widehat M^{++}-2\o_\m^{-+} \widehat M^{+-}\right)-\textrm{h.c.} \ ,\ee
with
\bea \left[\widehat M^{++}\right]^{\widehat N_+} & = & \tilde y^+  \tilde y^+ + \!\!\!\sum_{n=\pm1,\pm2,...}\!\!\!\!\!\!8(-1)^{n-\ft{1+\e}2}\oint_{C(\e)}\frac{d\eta}{2\pi i}\left(\frac{\y+1}{\y-1}\right)^n {\cal P}_1(\y E^L)\left( {\cal F}-{\cal G}\right)\tilde{\breve{a}}^+\tilde{\breve{a}}^+\ ,\\[5pt]
 \left[\widehat M^{--}\right]^{\widehat N_+} & = &  \tilde y^-  \tilde y^- + \!\!\!\sum_{n=\pm1,\pm2,...}\!\!\!\!\!\!8(-1)^{n-\ft{1+\e}2}\oint_{C(\e)}\frac{d\eta}{2\pi i}\left(\frac{\y+1}{\y-1}\right)^n {\cal P}_1(\y E^L)\left( {\cal F}-\cal{G}'\right)  \tilde{\breve{a}}^-\tilde{\breve{a}}^-\ ,\\[5pt]
\left[\widehat M^{+-}\right]^{\widehat N_+} & = &  \tilde y^+  \tilde y^- -  \!\!\!\sum_{n=\pm1,\pm2,...}\!\!\!\!\!\!8(-1)^{n-\ft{1+\e}2}\oint_{C(\e)}\frac{d\eta}{2\pi i}\left(\frac{\y+1}{\y-1}\right)^n{\cal P}_1(\y E^L)\left[ ({\cal Q}+{\cal F})  \tilde{\breve{a}}^+ \tilde{\breve{a}}^-  -\eta r({\cal P}+\cal{R})\right]\ ,\qquad\quad
\eea
where we have defined the following functions of the modified number operator $\tilde{\breve{a}}^+\tilde{\breve{a}}^-$, 
\bea  {\cal F}(\tilde{\breve a}^+\tilde{\breve a}^-;\eta;r) & = & \int^{1}_{-1}dt \,j^n(t)\,\frac{t+1}{\chi^3}\,e^{\ft{i\s_n (t-1)}{\chi} \tilde{\breve{a}}^+\tilde{\breve{a}}^-}  \ , \\[5pt] 
{\cal G}(\tilde{\breve{a}}^+\tilde{\breve{a}}^-;\eta;r) & = & \int^{1}_{-1}dt\, j^n(t)\int^{1}_{-1}dt' \,j^n(t')\,\frac{(t'-1)(1+\s_n)+(t-1)(1-\s_n)+2}{{\tilde \chi}^3} \,e^{\ft{i\s_n(\tilde{t}-1)}{\tilde \chi}\tilde{\breve{a}}^+  \tilde{\breve{a}}^-}  \ , \qquad\quad\\[5pt]
{\cal G}'(\tilde{\breve{a}}^+\tilde{\breve{a}}^-;\eta;r) & = & \int^{1}_{-1}dt\, j^n(t)\int^{1}_{-1}dt' \,j^n(t')\,\frac{(t-1)(1+\s_n)+(t'-1)(1-\s_n)+2}{{\tilde \chi}^3} \,e^{\ft{i\s_n(\tilde{t}-1)}{\tilde \chi}\tilde{\breve{a}}^+  \tilde{\breve{a}}^-}  \ ,\qquad \quad\\[5pt]
{\cal P}(\tilde{\breve a}^+\tilde{\breve a}^-;\eta;r) & = & \int^{1}_{-1}dt\, j^n(t)\,\frac{e^{\ft{i\s_n (t-1)}{\chi}\tilde{\breve{a}}^+  \tilde{\breve{a}}^-}}{\chi^2}\, \ , \\[5pt] 
{\cal Q}(\tilde{\breve a}^+\tilde{\breve a}^-;\eta;r) & = & \int^{1}_{-1}dt\, j^n(t)\int^{1}_{-1}dt' \,j^n(t')\,\frac{\tilde{t}+1}{{\tilde \chi}^3}e^{\ft{i\s_n\tilde{t}}{\tilde \chi}\tilde{\breve{a}}^+\tilde{\breve{a}}^-}   \ , \\[5pt]
{\cal R}(\tilde{\breve a}^+\tilde{\breve a}^-;\eta;r) & = & \int^{1}_{-1}dt\, j^n(t)\int^{1}_{-1}dt'\, j^n(t')\,\frac{e^{\ft{i\s_n(\tilde{t}-1)}{\tilde \chi}\tilde{\breve{a}}^+\tilde{\breve{a}}^-} }{{\tilde \chi}^2}  \ .
\eea
with $\chi:=t+1+i\s_n\eta r(t-1)$ and $\tilde\chi:=\tilde t+1+i\s_n\eta r(\tilde t-1)$. 
Performing the integrals over $(\eta,t,t')$ yields a generating function of spacetime gauge fields that is real-analytic in $Y$ and $Z$ at $Y=Z=0$ for positive $r$.  
At $r=0$, there are contributions from the integration close to $t=-1$ and $\tilde t=-1$ that are singular at $\tilde{\breve{a}}^\pm|_{r=0;\eta=1}\equiv \tilde z^\pm \pm i\tilde \yb^{\mp}=0$. 
We leave the issue of possible corresponding divergencies in the spacetime gauge fields, to be read off in the twistor gauge, for future studies.  
%


\scss{Weyl zero-form master field}\label{Sec:fullWeyl}


Using the gauge function \eq{gaugef}, the Weyl zero-form master field in \eq{Lrot} takes the following form in the case of the diagonal solutions given by \eq{3.6}, \eq{3.51}-\eq{diagansatz2}:
\be \widehat \Phi_{(K)}(x|Y,Z)~=~\sum_{\mathbf n }  \nu_{\mathbf{n}}P_{\mathbf{n}}^L(Y) \star\k_y\  \ ,\label{generalWeyl0f}\ee
where we use the notation $P_{\mathbf{n}}^L(Y)\equiv L^{-1}(x|Y)\star P_{\mathbf{n}}(Y)\star L(x|Y)$ introduced in \eq{fL} and we have used $\widehat L^{-1}_{(K)}\star P_{\mathbf{n}} \star \k_y\star \pi(\widehat L_{(K)})=L^{-1}(x|Y)\star \tilde L^{-1}_{(K)}(z|Z)\star P_{\mathbf{n}}(Y)\star \k_y\star \tilde L_{(K)}(z|Z)\star \pi_y(L(x|Y))=P_{\mathbf{n}}^L(Y)\star\k_y$. We note that in the $K$-gauge the full zero-form $\widehat \Phi_{(K)}$ does not depend on $Z^{\underline\a}$, \emph{viz.}
\be \widehat \Phi_{(K)}~=~\widehat \Phi_{(K)}|_{Z=0}~=:~ \Phi_{(K)}\ .\ee
%
%
Recalling the definitions in \eq{PL}-\eq{brevevarkL} and using the $\star$-product lemma
\bea e^{-\ft12 Y^{\underline \a}M_{\underline{\a\b}}Y^{\underline \b}}\,\star\,\k_y \ = \ \frac{1}{\sqrt{\vark^2}}\,\left[\exp\left\{\ft12 y^\a\vark^{-1}_{\a\b}y^\b-\ft12\yb^{\ad}(\bar{ \vark}_{\ad\bd}-\bar{ v}_{\ad}{}^\a \vark^{-1}_\a{}^{\b}v_{\b\bd})\yb^{\bd}+iy^\a\yb^{\bd} \vark^{-1}_{\a\b} v^\b{}_{\bd}\right\}\right]_{\textrm{Weyl}} \ ,\qquad\label{Weyl12}\eea
for matrices $M_{\underline{\a\b}}\in \msp(4;\Comp)$, which admit the decomposition \eq{kv}, and $\vark^2:=\det \vark=\ft12 \vark^{\a\b} \vark_{\a\b}$ and $\vark^{-1}_{\a\b}=-\vark^{-2}  \vark_{\a\b}$, one obtains 
\bea&& \left[\,\widehat \Phi_{(K)}\,\right]^{\textrm{Weyl}} \ = \ 4\sum_{n_1, n_2}\nu_{n_1 ,n_2} (-1)^{|n_1|+|n_2|-1} 
\oint_{C(\ve_1)} \frac{ds_1}{2\pi i}\frac{(s_1+1)^{n_1-\ft12}}{(s_1-1)^{n_1+\ft12}}\oint_{C(\ve_2)}\frac{ds_2}{2\pi i}\frac{(s_2+1)^{n_2-\ft12}}{(s_2-1)^{n_2+\ft12}} \frac{1}{\sqrt{(\breve
\vark^L)^2}} \ \times\nn\\[5pt]
&\times& \exp\left\{\ft12 y^\a(\breve\vark^L)^{-1}_{\a\b}y^\b-\ft12\yb^{\ad}\left[(\breve{\bar{\vark}}^L)_{\ad\bd}-(\breve{\bar{v}}^L)_{\ad}{}^\a (\breve\vark^L)^{-1}_\a{}^{\b} (\breve{v}^L)_{\b\bd})\right]\yb^{\bd}+iy^\a\yb^{\bd} (\breve\vark^L)^{-1}_{\a\b} (\breve{v}^L)^\b{}_{\bd}]\right\}\ , \ \qquad \label{generalPhi}\eea
where $\breve \vark^L_{\a\b}=\frac12 (s_2+s_1) (\vark^L_{(+)})_{\a\b}+\frac12(s_2-s_1) (\vark^L_{(-)})_{\a\b}$, \emph{idem} for $\breve{\bar{\vark}}^L_{\ad\bd}$ and $\breve v^L_{\a\bd}$. 

The expression of the Weyl zero-form master field simplifies for the special superpositions of axisymmetric solutions corresponding to the rank-$n$ projectors ${\cal P}_n(E)$ and ${\cal P}_n(J)$ defined in \eq{enhanced2} and \eq{enhanced}, yielding $\mso(2)\oplus\mso(3)$-symmetric and $\mso(2)\oplus\mso(2,1)$-symmetric solutions. These depend on a single $AdS$ generator, $K_{(+)}$ or $K_{(-)}$, the $L$-rotation of which maintains the property that $(K_{(\pm)}^2)_{\underline{\a\b}}=-C_{\underline{\a\b}}$, which means that the corresponding AdS Killing vector $v^L_{\a\ad}$ is hypersurface-orthogonal, as explained in Section \ref{gaugefansatz}. As shown in \cite{Didenko:2009tc} in a pure gravity context, this property carries over to the corresponding Killing vector of the black-hole solution obtained by consistent deformation of the AdS Killing equation -- and the so-obtained black hole is therefore static. Extending this criterion to the higher-spin theory, as in \cite{Didenko:2009td}, one can therefore refer to the solutions based on the symmetry-enhanced projectors as static. The corresponding Weyl master zero-form reads
\be 
\widehat \Phi_{(K)}\ = \  2\sum_{n=\pm 1,\pm2,...} (-1)^{n-\ft{1+\e}2}\n_{n}\oint_{C(\e)}\frac{d\eta}{2\pi i}\left(\frac{\eta+1}{\eta-1}\right)^{n} e^{-4\eta K^L_{(q)}}\star \k_y  \label{enhancedPhi} \ , \ee
where $\e=n/|n|$ and $q=\pm$. Using the lemma \eq{Weyl12} together with the hypersurface-orthogonality condition for $v^L_{\a\ad}$ (\emph{i.e.}, the second equation in \eq{K2=1cpbb}) this can be written as
\bea \left[\widehat \Phi_{(K)}\right]^{\textrm{Weyl}} & = &  \frac{2}{\sqrt{(
\vark_{(q)}^L)^2}}\sum_{n=\pm 1,\pm2,...} (-1)^{n-\ft{1+\e}2}\nu_{n}\oint_{C(\e)}\frac{d\eta}{2\pi i\eta}\left(\frac{\eta+1}{\eta-1}\right)^{n}  \ \times\nn\\[5pt]&\times&\exp\left\{\frac{1}{\eta} \left[\ft12 y^\a
(\vark_{(q)}^L)^{-1}_{\a\b}y^\b+\ft12\yb^{\ad}(\bar{\vark}_{(q)}^L)^{-1}_{\ad\bd}\yb^{\bd}+iy^\a\yb^{\bd}(\vark_{(q)}^L)^{-1}_{\a\b}(v_{(q)}^L)^\b{}_{\bd}\right]\right\}
\label{enhancedPhi2} \ . \eea
Note the dependence on the inverse square root of $(\vark_{(q)}^L)^2$, both in \eq{generalPhi} and \eq{enhancedPhi2}, appearing in the prefactor and in the exponent (through $(\vark_{(q)}^L)^{-1}_{\a\b}$). This recovers the result in \eq{deltaint}, in the sense that for the solutions based on $\pi$-odd principal Cartan generators ($E$ and $iP$), for which $(\vark_{(q)})_{\a\b}=0$ and the diagonal blocks in $K_{(q)}^L$ vanish for $x^\mu=0$, the internal, $x$-independent Weyl master zero-form $\widehat\Phi'$ has a delta-function-like behaviour in twistor space. The latter is thus softened by the spacetime dependence introduced via the gauge-function, and in particular $\sqrt{(\vark_{(q)}^L)^2}$ appears as the parameter of a limit representation of the delta function.


\scs{Weak-field Analysis: Weyl Tensors in Asymptotic Regions}\label{Sec:Weyl}

In regions where $(\widehat \Phi,\widehat V_{\underline\a})$ approach their vacuum values, \emph{i.e.} vanish, the full non-abelian theory can be approximated by its abelian free-theory limit, in which it makes sense to assign physical meaning to individual Weyl tensors of fixed spin, including a physical scalar field. One can check that the latter is always real, for both the axisymmetric and the symmetry-enhanced solutions, due to the reality properties imposed on the deformation parameters that are collected in \eq{Ms} and in Table \ref{Table3} (note that the reality conditions may also constrain the ranges of $(n_1,n_2)$ and $n$). In this section we extract the Weyl zero-form component fields in the $K$-adapted gauge and examine their nature, focusing on the solutions depending on $E$ and $J$.
We defer to a future publication a more thorough study of the individual Weyl tensors  as well as of the electric/magnetic duality connecting the solutions of the Type A and Type B models, in particular taking into account the effect of going from the $K$-adapted gauge to the twistor gauge \eq{twistorgauge}.
For notational simplicity, in what follows we shall suppress the label $(K)$ indicating the $K$-adapted gauge.

\scss{Bi-axisymmetric case: almost type-D Weyl tensors}

The generating function 
\be \left[ {\cal C}(x|y)\right]^{\textrm{Weyl}}~:=~\Phi|_{\yb=0}~=~ \sum_{s=0}^\infty \frac1{(2s)!} C^{(s)}_{\a(2s)}(x) \, y^{\a(2s)}\ ,
\ee 
of the self-dual Weyl tensors $C_{\a(2s)}(x)$ is found to be 
\be {\cal C}(y|x)~=~ \sum_{n_1, n_2}\nu_{n_1 ,n_2}{\cal C}_{n_1,n_2}(y|x)\ ,\qquad \left[{\cal C}_{n_1,n_2}(x|y)\right]^{\textrm{Weyl}}~=:~\sum_{s=0}^\infty \frac1{(2s)!} (C^{(s)}_{n_1,n_2})_{\a(2s)}y^{\a(2s)}\ ,
\ee\be {\cal C}_{n_1,n_2}(y|x)~=~  4(-1)^{|n_1|+|n_2|-1} 
\oint_{C(\ve_1)} \frac{ds_1}{2\pi i}\frac{(s_1+1)^{n_1-\ft12}}{(s_1-1)^{n_1+\ft12}}\oint_{C(\ve_2)}\frac{ds_2}{2\pi i}\frac{(s_2+1)^{n_2-\ft12}}{(s_2-1)^{n_2+\ft12}}
\frac{1}{\sqrt{(\breve
\vark^L)^2}}\,\left[e^{\ft12 y^\a
(\breve\vark^L)^{-1}_{\a\b} y^\b}\right]_{\textrm{Weyl}} \ . \label{generalWeyl}\ee

As explained in Section \ref{Sspher}, the Killing two-forms $(\vark^L_{(+)})_{\a\b}$ and $(\vark^L_{(-)})_{\a\b}$ are both type-$\{1,1\}$ but have different principal spinors. 
As a consequence, for fixed $n_1,n_2$ and generic $x^\mu$, the residues in \eq{generalWeyl} amount to powers of different combinations of $\vark_{(+)}^L$ and $\vark_{(-)}^L$, and
the spin-$s$ Weyl tensor $C^{(s)}_{n_1,n_2}(x)$ is algebraically general for $s\leq k:=|n_1|+|n_2|-1$ and type-$\{s-k,s-k,\underbrace{1,\dots,1}_{2k}\}$, which we refer to as ``almost type-D'', for $s> k$. 
For instance, if $K_{(+)}=E$ and $K_{(-)}=J$, and if $n_1,n_2>0$ or $-n_1,-n_2>0$, we recall that 
\be (\vark^L_{(E)})_{\a\b}~=~ r \widetilde{\cal D}_{\a\b}\ ,\qquad (\vark^L_{(J)})_{\a\b}~=~  i\cos\theta \,\widetilde{\cal D}_{\a\b}+ \sqrt{1+r^2}\sin\theta\,\widetilde{\cal I}_{\a\b}\ ,\ee
%
where thus $\tilde u^\pm_\a(x)$ are the principal spinors of $(\vark^L_{(E)})_{\a\b}$, and $\widetilde{\cal D}_{\a\b}$ and $ \widetilde{\cal I}_{\a\b}$ have been defined in \eq{widetildeD} (and we are suppressing the index $(E)$ on them).
For example, the generating functions for the Weyl tensors at the first excited level read
\be  \left[ {\cal C}_{\ft32,\ft12}(y|x)\right]^{\textrm{Weyl}} ~ = ~ \frac{4i}{r^2}\,e^{\ft{ y^\a \widetilde{\cal D}_{\a\b} y^\b}{r}}\,\left\{i\cos\theta-\frac{1}{2r}\,y^\a y^\b\left[(r-i\cos\theta)\widetilde{\cal D}_{\a\b}+\sqrt{1+r^2}\,\sin\theta\, \widetilde{\cal I}_{\a\b}\right]\right\} \ , \label{axisexc}\ee
\be \left[{\cal C}_{\ft12,\ft32}(y|x)\right]^{\textrm{Weyl}} ~=~ -\frac{4i}{r^2} \, e^{\ft{ y^\a \widetilde{\cal D}_{\a\b} y^\b}{r}}\left\{ i\cos\th + \frac{1}{2r}\,y^\a y^\b\left[(r+i\cos\th)\widetilde{\cal D}_{\a\b}-\sqrt{1+r^2}\sin\th\,\widetilde{\cal I}_{\a\b}\right]\right\}\ ,\label{axisexd}\ee
whose scalar component falls of like $r^{-2}$ rather than $r^{-1}$. 
If $n_1,-n_2>0$ or $n_1,-n_2<0$,  it is instead convenient to use a $J$-adapted spin-frame (see Eq. \eq{kLJ}), where $\tilde u^\pm_\a(x)$ are the principal spinors of $(\vark^L_{(J)})_{\a\b}$. 

\scss{Symmetry-enhanced cases: type-D Weyl tensors}



\scsss{Spherical symmetry: generalized electric and magnetic charges}

Specializing Eq. \eq{enhancedPhi} to the case of solutions based on the spherically-symmetric projectors ${\cal P}_n(E)$ and inserting the appropriate reality properties of the deformation parameters given in Table \ref{Table3}, the rotationally-invariant Weyl zero-forms are given in the non-minimal models by
\bea \widehat\Phi ~=~ 2\sum_{n=1}^{\infty}i^n\sum_{\e=\pm 1} (-1)^{\ft{1+\e}2(n-1)}\mu_{\e n}\oint_{C(\e)}\frac{d\eta}{2\pi i}\left(\frac{\eta+1}{\eta-1}\right)^{\e n} \left[e^{-4\eta E^L}\right]_{\rm Weyl}\star \k_y \ ,\eea
and in the minimal models by 
\bea \widehat\Phi(Y|x) \ = \ -2\sum_{n=1}^{\infty}(-i)^n \mu_n\sum_{\e=\pm 1}\oint_{C(\e)}\frac{d\eta}{2\pi i}\left(\frac{\eta+1}{\eta-1}\right)^{\e n} \left[e^{-4\eta E^L}\right]_{\rm Weyl}\star \k_y \ .\eea
As explained in Section \ref{Sspher}, the $L$-rotation of $E$ generates an element $E^L$ with components \eq{kLE}. It follows that $e^{-4\eta E^L}\star \k_y|_{\yb=0}= \ft1{\sqrt{(\eta \vark^L)^2}} \exp \ft1{2\eta} y^\a(\vark^L)^{-1}_{\a\b} y^\b$ with $({\vark}^L)^{-1}_{\a\b}=\ft1{r} \widetilde{\cal D}_{\a\b}$ and
$(\vark^L)^2=- r^2$, which yields the following Weyl-tensor generating
function in non-minimal models:
\bea {\cal C}(y|x) \ = \
\frac{2}{ir}\sum_{n=1}^{\infty}i^n\oint_{C(1)}\frac{d\eta}{2\pi
i\eta}\left(\frac{\eta+1}{\eta-1}\right)^{n}\sum_{\e=\pm 1}
(-1)^{\ft{1+\e}2(n-1)}\mu_{\e n}\,\left[e^{\ft{\e}{2\eta}\,
y^\a\vark_{L\,\a\b}^{-1}y^\b}\right]_{\textrm{Weyl}} \ ;\label{o3Weylnm}\eea
and in minimal models:
\bea {\cal C}(y|x) \ = \ \frac{4i}{r}\sum_{n=1}^{\infty}(-i)^n
\mu_n\oint_{C(1)}\frac{d\eta}{2\pi i\eta}\left(\frac{\eta+1}{\eta-1}\right)^{ n}
\left[\cosh\left(\frac{1}{2\eta}\,y^\a\vark_{L\,\a\b}^{-1}y^\b\right)\right]_{\textrm{Weyl}}\ .\label{o3Weylm}\eea
The sum over $\e=\pm 1$ in \eq{o3Weylnm}
includes the independent contributions of positive-energy ($\e=+1$)
and negative-energy ($\e=-1$) excitations, that need to be included in
the sum with equal coefficients in the case of the minimal model
\eq{o3Weylm}. In other words, the
spherically-symmetric solutions are built in terms of rank-$n$ projectors on combinations of states (with fixed energy and vanishing $J$)
in the subsectors $({\cal F}_1^+\otimes{\cal F}_2^+)\oplus({\cal
F}_1^-\otimes{\cal F}_2^-)$ of the whole Fock-space.

For a fixed projector ${\cal P}_n(E)$, expansion in $y$ and auxiliary integration yield the physical  scalar ($s=0$) and an infinite tower of spherically-symmetric Type-D Weyl tensors of spin $s\geqslant 1$ of the form (up to real $n$-dependent numerical factors)
\bea C^{(n)}_{\a(2s)}\ \sim \ \frac{i^{n-1}\mu_n}{r^{s+1}}\,(\tilde u^+
\tilde u^- )^s_{\a(2s)} \ ,\label{nsWeylspher}\eea
where the deformation parameter $\mu_n$ is real in the case of scalar singleton projectors ($n$ odd) and purely imaginary in the case of spinor singleton projectors($n$ even).
Although our solutions are not presented in the physical twistor gauge, one may argue that going to this gauge will alter the leading behaviors of the Weyl tensors only by higher orders in $\mu_n$ (it may also affect the asymptotic anti-de Sitter radius itself). 
Thus, to the leading order in $\mu_n$, asymptotically defined spin-$s$ charges for $s\geqslant 1$ can be read off by comparing with the linearized gauge-field equations. 
Whether it is possible to invert the relation between these charges and the deformation parameters remains to be clarified: if possible then one could in principle choose the deformation parameters as to switch off all spins except one. 

The asymptotic charges depend on the parameter $b$. Drawing on the
analogy with the general form of the spin-$1$ Faraday tensor and the
spin-$2$ Weyl tensor of an $AdS_4$ black hole \cite{Didenko:2009tc},
one can thus regard the deformation parameters of the solutions based on the scalar singleton
as generalized electric charges (or generalized masses) in the Type A model and generalized magnetic charges (or generalized NUT charges) in the Type B model, and, conversely, those of the solutions based on the spinor
singleton as magnetic-like charges in the Type A model and electric-like charges in the Type B model. 
In this sense, the solutions of the Type A and Type B minimal bosonic models are related by a generalized electromagnetic duality\footnote{The fact that the Type A and Type B models, respectively, with perturbative spectra consisting of the symmetrized tensor product of two scalar singletons (containing a parity-even scalar in $\mathfrak D(1;(0))$) and the anti-symmetrized product of two spinor singletons (containing a parity-odd scalar in $\mathfrak D(2;(0))$) have been conjectured to be dual to free scalars and free fermions \cite{Sezgin:2003pt} implies that the electric-like solutions should have a direct interpretation in terms of the free holographic conformal field theory (perhaps in terms of thermal properties).}.

In particular, by setting $\mu_n=\delta_{n,1}\mu $ in \eq{o3Weylnm}, one obtains the static BPS
solution (of the non-minimal model) found in \cite{Didenko:2009td}\footnote{As mentioned in the
Introduction, the exact solution of
\cite{Didenko:2009td} was found not via the gauge-function
method, but rather by first solving the equations in first-order
approximation and then checking that the non-linear corrections
vanish identically (due to the Kerr-Schild form of the solution).}
\bea \left[\Phi_{\ft12,\ft12}(Y|x)\right]^{\textrm{Weyl}} \ = \  \frac{4\m}{r}\,\exp\left\{\ft12 y^\a
(\vark^L)^{-1}_{\a\b}y^\b+\ft12\yb^{\ad}(\bar{\vark}^L)^{-1}_{\ad\bd}\yb^{\bd}+iy^\a\yb^{\bd}(\vark^L)^{-1}_{\a\b}(v^L)^\b{}_{\bd}\right\}
\ ,\label{sphericvac}\eea
which is therefore based on the scalar singleton vacuum-to-vacuum
projector $4e^{-4E}$. Its spin-$2$ sector contains the Weyl tensor
of an AdS-Schwarzschild black hole of mass $\mu$, and, as remarked in
\cite{Didenko:2009td}, all the Weyl tensor in \eq{sphericvac}
correspond to spin-$s$ gauge fields of Kerr-Schild type.

The singularities in the individual Weyl tensors at $r=0$ is resolved at the level of the full master-field in twistor space, \emph{e.g.} \
\be \widehat \Phi_{\ft12,\ft12}~\stackrel{r\rightarrow 0}{\longrightarrow} ~\widehat \Phi'_{\ft12,\ft12} ~=~\nu_{\ft12,\ft12}\kappa_{y-i\s_0\bar y}~=~2\pi \nu_{\ft12,\ft12} \left[\delta^2(y-i\s_0 \yb)\right]_{\textrm{Weyl}}\ ,\label{Phidelta} \ee
which is actually a well-defined distribution viewed as the symbol of an operator (see Conclusions). For a general $\pi$-odd projector $4 e^{-y^\a v_{\a\ad}\bar y^{\ad}}$ we have
\bea e^{-y^\a v_{\a\ad}\bar y^{\ad}}\star \kappa_y \ = \ 2\pi\,\left[\d^{2}(y-iv\bar y)\right]_{\textrm{Weyl}} \ = \ \kappa_{y-iv\bar y} \ .\eea
In this sense, the radial coordinate $r$ appears in the generating functions \eq{o3Weylnm} and \eq{o3Weylm} as the parameter of a limit representation of a $\d$-function in twistor space.

\scss{Cylindrical symmetry: electric and magnetic Weyl tensors}\label{Sec:Weylcyl}

We now turn to symmetry-enhanced projectors that depend on the difference $w_2-w_1$ of the number operators $w_1=E-J$, $w_2=E+J$ (\emph{i.e.}, we set $\e_1=-\e_2$ in \eq{enhanced}), thus getting the projectors ${\cal P}_n(J)$. Using Eqs. \eq{nuenhanced} and \eq{enhanced} the corresponding Weyl zero-form reads
\bea \widehat \Phi \ = \ 2\sum_{n=1}^{\infty}i^{n+1}\sum_{\e=\pm 1}(-1)^{\ft{\e+1}2 (n-1)}\m_{\e n}\oint_{C(\e)}\frac{d\z}{2\pi i}\left(\frac{\z+1}{\z-1}\right)^{\e n}\left[e^{-4\z J^L}\right]_{\rm Weyl}\star \k_y \ ,\eea
for the non-minimal model, and
\bea \widehat \Phi \ = \ 2\sum_{n=1}^{\infty}(-i)^{n+1}\m_n\sum_{\e=\pm
1}\oint_{C(\e)}\frac{d\z}{2\pi i}\left(\frac{\z+1}{\z-1}\right)^{\e
n}\left[e^{-4\z J^L}\right]_{\rm Weyl}\star \k_y \ ,\eea
in the case of the minimal model, with $e^{-4\z J^L}=L^{-1}\star
e^{-4\z J}\star L$. One can repeat the same steps of the
spherically-symmetric case, beginning this time with the generator
$J=-\ft18\left[y^\a(\s_{12})_{\a\b}y^\b+\yb^{\ad}(\bar{\s}_{12})_{\ad\bd}\yb^{\bd}\right]$,
\emph{i.e.}, with a $\vark$-type $K$ matrix, with
$\vark_{\a\b}=i(u^+_\a u^-_\b+u^-_\a u^+_\b)$. The $L$-rotation
gives rise to $\vark^L$ and spacelike $v^L$ (which corresponds, in
fact, to $\ft\partial{\partial\varphi}$) that, at any spacetime
point, can be brought to the form
\bea v^L_{\a\bd} & = & r\sin\theta\,(\tilde u^+_a\bar{\tilde u}^-_{\bd}+\tilde u^-_\a\bar{\tilde u}^{+}_{\bd}) \ ,\label{vLJ}\\[5pt]
\vark^L_{\a\b} & = & i\sqrt{1+r^2\sin^2\theta}\,(\tilde u^+_\a \tilde u^-_\b+\tilde u^-_\a \tilde u^+_\b) \ , \label{kLJ}\eea
in global coordinates and on a properly chosen, $J$-adapted spinor basis
$\tilde u^{\pm}_\a(x)$ (see Appendix \ref{App:L-rot}).  One has $e^{-4\zeta J^L}\star \k_y|_{\yb=0}= \ft1{\sqrt{(\zeta \vark^L)^2}} \exp \ft1{2\zeta} y^\a(\vark^L)^{-1}_{\a\b} y^\b$ with $({\vark}^L)^{-1}_{\a\b}=-i(1+r^2\sin^2\th)^{-1/2} \,\widetilde {\cal D}_{\a\b}$ and
$(\vark^L)^2=1+r^2\sin^2\theta$, from which it follows that the Weyl tensor
generating function reads
\bea   C(y|x) \ = \
\frac{2}{\sqrt{1+r^2\sin^2\theta}}\sum_{n=1}^{\infty}i^{n+1}\oint_{C(1)}\frac{d\z}{2\pi
i\z}\left(\frac{\z+1}{\z-1}\right)^{n}\sum_{\e=\pm 1}
(-1)^{\ft{1+\e}2(n-1)}\m_{\e n}\,\left[e^{\ft{\e}{2\z}\,
y^\a\vark_{L\,\a\b}^{-1}y^\b}\right]_{\textrm{Weyl}} \ ,\qquad\label{o21Weylnm}\eea
for solutions of the non-minimal model, and
\bea C(y|x) \ = \
\frac{4}{\sqrt{1+r^2\sin^2\theta}}\sum_{n=1}^{\infty}(-i)^{n+1}
\m_n\oint_{C(1)}\frac{d\z}{2\pi i\z}\left(\frac{\z+1}{\z-1}\right)^{
n} \left[\cosh\left(\frac{1}{2\z}\,y^\a\vark_{L\,\a\b}^{-1}y^\b\right)\right]_{\textrm{Weyl}}\
,\label{o21Weylm}\eea
for solutions of the minimal one. Since the Killing two-form is imaginary, for every fixed $n$ the electric/magnetic type of the type-D Weyl tensors flips according to whether the spin is even/odd, for $n$ odd, vicecersa for $n$ even,
\bea C^{(n)}_{\a(2s)} \ \sim \ \frac{i^{n+s+1}\m_n}{(1+r^2\sin^2\theta)^{\ft{s+1}2}}\,(\tilde u^+ \tilde u^-)^s_{\a(2s)} \ . \eea
Note that such Weyl tensors do not blow up anywhere and do not vanish at spatial infinity (they are constant along the $z$ axis, with a behaviour similar to that of the Melvin solution in General Relativity \cite{Melvin}). As anticipated, these solutions are $\mso(2)_J\oplus\mso(2,1)_{E, M_{03},P_3}$-symmetric, and are built on the spacelike AdS Killing vector $\partial/\partial\varphi$ in the same way as the spherically-symmetric ones are based on the timelike vector $\partial/\partial t$, \emph{i.e.}, the $\mso(2,1)$ is the stability subalgebra of $\partial/\partial\varphi$. In other words, here the roles of $E$ and $J$ are exchanged, with respect to the rotationally-invariant case, and the corresponding solutions are based on projectors onto combination of states belonging to non-unitary analogues of the (anti-)supersingleton of fixed $J$ and vanishing energy. Such states belong to the subspace $({\cal F}_1^-\otimes{\cal F}_2^+)\oplus({\cal
F}_1^+\otimes{\cal F}_2^-)$ of the full Fock space.

Note also that the regularity of such Weyl tensors corresponds, at the level of the internal solution $\Phi'$, to the regularity of the product of a $\pi$-even projector with $\k_y$, that in fact reproduces the projector itself up to a sign,
\bea e^{\mp\ft12y^\a\vark_{\a\b}y^\b+\textrm{h.c.}}\star\kappa_y \ = \ \mp\, e^{-\ft12y^\a\vark_{\a\b}y^\b+\textrm{h.c.}} \ .\eea

\scs{Strong-field Analysis: Zero-form Charges}\label{Sec:0f}

Given the huge gauge symmetry of the theory, it is extremely important to have some quantities that are invariant under the full set of gauge transformations of the theory and that can be evaluated on the solutions. Such invariants enable one to distinguish gauge-inequivalent field configurations and to characterize them physically even in regions of spacetime where the curvatures are large and consequently the weak-field analysis is not reliable.  We shall focus on the evaluation of the zero-form charges \eq{fullamp}, that, as anticipated in Section \ref{Sec:0charges}, are finite on the solutions at hand.

Inserting the general expression of the Weyl zero-form \eq{generalWeyl0f} and using that $\k_y\star\k_y=1$ and the orthogonality and idempotency of the projectors, one gets
\bea {\cal I}_{2N}&:=& \widehat Tr_{\Real}[(\widehat\Phi\star\pi(\widehat\Phi))^{\star N}\star\widehat\kappa\widehat{\bar\kappa}]  \ = \ \widehat Tr_{\Real}[(\widehat\Phi'\star\pi(\widehat\Phi'))^{\star N}\star\widehat\kappa\widehat{\bar\kappa}]  \ = \ \sum_{\mathbf{n}\in (\integ +\ft12)^2}\nu_{\mathbf{n}}^{2N}\left.P_{\mathbf{n}}\right|_{Y=0}\ ,\label{chargesprime} \eea
for the axisymmetric projectors and analogously, substituting the double index $\mathbf{n}$ with the single index $n=\pm1,\pm2,...$  everywhere, for the symmetry-enhanced projectors ${\cal P}_n$. From the forms \eq{diagproj} and \eq{enhanced2} (equivalently \eq{intproj} and  \eq{enhanced})  of the projectors it thus follows that
\bea {\cal I}_{2N}(K_{(+)},K_{(-)}) \ = \  4\sum_{\mathbf{n}\in (\integ +\ft12)^2}(-1)^{n_1+n_2-1} \nu_{\mathbf{n}}^{2N}  \label{0finvo2} \ ,\eea
for the axisymmetric solutions based on a given Cartan pair $(K_{(+)},K_{(-)})$, and
\bea {\cal I}_{2N}(K_{(q)}) \ = \  4\sum_{n=\pm1,\pm2,...}(-1)^{n-1}|n| \nu_{n}^{2N} \ ,\label{0finvenh}\eea
for the symmetry-enhanced ones, where we recall that the relation between $(n_1,n_2)$ and $n$ is $n:=q n_1+n_2$ (see also Appendix \ref{AppProj} for the notation concerning projectors). Specifying to the first Cartan pair $(K_{(+)}=E$, $K_{(-)}=J)$ and recalling the corresponding reality conditions on the deformation parameters (see \eq{Ms} Table \ref{Table3}) one gets
\be {\cal I}_{2N}(E,J) \ = \ -4q\sum_{\mathbf{n}\in (\integ +\ft12)^2}(-1)^{(N+1)(n_1+n_2)}\m^{2N}_{\mathbf{n}} \ , \ee
and 
\be {\cal I}_{2N}(K_{(q)})  \ = \ -4q\sum_{n=\pm1,\pm2,...}|n|(-1)^{(N+1)(n+\ft{1-q}2)}\m^{2N}_{n} \ . \ee

The conclusion is that the zero-form invariants ${\cal I}_{2N}$ extract, in general, a linear combination of powers of the deformation parameters $\nu_{\mathbf{n}}$ that characterize every solution, and that can be thought of as the eigenvalues of the expansion of the solution on the (anti-)supersingleton basis of projectors. For solutions based on a single projector (such as, for example, the BPS solution \eq{sphericvac} of \cite{Didenko:2009td}), these local invariants capture  (even powers of) the unique deformation parameter sitting in front of the spin-two Weyl tensor as well as of its higher and lower-spin partners,  formally resembling the ADM mass. Note also that, for any odd N, there is a sign difference between the invariants referred to solutions based of (anti-)supersingleton projectors (those with $\e_1=\e_2$, \emph{i.e.}, projectors on states belonging to the Fock space sectors $({\cal F}^+\otimes{\cal F}^+)\oplus({\cal F}^-\otimes{\cal F}^-)$)  and on its ``non-compact'' counterpart ($\e_1=-\e_2$, \emph{i.e.}, projectors on states in $({\cal F}^+\otimes{\cal F}^-)\oplus({\cal F}^-\otimes{\cal F}^+)$), which is related to the opposite reality properties of the deformation parameters. In any case, solutions with deformation parameters $\nu_{\mathbf{n}}$ such that the combinations \eq{0finvo2}, \eq{0finvenh} give different results are gauge-inequivalent\footnote{It is possible, however, to find non-polynomial parameters that transform any projector $P_n$ into any other $P_m$ (for example, the element $g = 1\!\!\!\!1 - \ket{n}\bra{n}-\ket{m}\bra{m}+\ket{n}\bra{m}+\ket{m}\bra{n}$ acts as $g^{-1}\star\ket{n}\bra{n}\star g=\ket{m}\bra{m}$), and that therefore alter the value of the zero-form charges. We shall insist on this set of invariants to distinguish gauge-inequivalent solutions, and consequently restrict the class of allowed gauge transformations to the set of ``small'' gauge parameters that do not permute projectors. The distinction is similar to the well-known one in Yang-Mills theories between``small''  gauge parameters, that do not connect different topological sectors of the theory, and ``large'' gauge transformations, that map sectors with different winding number into each other. The difference is that in the context of higher-spin gravity the ``large'' gauge transformations altering the zero-form charges pick boundary terms in twistor-space rather than in spacetime.}. 

It is interesting to notice that \eq{0finvenh} is \emph{not} divergent for any choice of (finite) deformation parameters, at least as long as the examined solution is based on finitely many projectors. This means that, for instance, although the rotationally-invariant Weyl curvatures \eq{nsWeylspher} asymptotically (where all Weyl tensors are weak and fields of different spins decouple from each other) resemble those of a collection of ``higher-spin Schwarzschild black holes'', the apparent singularity in $r=0$ (\emph{i.e.}, in the strong-curvature region, where the pure spin-$2$ curvature invariants are no longer good observables) of the individual Weyl tensors does not actually lead to divergent higher-spin invariant zero-form charges. 

Let us note also that, under the same conditions, the interaction ambiguity \eq{calB}-\eq{Theta} is also well-defined on the solutions here presented. The key point is again that any product of the basic building block $\widehat \Phi\star\pi(\widehat \Phi)$ collapses to a single power of the projectors, due to the orthogonality properties of the latter. Indeed, assuming for example \eq{theta}, one has 
\be \theta[\widehat \Phi\star\pi(\widehat \Phi)] \ = \ \sum_{k=0}^{\infty}\,\sum_{\mathbf{n}\in (\integ +\ft12)^2} \theta_{2k}({\cal I}_{2N})\n_{\mathbf{n}}^{2k} P_{\mathbf{n}} \ ,\ee
where the coefficients $\theta_{2k}({\cal I}_{2N})$ reduce to functions of the deformation parameters $\n_{\mathbf{n}}$ according to \eq{0finvo2}.

\scs{Conclusions, Comments and Outlook}\label{Sec:concl}

\scss{Summary and comments}

In this paper we have presented six infinite families of exact solutions to Vasiliev's four-dimensional higher-spin field equations. The solutions are obtained by combining the gauge-function method, previously used for other exact solutions \cite{Sezgin:2005pv,Iazeolla:2007wt,Didenko:2006zd}, with an internal Ansatz generalizing that of \cite{Didenko:2009td}, based on the separation of the dependence of the master-fields on $Y$ and $Z$ twistor variables. The resulting solutions are organized in three pairs, each pair characterized by a biaxial isometry group $\mso(2)\oplus\mso(2)$ embedded into $\msp(4;\Real)$ in three inequivalent ways. One of the families contains a subset of solutions in which one of the two $\mso(2)$ enhances to $\mso(3)$, while in the remaining families the enhanced symmetry algebra is $\mso(2,1)$. In all of our solutions, all spins are activated for generic choices of deformation parameters. While each of the symmetry-enhanced solutions contains exclusively Petrov type-D Weyl tensors, the Weyl tensors of the biaxially symmetric solutions are not type-D but still algebraically special for large enough spin, such that one may refer to them as ``almost type-D''. 

Given the high complexity of higher-spin gravity -- a theory describing infinitely many fields coupled through infinitely non-linear and non-local interactions -- we find it rather interesting that Vasiliev's remarkable formulation facilitates the systematic construction of non-trivial exact solutions. This is essentially due to the formal simplicity of Vasiliev's equations formulated as an unfolded system in correspondence spaces $T^\ast{\cal X}\times {\cal T}$, where $T^\ast{\cal X}$ contains spacetime and ${\cal T}$ is a twistor space.  The gauge-function solution method, which locally strips the spacetime dependence off the master-fields, is particularly natural for this type of unfolded equations that relate the $x$-dependence of the fields to their internal, twistor-space behaviour. In this fashion, it has been shown in the paper how the spacetime properties of the solutions are to a large extent inherited from those of the twistor-space projectors they are built on. 

For example, the solutions with $\mso(2)_E\oplus\mso(2)_J$-symmetry that admit spherically-symmetric enhancements are based on non-polynomial fibre elements that are projectors $P_{n_1,n_2}(Y)$, with $n_1n_2>0$, onto (anti-)supersingleton states (as shown in \cite{Iazeolla:2008ix}), which are characterized by $|E|>|J|$. More precisely, having absorbed the spacetime dependence into gauge functions, the remaining internal, $x$-independent master-fields are expanded over such a basis of projectors, with eigenvalues $\n_{n_1,n_2}$ playing the role of deformation parameters. The resulting $x$-independent Weyl zero-form turns out to be a distribution in twistor-space (see for example the first of Eqs. \eq{deltaint}). Reinstating the $x$-dependence using a specific, convenient gauge function, its singular behaviour is softened: the radial coordinate $r$ indeed appears as the parameter of a limit representation of a twistor-space delta function.  The resulting individual Weyl tensors are finite for $r>0$ and, in fact, in the chosen gauge coincide with the linearized Weyl tensors --- thereby extending the well-known Kerr-Schild property of black-hole gravity solutions to the higher-spin context --- depending on $r$ as $r^{-s-1}$, where $s$ is their spin. In particular, the spin-$2$ Weyl tensor exhibits both the singular spacetime behaviour at $r=0$ and the algebraic structure of that of an AdS-Schwarzschild black-hole solution. 

On the other hand, the projectors with  $n_1n_2<0$, \emph{i.e.} the sector where $|E|<|J|$, give rise to an internal, $x$-independent Weyl zero-form master-field that is a regular function on twistor space, corresponding, after reinstating the $x$-dependence, to spacetime curvatures which are regular everywhere and exhibit cylindrical symmetry --- in a fashion that is reminiscent of the Melvin solution of General Relativity (though the fall-off behaviour with the cylindrical radius is different).  

Instead of looking at individual Weyl tensors, in order to characterize the solutions in strong curvature regions one may instead examine higher-spin invariant observables. As we have seen, no divergence occurs in the higher-spin invariant zero-form charges that we have studied in this paper, even in the cases in which the individual spin-$s$ Weyl tensors blow up. In this sense, the spacetime singularities may be resolved at the level of master-fields living in correspondence space. Let us examine, as a concrete example, the behaviour of the Weyl zero-form master-field of the BPS solution \eq{sphericvac}. As can be seen from Eq. \eq{Phidelta}, at $r=0$ the Weyl-ordered symbol of this master-field is a distribution in twistor space. However, by moving to normal-ordering the resulting symbol becomes a regular, gaussian function, as discussed in Appendix \ref{App:A} (see also Eq. \eq{kznorm}).  This resolution of component-field singularities is tied to the fact that the fibre-space of higher-spin gravity is infinite-dimensional, which implies that a change of ordering in general gives rise to an infinite ``vacuum energy'' that may cancel singularities. Therefore, in this sense it is conceivable that the coupling of an infinite tower of gauge fields of all spins results in that the singularities of the individual Weyl tensors are actually artefacts of the choice of ordering. Indeed, the zero-form invariants that we have tested are, at least formally, not only invariant under higher-spin gauge transformations but also under change of ordering (that reduce to total derivatives in twistor space, see Appendix \ref{App:A}).  

\scss{Outlook}

Studying exact solutions to higher-spin gravity poses many stimulating challenges. From a conceptual point of view, one needs to develop tools and methods to analyze their physical properties, and in particular the geometry of a theory that, due to the non-locality of the interactions, represents a departure from the familiar framework of General Relativity, including perturbative stringy corrections. 

A natural question induced by the non-locality of higher-spin interactions is whether it can suppress short-distance singularities. Rather than examining individual component fields, it makes more sense to investigate this issue at the level of various higher-spin observables such as those described in Section \ref{Sec:0charges} (see also \cite{Colombo:2010fu} and \cite{Sezgin:2011hq}). 
The analysis via the zero-form invariants ${\cal I}_{2N}$ carried out in this paper is, however, not conclusive: apart from subtleties related to the treatment of boundary terms in twistor space, arising from changing the ordering prescription, the singular nature of the solutions at hand remains to be tested with other invariants, some of which resemble more closely the non-local observables familiar from General Relativity. Particularly interesting in this respect is the integral of the on-shell closed two-form \eq{abelianpform} (with $p=2$) over a closed surface surrounding the singularity at $r=0$ of the solutions belonging to our first family. Moreover, higher-spin invariants involving Weyl curvatures, such as ${\cal I}_{2N}$, may turn out to be the proper quantities for a generalization of the classification criteria of purely gravitational solutions, such as the Petrov classification. 

It is actually possible to have divergent zero-form invariants for solutions of the type constructed here, possibly signaling a physical singularity, provided they are based on infinitely many projectors with not too small deformation parameters. Another noteworthy fact that we have observed about such solutions is that, at least for certain choices of the deformation parameters $\m_n$, the poles of the corresponding Weyl tensors generating function, inherited from the integral realization of the projectors, acquire an imaginary part and move away from the real axis. This migration of poles may turn out to have interesting physical effects, not unrelated to the divergence of the zero-form charges. For instance, in the case of the axisymmetric $(E,J)$ solutions based on supersingleton projectors, this mechanism implies that the angular dependence in the corresponding Killing two-form $\breve \vark$ is no longer weighted by an evanescent parameter $\zeta$ (which would inevitably fix the singularity of the Weyl tensors in the origin, as in Eqs. \eq{axisexc} and \eq{axisexd}, not differently from the spherically-symmetric case), but rather by a complex, non-vanishing quantity. This gives rise to an angular dependence of the Weyl-tensor singularities, and the solutions may thus acquire a non-trivial angular momentum. We plan to report on this effect in a future publication. 

It would also be interesting to study in detail the spacetime behaviour of the gauge-fields generating function $W_\mu$, and in particular the contribution of the $Z$-space projectors resulting from the inclusion of $\circ$-product projectors in $j_{\mathbf{n}}^\pm(t)$, especially in the light of the results of \cite{Iazeolla:2007wt} where it was shown that certain specific choices of the parameters $\th_k$ (see Appendix \ref{App:D}) would lead to a degenerate spin-$2$ component field. While in the present paper we have mostly been working in what we called $K$-gauge, where the alignment induced by $\tilde L_K$ (see Section \ref{gaugefansatz}) ensures that singularities in $\widehat S'_{\underline\a}$ at $Z=0$ are resolved in $\widehat S_{K;\underline\a}$, it would be important to be able to study the solutions in different gauges -- in particular in the twistor gauge, that can be reached by constructing an appropriate element $\widehat G^v_K$ acting on the gauge-function as in \eq{Gnew}. As far as the spacetime gauge fields are concerned, based on properties of known exact solutions\footnote{In \cite{Sezgin:2005pv} it was found that in $\mso(3,1)$-invariant solutions with deformation parameter $\nu$, the metric $ds^2_{\nu}$ approaches an $AdS_4$ metric $ds^2_{\l(\nu)}$ with deformed inverse radius $\l(\nu)\neq \l(0)\equiv \l$ in the asymptotic region where the Weyl zero-form $\widehat \Phi$ goes to zero. 
In other words, the fact that $\widehat \Phi$ falls off in an asymptotic region does not imply that the weak-field expansion in this region is around the undeformed $AdS_4$ vacuum with metric $ds^2_{\l}$.}, 
we expect that the deviation in the asymptotic behavior of $\widehat W_K=\widehat L^{-1}_K\star d\widehat L_K-\widehat K_K$ from that of $\widehat W_v=\widehat L^{-1}_v\star d\widehat L_v-\widehat K_v$ may remain finite at $Z=0$, that is, $W_K=\widehat W_K|_{Z=0}$ and $W_v=\widehat W_v|_{Z=0}$ may exhibit different asymptotic behaviors, possibly amplified by the aforementioned singularities in $\widehat S'_{\underline\a}$ at $Z=0$. 
We plan to return to these issues, and the construction of the gauge function $\widehat G^v_K$ in future studies.

Another open question is whether the candidate higher-spin black-hole solutions possess horizons and to investigate the associated thermodynamics. This is a subtle issue in this context, essentially due to the fact that the familiar general-relativistic concepts involved -- the standard metric tensor, invariant length interval, trapped surfaces, etc. -- are not higher-spin invariant, and suitable generalizations need to be defined to probe the strong-field regions. A class of objects that can be useful in this sense is the set of higher-spin metrics $G_{M_1\dots M_{s}}= \widehat{Tr}_{\Real}\left[\widehat\kappa\widehat{\bar\kappa}\star \widehat E_{(M_1}\star \cdots \star \widehat E_{M_s)}\right]$ ($s=2,4,\dots$), first proposed in \cite{Sezgin:2011hq}. It would be interesting to evaluate such generalized metrics on our solutions and to study their behaviour. 

Moreover, a natural direction to explore, in the light of the higher-spin gravity/O(N)-vector models correspondence, is the study of the boundary duals of such solutions -- a direction which would not only be worth pursuing in its own right but may also possibly shed some light on some of the above-mentioned issues, including the thermodynamics of such systems. It is an interesting fact, in this sense, that some of our solutions are directly related, through the projectors $P_{n_1,n_2}(E,J)$, to supersingleton states, \emph{i.e.} to the modes of boundary conformal scalar and fermion fields. 

Due to the non-locality of higher-spin interactions, further surprises may be kept in store in the context of the generalization to multi-body solutions obtained by dressing linear combinations of single-body solutions centered at spatially well-separated points. While in supergravity the existence of such solutions is intuitively physically clear, due to the locality of the theory, whether or not a large spatial separation leads to negligible corrections in the higher-spin context is a more non-trivial question. A natural tool at our disposal to study this issue are the zero-form invariants ${\cal I}_{2N}$, that one may think of as some sort of correlation functions among soliton-like objects and that can be used to test a kind of cluster decomposition principle (for a related discussion, see also \cite{Colombo:2010fu}). For example, for a hypothetical two-body solution of the form $\widehat\Phi = \widehat\Phi_1+ \widehat\Phi_2+\widehat \Phi_{12}$, where $\widehat\Phi_i$ are two spherically-symmetric solutions \eq{sphericvac} the centers of which are a distance $r_{12}$ apart, one has that ${\cal I}_2(\Phi)=\widehat{Tr}_{\Real}\left[\widehat\Phi\star\pi(\widehat\Phi)\star\k\bar\k\right]$ is given by ${\cal I}_{2}(\widehat\Phi_1)+{\cal I}_{2}(\widehat\Phi_2)$ plus small cross-terms $\widehat{Tr}_{\Real}\left[\widehat\Phi_1\star\pi(\widehat\Phi_2)\star\k\bar\k\right]$  that decay as $(1+r_{12}^2)^{-1/2}$ (plus small contributions from $\Phi_{12}$ that we have not analyzed).

Furthermore, we claim it is possible to further exploit the Ans\"atze based on Fock-space projectors to obtain other types of exact solutions, with different physical and algebraic properties. For instance, one possibility is to study the inclusion of other Weyl zero-form moduli such as massless particle states (belonging to the $\msp(4;\Real)$ representation $\mD^{(\pm)}$, mentioned in Section \ref{gaugefmethod}, spanned by the twisted-adjoint action on the lowest-weight state projector ${\cal P}_1(E)$, as shown in \cite{Iazeolla:2008ix}) in the solutions along with the soliton-like, coherent states here treated. Moreover, suitable limits of the Ans\"atze here presented could be studied, and in particular the possibility of having solutions of Petrov-type N seems within reach, leading to field configurations where the distributions in twistor-space appearing in the Ansatz actually lead to distributions in spacetime and to generalizations of impulsive-wave solutions of General Relativity.

\vspace{2.5cm}

{\bf Acknowledgments}

We are grateful to G.~Barnich, F.~Bastianelli, N.~Boulanger, D.~Chialva, N.~Colombo, O.~Corradini, M.~R.~Douglas, D.~Fioravanti, C.~Maccaferri, A.~Maloney, L.~Mazzucato, O.~Pujolas, S.~S.~Razamat, A.~Sagnotti, Ph.~Spindel, W.~Taylor, A.~Waldron, Xi Yin and especially to V.~E.~Didenko and M.~A.~Vasiliev  for stimulating discussions. C.~I. wishes to thank the Simons Center for Geometry and Physics at Stony Brook University for kind hospitality during the final stage of this work, and gratefully acknowledges the ``Universit\`a degli Studi `G. Marconi' '' in Rome for partial support. 

\vspace{2cm}

\begin{appendix}


\scs{Spinor conventions and $AdS_4$ Background}\label{App:conv}


We use the conventions of \cite{Iazeolla:2008ix} in which $SO(3,2)$ generators $M_{AB}$ with $A,B=0,1,2,3,0'$ obey
\be [M_{AB},M_{CD}]\ =\ 4i\y_{[C|[B}M_{A]|D]}\ ,\qquad
(M_{AB})^\dagger\ =\ M_{AB}\ ,\label{sogena}\ee
which can be decomposed using $\eta_{AB}~=~(\eta_{ab};-1)$ with $a,b=0,1,2,3$ as 
\be [M_{ab},M_{cd}]_\star\ =\ 4i\y_{[c|[b}M_{a]|d]}\ ,\qquad
[M_{ab},P_c]_\star\ =\ 2i\y_{c[b}P_{a]}\ ,\qquad [P_a,P_b]_\star\ =\
i\lambda^2 M_{ab}\ ,\label{sogenb}\ee
where $M_{ab}$ generate the Lorentz subalgebra $\mso(3,1)$, and $P_a=\l M_{0'a}$ with $\l$ being the inverse $AdS_4$ radius related to the cosmological constant via $\L=-3\l^2$. Decomposing further under the maximal compact subalgebra, the $AdS_4$ energy generator $E=P_0=\l M_{0'0}$ and the spatial $\mso(3)$ rotations are generated by $M_{rs}$ with $r,s=1,2,3$.
In terms of the oscillators $Y_{\underline\a}=(y_\a,\yb_{\ad})$ defined in \eq{oscillators}, their realization is taken to be
\be M_{AB}~=~ -\ft18  (\C_{AB})_{\underline{\a\b}}\,Y^{\underline\a}\star Y^{\underline\b}\ ,\label{MAB}\ee
 \be
 M_{ab}\ =\ -\frac18 \left[~ (\s_{ab})^{\a\b}y_\a\star y_\b+
 (\sb_{ab})^{\ad\bd}\bar y_{\ad}\star \yb_{\bd}~\right]\ ,\qquad P_{a}\ =\
 \frac{\l}4 (\s_a)^{\a\bd}y_\a \star \yb_{\bd}\ ,\label{mab}
 \ee
using Dirac matrices obeying $(\C_A)_{\underline\a}{}^{\underline\b}(\C_B C)_{\underline{\b\c}}=
\eta_{AB}C_{\underline{\a\c}}+(\C_{AB} C)_{\underline{\a\c}}$, and van der Waerden symbols obeying 
 \be
  (\s^{a})_{\a}{}^{\ad}(\sb^{b})_{\ad}{}^{\b}~=~ \y^{ab}\d_{\a}^{\b}\
 +\ (\s^{ab})_{\a}{}^{\b} \ ,\qquad
 (\sb^{a})_{\ad}{}^{\a}(\s^{b})_{\a}{}^{\bd}~=~\y^{ab}\d^{\bd}_{\ad}\
 +\ (\sb^{ab})_{\ad}{}^{\bd} \ ,\label{so4a}\ee\be
 \ft12 \e_{abcd}(\s^{cd})_{\a\b}~=~ i (\s_{ab})_{\a\b}\ ,\qquad \ft12
 \e_{abcd}(\sb^{cd})_{\ad\bd}~=~ -i (\sb_{ab})_{\ad\bd}\ ,\label{so4b}
\ee
\be ((\s^a)_{\a\bd})^\dagger~=~
(\sb^a)_{\ad\b} ~=~ (\s^a)_{\b\ad} \ , \qquad ((\s^{ab})_{\a\b})^\dagger\ =\ (\sb^{ab})_{\ad\bd} \ .\ee
and raising and lowering spinor indices according to the
conventions $A^\a=\epsilon^{\a\b}A_\b$ and $A_\a=A^\b\epsilon_{\b\a}$ where
\be \e^{\a\b}\e_{\c\d} \ = \ 2 \d^{\a\b}_{\c\d} \ , \qquad
\e^{\a\b}\e_{\a\c} \ = \ \d^\b_\c \ ,\qquad (\e_{\a\b})^\dagger \ = \ \e_{\ad\bd} \ .\ee
The $\mso(3,2)$-valued connection 
 \be
  \O~:=~-i \left(\frac12 \omega^{ab} M_{ab}+e^a P_a\right) ~:=~ \frac1{2i}
 \left(\frac12 \omega^{\a\b}~y_\a \star y_\b
 +  e^{\a\dot\b}~y_\a \star {\bar y}_{\dot\b}+\frac12 \bar{\omega}^{\dot\a\dot\b}~{\bar y}_{\dot\a}\star {\bar y}_{\dot\b}\right)\
 ,\label{Omega}
 \ee
  \be
 \o^{\a\b}~=~ -\ft14(\s_{ab})^{\a\b}~\o^{ab}\ , \qquad \omega_{ab}~=~\ft12\left( (\s_{ab})^{\a\b} \o_{\a\b}+(\bar\s_{ab})^{\ad\bd} \bar\o_{\ad\bd}\right)\ ,\ee
 \be e^{\a\dot\a}~=~ \ft{\lambda}2(\s_{a})^{\a \dot\a}~e^{a}\ , \qquad e_a~=~ -\l^{-1} (\s_a)^{\a\ad} e_{\a\ad}\ ,\label{convert}\ee
and field strength 
\be {\cal R}~:=~ d\O+\O\star \O~:=~-i \left(\frac12 {\cal R}^{ab}M_{ab}+{\cal R}^a P_a\right) ~:=~ \frac1{2i}
 \left(\frac12 {\cal R}^{\a\b}~y_\a \star y_\b
 +  {\cal R}^{\a\dot\b}~y_\a \star {\bar y}_{\dot\b}+\frac12 \bar{\cal R}^{\dot\a\dot\b}~{\bar y}_{\dot\a}\star {\bar y}_{\dot\b}\right)\
 ,\label{calRdef}\ee
\be
 {\cal R}^{\a\b}\ =\ -\ft14(\s_{ab})^{\a\b}~{\cal R}^{ab}\ ,
 \qquad {\cal R}_{ab}~=~\ft12\left( (\s_{ab})^{\a\b} {\cal R}_{\a\b}+(\bar\s_{ab})^{\ad\bd} \bar{\cal R}_{\ad\bd}\right)\ ,\ee
 \be 
 {\cal R}^{\a\dot\a}\ =\ \ft{\lambda}2(\s_{a})^{\a \dot\a}~{\cal R}^{a}\ ,
 \qquad {\cal R}_a~=~ -\l^{-1} (\s_a)^{\a\ad} {\cal R}_{\a\ad}\ .\ee
In these conventions, it follows that  
 \be
 {\cal R}_{\a\b}~=~ d\o_{\a\b} -\o_{\a}^{\c}\o_{\c\b}-
 e_{\a}^{\cd}\bar e_{\cd\b}\ ,\qquad 
 {\cal R}_{\a\dot\b}~=~  de_{\a\bd}+ \o_{\a\c}\wedge
 e^{\c}{}_{\bd}+\bar{\o}_{\bd\dd}\wedge e_{\a}{}^{\dd}\
 ,\ee\be 
 {\cal R}^{ab}~=~ R_{ab}+\lambda^2
 e^a\wedge e^b\ ,\qquad R_{ab}~:=~d\o^{ab}+\o^a{}_c\wedge\o^{cb}\ ,\ee\be
 {\cal R}^a~=~ T^a ~:=~d e^a+\o^a{}_b\wedge e^b\ ,
 \label{curvcomp} \ee
where $R_{ab}:=\frac12 e^c e^d R_{cd,ab}$ and $T_a:=e^b e^c T^a_{bc}$ are the Riemann and torsion two-forms.
The metric $g_{\mu\nu}:=e^a_\mu e^b_{\nu}\eta_{ab}$. The $AdS_4$ vacuum solution $\O_{(0)}=e_{(0)}+\o_{(0)}$ obeying $d\O_{(0)}+\O_{(0)}\star\O_{(0)}=0$, with Riemann tensor $ R_{(0)\m\n,\r\s}=
 -\lambda^2 \left( g_{(0)\mu\rho} g_{(0)\nu\sigma}-
  g_{(0)\nu\rho} g_{(0)\mu\sigma} \right)$ and vanishing torsion, can be expressed as $\O_{(0)}=L^{-1}\star dL$ where the gauge function $L\in SO(3,2)/SO(3,1)$. The stereographic coordinates $x^\mu$ defined by \eq{L}, are related to the coordinates $X^A$ of the five-dimensional embedding space with metric
$ds^2  =dX^A dX^B\eta_{AB}$, 
in which $AdS_4$ is embedded as the hyperboloid
$X^A X^B \eta_{AB}=  -\ft1{\l^2} $,
as
\bea x^\m \ = \ \frac{X^\m}{1+\sqrt{1+\l^2 X^\m X_\m}} \ , \qquad X^\m \ = \ \frac{2x^\m}{1-\l^2 x^2}\ , \qquad \m \ = \ 0,1,2,3 \ .\label{A.15}\eea
%
The global spherical coordinates $(t,r,\theta,\phi)$ in which the metric reads
\bea  ds^2 \ = \ -(1+\l^2r^2)dt^2+\frac{dr^2}{1+\l^2 r^2}+r^2(d\theta^2+\sin^2\theta d\phi^2) \ ,\label{metricglob}\eea
are related locally to the embedding coordinates by
\bea & X_0 \ = \ \sqrt{\l^{-2}+r^2}\sin t \ , \qquad X_{0'} \ = \ \sqrt{\l^{-2}+r^2}\cos t \ , & \nn\\[5pt]
& X_1 \ = \ r\sin\theta\cos\phi \ , \quad  X_2 \ = \ r\sin\theta\sin\phi \ , \quad X_3 \ = \ r\cos\theta \ ,& \label{AdSspherical}\eea
providing a one-to-one map if $t\in [0,2\pi)$, $r\in[0,\infty)$, $\theta\in[0,\pi]$ and $\phi\in[0,2\pi)$ defining the single cover of $AdS_4$. This manifold can be covered by two sets of stereographic coordinates, $x^\mu_{(i)}$, $i=N,S$, related by the inversion $x^\m_N = -x^\m_S/(\l x_S)^2$ in the overlap region $\l^2 (x_N)^2, \l^2 (x_S)^2  <  0$, and the transition function $T_N^S=(L_N)^{-1}\star L_S\in SO(3,1)$. The map $x^\mu \rightarrow -x^\m/(\l x)^2$ leaves the metric invariant, maps the future and past time-like cones into themselves and exchanges the two space-like regions $0<\l^2 x^2< 1$ and $\l^2 x^2 > 1$ while leaving the boundary $\l^2 x^2 =1$ fixed. It follows that the single cover of $AdS_4$ is formally covered by taking $x^\mu\in \Real^{3,1}$.

Petrov's invariant classification of spin-2 Weyl tensors \cite{Petrov:2000bs,PenroseRindler} is based on their algebraic properties at any spacetime point. Generalized to the higher-spin context and by making use of spinor language, it amounts to study the roots of the degree-$2s$ polynomial $\O(\zeta):=C_{\a(2s)}\z^{\a_1}\ldots\z^{\a_{2s}}$, where $C_{\a(2s)}\equiv C_{\a_1\a_2\ldots\a_{2s}}=C_{(\a_1\a_2\ldots\a_{2s})}$ is the self-dual part of the Weyl tensor and $\z^{\a}$ an arbitrary non-vanishing two-component spinor. Factorizing the polynomial in terms of its roots defines a set of $2s$ spinors which one refers to as \emph{principal spinors},  \emph{viz.} $\O(\zeta)=u^1_{\a_1}\z^{\a_1}\ldots u^{2s}_{\a_{2s}}\z^{\a_{2s}}$, so $C_{\a(2s)}=u^1_{(\a_1}\ldots u^{2s}_{\a_{2s})}$. If $\O(\z)$ has multiple roots, the corresponding principal spinors are collinear. The classification then amounts to distinguish how many different roots $\O(\z)$ has, i.e., how many non-collinear principal spinors enter the factorization of the spin-$s$ Weyl tensor. Clearly, this classification can be given in terms of the partitions $\{p_1,...,p_k\}$ ($k\leq 2s$) of $2s$ in integers obeying $p_1+p_2+...+p_k=2s$ and $p_i\geqslant p_{i+1}$. In the spin-2 case, this singles out the familiar six different possibilities: $\{1,1,1,1\}$ (type I in Petrov's original terminology); $\{2,1,1\}$ (type II); $\{2,2\}$ (type D); $\{3,1\}$ (type III); and $\{4\}$ (type N) plus the trivial case of a vanishing Weyl tensor (type O). The type-D case is related to gravitational field configurations surrounding isolated massive objects; for arbitrary spin-$s$, we refer to the type $\{s,s\}$ as \emph{generalized type D}.


\scs{Properties of the Doubled Oscillator Algebra}\label{App:A}


\scss{Orderings and symbols}

The algebra $\widehat\O^{[0]}({\cal Y}\times {\cal Z})$ in which the master fields takes their values consist of non-polynomial completions of the enveloping algebra 
${\cal U}[Y,Z]={\cal U}[Y]\otimes {\cal U}[Z]$,
consisting of arbitrary $\star$-polynomials $\widehat P(Y,Z)$ subject to the commutation rules
\be [Y_{\underline\a},Y_{\underline\b}]_\star~=~2iC_{\underline{\a\b}}\ ,\quad [Y_{\underline\a},Z_{\underline\b}]_\star~=~0\ ,\quad [Z_{\underline\a},Z_{\underline\b}]_\star~=~-2iC_{\underline{\a\b}}\ .\ee
Such completions can be analyzed using different ordering prescriptions, that is, different bases
$B=\left\{\widehat E^i_B(Y,Z)\right\}_{i\in{\cal S}}$
for ${\cal U}[Y,Z]$ consisting of basis elements $\widehat E^i_B$ that are $\star$-polynomials in $Y$ and $Z$ labelled by discrete indices $i$. These elements can be expanded in terms of totally symmetric $\star$-monomials as
\bea \widehat E^i_B(Y,Z)&=&\widehat M^i_B(Y,Z)+\widehat L^i_B(Y,Z)\ ,\eea
where $\widehat M^i_B$ denotes the monomial of maximal degree and $\widehat L^i_B$ consists of the remainders. We call $B$ symbolizable if $\widehat M^i_B\neq \widehat M^j_B$ for $i\neq j$ so that there exists a linear map $[\cdot]_B:{\cal U}[Y,Z]\rightarrow {\cal U}[Y,Z]$, referred to as the Wigner map, defined by 
\bea \mbox{Symbolizable ordering}&:& \left[\widehat M^i_B(Y,Z)\right]_B~:=~\widehat E^i_B(Y,Z)\  .\eea
In particular, the totally-symmetric, or Weyl-ordered, basis for ${\cal U}[Y,Z]$ is defined by $\widehat L^i_{\rm Weyl}(Y,Z)=0$, \emph{i.e.}
\be \widehat E^i_{\rm Weyl}(Y,Z)~\equiv~\left[\widehat M^i_{\rm Weyl}(Y,Z)\right]_{\rm Weyl}~=~\widehat M^i_{\rm Weyl}(Y,Z)\ \ee
The symbol $[\widehat P]^B\in{\cal U}[Y,Z] $ of $\widehat P\in{\cal U}[Y,Z]$ is the Weyl-ordered element defined by the inverse Wigner map, \emph{viz.}
\bea \left[\left[\widehat P(Y,Z)\right]^B\right]_B&:=&\ \widehat P(Y,Z)\ ,\eea
that is, if $\widehat P=\sum_{i\in{\cal S}} P^B_i \widehat E^i_B$, with $P^B_i\in \Comp$, then $\widehat P=\sum_{i\in{\cal S}} P^B_i [\widehat M^i_B]_B = \left[\sum_{i\in{\cal S}}P^B_i \widehat M^i_B\right]_B$ from which it follows that $[\widehat P]^B=\sum_{i\in {\cal S}}P^B_i\widehat M^i_B$. We note that $\left[\widehat E^i_B\right]^B=\widehat M^i_B$ and that if $B$ and $B'$ are two bases related by 
\be \widehat E^i_B~=~(t^{B'}_{B})^i{}_{j}\widehat E^{j}_{B'}\ ,\qquad \widehat M^i_B~=~\widehat M^i_{B'}~=:~\widehat M^i\ ,\label{tbb'}\ee
with $(t^{B'}_{B})^i{}_{j}\in\Comp$, then 
\be \left[\widehat E^i_{B}\right]^{B'}~=~(t_{B}^{B'})^i{}_{j}\widehat M^j\ .\ee
The $\star$-product of ${\cal U}[Y,Z]$, which does not refer to any specific order, induces a composition rule between symbols also denoted by $\star$, \emph{viz.}
\bea \left[[\widehat P]^B{\star}[\widehat Q]^B\right]_B&:=& \widehat P\star\widehat Q\ .\eea

\scss{Universal orders}

An ordering $B$ is said to be universal if
\bea \mbox{Universal ordering}&:&\left[U^{\underline\a}Y_{\underline\a}+V^{\underline\a}Z_{\underline\a}~,~\widehat P\right]_\star\ =\ 2i\left[(U^{\underline\a}\partial^{(Y)}_{\underline\a}- V^{\underline\a}\partial^{(Z)}_{\underline\a})\widehat P^B\right]_B\eea
for all $\widehat P\in {\cal U}[Y,Z]$ and classical spinors $(U^{\underline\a},V^{\underline{\a}})$.  The transition \eq{tbb'} between two universal orderings $B$ and $B'$ are generated by symmetric bi-vector fields, \emph{viz.}
\bea \left[\widehat E^i_{B}\right]^{B'}&=& t^{B'}_{B}(\D) \widehat M^{i}\ ,\qquad t^{B'}_{B}(\D)\ =\ 1+\sum_{n=1}^\infty (t^{B'}_{B})_n \D^n\ ,\qquad (t^{B'}_{B})_n~\in\ \Comp\ ,\\[5pt] \widehat\D&=&  \D^{\hat{\underline\a},\hat{\underline\b}}\partial_{\hat{\underline \a}}\partial_{\hat{\underline\b}}\ ,\quad \partial_{\hat{\underline \a}}\ :=\ (\partial^{(Y)}_{\underline\a},\partial^{(Z)}_{\underline\a})\ .\eea
Requiring $\msp(4,\Real)_Y\times \msp(4,\Real)_Z$-invariance implies the Weyl order, while requiring $\msp(4,\Real)_{\rm diag}$-invariance leaves a family of 
\bea \mbox{diagonal Weyl orders}&:&\D\ =\ C^{\underline\a\underline\b} ~ \partial^{(Y)}_{\underline\a}\partial^{(Z)}_{\underline\b}\ ,\eea
that reduces to the Weyl order in ${\cal U}[Y]$ or ${\cal U}[Z]$, \emph{viz.}
\bea [P(Y)]^{{\rm Weyl}_{\rm diag}}&=& [P(Y)]^{\rm Weyl}\ ,\qquad [P(Z)]^{{\rm Weyl}_{\rm diag}}\ =\ [P(Z)]^{\rm Weyl}\ .\eea
The outer anti-automorphism $\tau$ induces a local action on symbols diagonal Weyl orders, \emph{viz.}
\bea \tau([\widehat P(Y,Z)]_{{\rm Weyl}_{\rm diag}})&=&[\widehat P(iY,-iZ)]_{{\rm Weyl}_{\rm diag}} \ .\eea
Two particular diagonal Weyl orders are the normal order $\widehat{\rm N}_+$ and the anti-normal order $\widehat{\rm N}_-$ with respect to the complexified Heisenberg algebra
\bea [\widehat A^-_{\underline\a},\widehat A^{+\underline\b}]_\star&=& \delta_{\underline\a}^{\underline\b}\ ,\quad \widehat A^-_{\underline\a}\ =\ \frac12(Y_{\underline\a}+Z_{\underline\a})\ ,\quad \widehat A^+_{\underline\a}\ =\ \frac1{2i}(Y_{\underline\a}-Z_{\underline\a})\ .\label{ComplexHeisenberg}\eea
In terms of $(Y_{\underline\a},Z_{\underline\a})$, one has 
\bea Y_{\underline\a} \star Y_{\underline\b} &=& [Y_{\underline\a} Y_{\underline\b}]_{\widehat{\rm N}_\pm} \pm iC_{\underline\a\underline\b}\ ,\quad Y_{\underline\a} \star Z_{\underline\b} \ =\ [Y_{\underline\a} Z_{\underline\b}]_{\widehat{\rm N}_\pm} \mp iC_{\underline\a\underline\b}\ ,\label{Nplus1}\\[5pt]
Z_{\underline\a} \star Y_{\underline\b} &=& [Z_{\underline\a} Y_{\underline\b} ]_{\widehat{\rm N}_\pm} \pm iC_{\underline\a\underline\b}\ ,\quad Z_{\underline\a} \star Z_{\underline\b} \ =\ [Z_{\underline\a} Z_{\underline\b}]_{\widehat{\rm N}_\pm} \mp iC_{\underline\a\underline\b}\ ,\label{Nplus2}\eea
which decompose under $SL(2,\Comp)$, using $Y_{\underline\a}=(y_\a,\bar y_{\ad})$, $Z_{\underline\a}=(z_\a,-\bar z_{\ad})$ and $C_{\underline\a\underline\b}=\left(\ba{cc}\e_{\a\b}&0\\0&\e_{\ad\bd}\ea\right)$ into
\bea y_{\a} \star y_{\b} &=& [y_{\a} y_{\b}]_{\widehat{\rm N}_\pm} \pm i\e_{\a\b}\ ,\quad y_{\a} \star z_{\b} \ =\ [y_{\a} z_{\b}]_{\widehat{\rm N}_\pm} \mp i\e_{\a\b}\ ,\label{Normal1}\\[5pt]
z_{\a} \star y_{\b} &=& [z_{\a} y_{\b} ]_{\widehat{\rm N}_\pm} \pm i\e_{\a\b}\ ,\quad z_{\a} \star z_{\b} \ =\ [z_{\a} z_{\b}]_{\widehat{\rm N}_\pm} \mp i\e_{\a\b}\ ,\\[8pt]
\bar y_{\ad} \star \bar y_{\bd} &=& [\bar y_{\ad} \bar y_{\bd}]_{\widehat{\rm N}_\pm} \pm i\e_{\ad\bd}\ ,\quad \bar y_{\ad} \star \bar z_{\bd} \ =\ [\bar y_{\ad} \bar z_{\bd}]_{\widehat{\rm N}_\pm} \pm i\e_{\ad\bd}\ ,\\[5pt]
\bar z_{\ad} \star \bar y_{\bd} &=& [\bar z_{\ad} \bar y_{\bd} ]_{\widehat{\rm N}_\pm} \mp i\e_{\ad\bd}\ ,\quad \bar z_{\ad} \star \bar z_{\bd} \ =\ [\bar z_{\ad} \bar z_{\bd}]_{\widehat{\rm N}_\pm} \mp i\e_{\ad\bd}\ .\label{Normal4}\eea

\scss{Chiral integration domain}\label{App:chiraldomain}

Upon presenting the induced $\star$-products among symbols using auxiliary integration variables, a generic order requires $16$ real variables while the $\widehat{\rm N}_\pm$-order is distinguished by only requiring $8$ real variables that can be taken to be either complex or chiral as follows:
\bea &&\left[\widehat F_1(Y,Z)\right]^{\widehat{\rm N}_\pm}{\star} \left[\widehat F_2(Y,Z)\right]^{\widehat{\rm N}_\pm}\nn\\[5pt]&&=~ \int_{{\cal R}_C} {d^4U d^4V\over (2\pi)^4} e^{\pm i V^{\underline\a} U_{\underline\a}} \left.\left[\widehat F_1(Y,Z)\right]^{\widehat{\rm N}_\pm}\right|_{(Y,Z)\rightarrow (Y+U,Z+U)} \left.\left[\widehat F_2(Y,Z)\right]^{\widehat{\rm N}_\pm}\right|_{(Y,Z)\rightarrow (Y+V,Z-V)} \ ,\qquad\qquad\label{star1}\eea
with 
\bea \mbox{Complex ($C=\Comp$)}&:&  {\cal R}_{\mathbb C}\ =\ \{~(U_{\underline\a},V_{\underline\a})~=~(u_\a,\bar u_{\ad};v_\a,\bar v_{\ad})~:~ (u_\a)^\dagger~=~\bar u_{\ad}~,\ (v_\a)^\dagger~=~\bar v_{\ad}~\}\ ,\qquad\label{complexdomain}\\[5pt]
\mbox{Chiral ($C=\Real$)}&:& {\cal R}_{\mathbb R}\ =\ \{~(U_{\underline\a},V_{\underline\a})~=~(u_\a,\bar u_{\ad};v_\a,\bar v_{\ad})~:~ (U_{\underline\a},V_{\underline\a})^\dagger~=~(U_{\underline\a},V_{\underline\a})~\}\ .\label{chiraldomain}\eea
These presentations are equivalent for $\widehat F_{1,2}\in {\cal U}[Y,Z]$ while they may give different results for the composition of non-polynomial elements.

Upon splitting into $\msl(2,\Comp)$-doublets one has
\be\left[\widehat F_1\star \widehat F_2\right]^{\,\widehat {\rm N}_\pm}~=~ \int_{{\cal R}_C} {d^4U d^4V\over (2\pi)^4} e^{\pm i( v^\a u_\a+\bar v^{\ad} \bar u_{\ad}) }\left[\widehat F_1\right]^{\widehat {\rm N}_\pm}(y+u,\bar y+\bar u;z+u,\bar z-\bar u)\left[\widehat F_2\right]^{\widehat {\rm N}_\pm}(y+v,\bar y+\bar v;z-v,\bar z+\bar v)\ .\label{E28}\ee
Correspondingly, there are two trace operations
\be \widehat{\rm Tr}_{C}[\widehat {\cal O}]~=~ \int_{{\cal R}_C} {d^4U d^4V\over (2\pi)^4} [\widehat {\cal O}]^B(U,V)\ ,\label{chiraltrace}\ee
which are formally independent of $B$, while the two choices for $C$ are not equivalent in general. 

In this paper, we always use the chiral integration domain, deferring the issue of the physical meaning of the two choices of $C$ for future studies.

\scss{Chiral delta functions and inner Kleinians}\label{C4}

Working with the chiral integration domain ${\cal R}_{\mathbb R}$, it makes sense to define the following real-analytic delta functions ($M_{\a}^{\b}\in GL(2;\mathbb C)$):
\bea \d^2(y_\a)&:=&  \d(y_1)\d(y_2)\ ,\qquad \d^2((My)_\a)\ =\ \frac{1}{\det M} \d^2(y_\a) \label{deltay}\\[5pt]
\d^2(z_\a)&:=&  \d(z_1)\d(z_2)\ ,\qquad \d^2((Mz)_\a)\ =\ \frac{1}{\det M} \d^2(z_\a)\ .\label{deltaz}\eea
Their hermitian conjugates are defined by 
$\d^2(\bar y_{\ad})= (\d^2(y_\a))^\dagger$ and $\d^2(\bar z_{\ad})=
(\d^2(z_\a))^\dagger$.
By splitting $y_\a$ and $z_\a$ into a complexified Heisenberg algebras
\bea [y^-,y^+]_\star&=&1\ ,\quad y^\pm~=~u^{\pm}\cdot  y\ ,\quad u^+\cdot u^-~=~-\frac{i}2\ ,\\[5pt]
[z^-,z^+]_\star&=&1\ ,\quad z^\pm~=~v^{\pm}\cdot  z\ ,\quad v^+\cdot z^-~=~\frac{i}2\ ,\eea
one can define idempotent inner Kleinian operators 
\bea \k_y&:=&(-1)_\star ^{N_y}\ ,\quad N_y~:=~ y^+\star y^-\ ,\\[5pt]
\k_z&:=&(-1)_\star ^{N_z}\ ,\quad N_z~:=~ z^+\star z^-\ ,\eea
using the notation ($c\in\Comp$)
\bea c_\star^{\widehat P}&=&\exp_\star (\widehat P\log c)\ ,\quad \exp_\star \widehat P\ =\ \sum_{n=0}^\infty \frac1{n!} \widehat P^{\star n}\ ,\quad \widehat P^{\star n}\ =\ \underbrace{\widehat P\star\cdots\star\widehat P}_{\tiny \mbox{$n$ times}}\ ,\eea
and representing $(-1)_\star ^{N_y}$ as
\bea (-1)_\star ^{N_y}&=& \lim_{\e\rightarrow 0} \exp_\star(i(\pi+\e)N_y)~=:~[\k_{y,B}(\e)]_B\ ,\eea
\emph{idem} $(-1)_\star ^{N_z}$. The broken $SL(2,\Comp)$-invariance is restored in the limit $\e\rightarrow 0$ provided $B$ is a diagonal Weyl order, such as $B={\rm Weyl}\,,\ \widehat{\rm N}_\pm$, which reduces to Weyl order for operators depending only on $Y$ or $Z$, \emph{i.e.}
\bea \k_y&=& [2\pi\delta^2(y_\a)]_{\rm Weyl}\ ,\quad \k_z~=~[2\pi\delta^2(z_\a)]_{\rm Weyl}\ .\eea
We also define
\bea \bar\k_{\yb}&:=&(\k_y)^\dagger\ =\ (-1)^{\bar N_\yb}_\star~=~ [2\pi\delta^2(\yb_{\ad})]_{\rm Weyl}\ ,\\[5pt]
\bar\k_{\zb}&:=&(\k_z)^\dagger\ =\ (-1)^{\bar N_\zb}_\star~=~ [2\pi\delta^2(\zb_{\ad})]_{\rm Weyl}\ ,\eea
using
\bea \bar N_{\yb}&:=& (N_y)^\dagger~=~\yb^+\star \yb^-\ ,\quad \yb^\pm~:=~(y^\mp)^\dagger~=~\bar u^\pm\cdot \yb\ ,\quad \bar u^\pm_{\ad}~:=~(u^\mp_\a)^\dagger\ ,\\[5pt]
\bar N_{\zb}&:=& (N_z)^\dagger~=~\zb^+\star \zb^-\ ,\quad \zb^\pm~:=~(z^\mp)^\dagger~=~\bar v^\pm\cdot \zb\ ,\quad \bar v^\pm_{\ad}~:=~(v^\mp_\a)^\dagger\ ,\eea
such that $[\yb^-,\yb^+]_\star=[\zb^-,\zb^+]_\star=1$ and $\bar u^+\cdot \bar u^-=-\bar v^+\cdot \bar v^-=-\frac{i}2$. The inner Kleinian elements generate the involutive automorphisms
\bea \pi_y(\widehat F)&:=& \k_y\star\widehat F\star \k_y\ ,\qquad \pi_z(\widehat F)\ :=\ \k_z\star\widehat F\star \k_z\ ,\\[5pt]
\bar\pi_{\yb}(\widehat F)&:=& \bar\k_{\yb}\star\widehat F\star \bar\k_{\yb}\ ,\qquad \pi_z(\widehat F)\ :=\ \bar\k_{\zb}\star\widehat F\star \bar\k_{\zb}\ ,\eea
acting locally on $\star$-composites, \emph{viz.}
\bea \pi_y(\widehat F(y,\bar y;z,\bar z))&=&\widehat F(-y,\bar y;z,\bar z)\ ,\eea
\emph{idem} $\k_z$, $\bar\k_{\yb}$ and $\bar\k_{\zb}$. The induced action on symbols, defined by
$\pi_y([\widehat F]^B):=[\pi_y(\widehat F)]^B$,
\emph{idem} $\k_z$, $\bar\k_{\yb}$ and $\bar\k_{\zb}$, acts locally in Weyl order, \emph{viz.}\footnote{In diagonal Weyl orders one has $\pi_y([\widehat F]^{{\rm Weyl}_{\rm diag}}(y,\bar y;\bar z)=[\widehat F]^{{\rm Weyl}_{\rm diag}}(-y,\bar y;\bar z)$, while the action of $\pi_y$ on a symbol is non-local if the symbol depends non-trivially on both $y_\a$ and $z_\a$.}
\bea \pi_y([\widehat F]^{\rm Weyl}(y,\bar y;z,\bar z))&=&[\widehat F]^{\rm Weyl}(-y,\bar y;z,\bar z)\ .\eea
\emph{idem} $\k_z$, $\bar\k_{\yb}$ and $\bar\k_{\zb}$. 
The inner automorphisms $\pi= \pi_y\pi_z$ and $\bar\pi=\bar\pi_{\bar y}\bar\pi_{\bar z}$
act locally in general ${\rm Weyl}_{\rm diag}$-orders, \emph{viz.}
\bea \pi([\widehat F]^{{\rm Weyl}_{\rm diag}}(y,\bar y;z,\bar z))&=&[\widehat F]^{{\rm Weyl}_{\rm diag}}(-y,\bar y;-z,\bar z)\ .\eea
This action is generated by conjugation by the elements
\bea
\widehat\kappa&=& \k_y\star \k_z\ ,\qquad \widehat{\bar\kappa}\ =\ \k_{\bar y}\star \k_{\bar z}\ .\eea
Their Weyl-ordered and normal-ordered symbols are given by
\bea [\widehat\kappa]^{\rm Weyl}&=& (2\pi)^2\d^2(y)\d^2(z)\ ,\quad [\widehat{\bar\kappa}]^{\rm Weyl}\ =\ (2\pi)^2 \d^2(\yb)\d^2(\zb)\ ,\\[5pt]
[\widehat\kappa]^{\widehat{\rm N}_+}&=& e^{iy^\a z_\a}\ ,\qquad [\widehat{\bar\kappa}]^{\widehat {\rm N}_+}\ =\ e^{-i\bar y^{\ad} \bar z_{\ad}}\ ,\eea
where we note the fact that one and the same operator can be completely factorized over ${\cal U}[Y]\otimes {\cal U}[Z]$ in one order and completely entangled in another order.


\scs{Terminology and Basic Properties of Unfolded Systems}\label{terminology}


Vasiliev's formalism \cite{vasiliev} provides a fully nonlinear and background-independent unfolded description of classical higher-spin gravities\footnote{For a review of the four-dimensional theory and its lower-dimensional avatars, see \cite{Vasiliev:1999ba,Vasiliev:1995dn}; for the natural generalization to symmetric tensor gauge fields in higher dimensions, see \cite{Vasiliev:2003ev} for the original work and \cite{Bekaert:2005vh} for a review.} in a certain duality picture; for a maximal duality extension in the case of four-dimensional bosonic models related to an action principle with non-trivial Poisson structures, see \cite{Boulanger:2011dd}. The aforementioned statements are to a large extent drawn from basic properties of unfolded dynamics, for which we use the following terminology:

Unfolded dynamics is the formulation of field theory based on free-differential algebras (FDAs), $\widehat{\cal A}$ (see for example \cite{Vasiliev:1988xc,Vasiliev:1988sa,Vasiliev:1999ba,Bekaert:2005vh} and references therein). Such an algebra is an $\mathbb N$-graded space of differential forms that remain invariant under the composition under degree-preserving $n$-ary products, possibly modulo further algebraic constraints. The latter can be supplemented either by hand or by compatibility requirements in which cases the algebra is referred to as being constrained or quasi-free, respectively. Depending on the level of complexity exhibited by the $n$-ary products, one distinguishes between FDAs that are graded-commutative (or exterior for short), associative and strongly-homotopy associative (or sh-associative for short); the former two are of relevance to supergravities and Vasiliev's higher-spin gravities, respectively, and one may expect that the latter are of relevance to extensions of Vasiliev's theories by mixed-symmetry fields and to tensionless closed strings\footnote{A key feature of sh-associative algebras is that the binary product $[\cdot,\cdot]_2$ may have an internal \emph{negative} degree (a first-quantized ghost number), say $1-\hat p$, such that the basic Yang-Mills-like curvature takes the form $\widehat d\widehat A+\frac12 [\widehat A,\widehat A]_2+\mbox{$n$-ary corrections}$ where thus $\widehat A$ is a $\hat p$-form; for example, in the application to first-quantized open membranes and second-quantized five branes one naturally has $\hat p=2$.}. Loosely speaking, sh-associative FDAs have perturbative expansions in terms of associative FDAs each of which in its turn has perturbative expansions in terms of exterior FDAs, leading to the notion of dualities.

Associative FDAs are algebras of elements, referred to as differential forms, closed under i) an associative non-commutative binary product $\star$, \emph{i.e.} $\widehat{\cal A}\star\widehat{\cal A}\subseteq \widehat{\cal A}$; 
and ii) the action of an exterior derivative $\widehat d$ assumed to obey Leibniz' rule, \emph{i.e.}\footnote{The algebra $\widehat{\cal A}$ is called minimal if $\widehat d\widehat{\cal A}\subseteq \widehat{\cal A}\star\widehat{\cal A}$.} $\widehat d\widehat{\cal A}\subseteq \widehat{\cal A}$ where $\widehat d(\widehat f\star\widehat g)=(\widehat d\widehat f)\star\widehat g+(-1)^{\textrm{deg}\widehat f}\widehat f\star(\widehat d\widehat g)$.
Their generating elements, say $\widehat Z^{\hat i}$, are thus differential forms, referred to as master fields for short, obeying $\widehat R^{\hat i}:=\widehat d\widehat Z^{\hat i}+Q^{\hat i}_\star(\widehat Z^{\hat j})=0$, referred to as generalized-curvature constraints and which one may think of as the fundamental equations of motion of the unfolded system (possibly in a given duality picture). The structure functions $Q^{\hat i}$, which are built using $\star$-product compositions, define a $Q$-structure $\overrightarrow Q:=Q^{\hat i}\overrightarrow\partial_{\hat i}$, which is a $\star$-vector field of degree one acting on $\widehat{\cal A}$. Compatibility, sometimes referred to Cartan integrability, requires $( Q^{\hat i}\overrightarrow \partial_{\hat i})\star Q^{\hat j}$, that is $({\cal L}_{\overrightarrow Q})^{\star 2}$ or $\overrightarrow Q^{\star 2}$ for short, to vanish; depending on whether integrability requires additional purely algebraic constraints or not, $\widehat{\cal A}$ is referred to as being quasi-free or free, respectively; if integrability does not require any truncation in form degree from above, the (quasi-)FDA is referred to as being universal. 
A variant of the quasi-free case is when algebraic constraints are supplemented by hand, referred to as constrained FDAs. 

A particular type of associative quasi-FDAs are those in which $\widehat Z^{\hat i}=(\widehat Z^i,\widehat J^r)$ where $\widehat J^r$ are central and closed elements of strictly positive (and even) degree, that is, $\widehat R^i=\widehat d\widehat Z^i+Q^i_\star(\widehat Z^j;\widehat J^r)=0$, $\widehat d\widehat J^r=0$ and $\widehat J^r\star \widehat Z^i=\widehat Z^i\star \widehat J^r$.
In the latter case, which contains the free case, the locally-defined solution space can be made explicit as \cite{Boulanger:2011dd}\be \widehat Z^i_{Z';\widehat \l} ~=~\left.\left[(\exp_\star  \overrightarrow T_{\widehat Z;\widehat \l})\star \widehat Z^i\right]\right|_{\widehat Z=Z'}\ ,\ee
where $Z'$ are reference solutions which represent local degrees of freedom; $\widehat \l^i$ are gauge functions for the fields in strictly positive form degree; and the $\star$-vector field $ \overrightarrow T_{\widehat Z;\widehat \l}:= (\widehat d\widehat \l^i-\overrightarrow \l\star Q^i)\overrightarrow \partial_i$ with $\overrightarrow \l=\widehat \l^i\overrightarrow \partial_i$ is the generator of the Cartan gauge algebra $\widehat{\mathfrak g}$ (represented softly in $\widehat{\cal A}$).

Given a generalized structure group $\widehat G=\exp_\star \widehat{\mathfrak t}$ generated by an unbroken sub-algebra $\widehat{\mathfrak t}\subseteq \widehat{\mathfrak g}$ (with $=$ in the unbroken phase) and a non-commutative symplectic manifold $\mathfrak{C}=\bigcup_I \mathfrak{C}_I$ with two-form $\C$ consisting of charts $\mathfrak{C}_I$ and with boundary $\partial\mathfrak{C}$, globally-defined solutions are obtained by using transition functions $\widehat T_I^{I'}\in \widehat G$ to glue together a set of locally-defined configurations $\{\widehat Z^i_I\}$, and factoring out $\widehat G_I$ from each local configuration space leading to gauge equivalence classes $[\widehat T_I^{I'}]_{\widehat G}$, $[Z^{\prime i}_I]_{\widehat G}$ and $[\widehat \l^i_I]_{\widehat G}$ of which the latter form (generalized) sections. Classical observables, in the form of intrinsically-defined functionals ${\cal O}[\{\widehat Z^{\hat i}_I\};\widehat T_I^{I'}]$, depend on moduli of three types: i) local moduli in the form of the reference solutions $\{Z^{\prime i}_I\}$ (and $\widehat J^r$); ii) the boundary values $\left([\widehat \l^i]_{\widehat G}\right)|_{\partial \widehat{\cal C}}$; and iii) global data contained in $[\widehat T_I^{I'}]_{\widehat G}$, monodromies and other constructions.

Thinking of $\mathfrak{C}$ as a sort of multiple fibration, or correspondence space, projections down to symplectic sub-manifolds provide unfolded sub-systems in different dual pictures. 
In the free case, where $Z'$ can be taken to be integration constants for the zero-forms, such projections preserve $Z'$, \emph{i.e.} the local degrees of freedom of the system, while truncating the moduli associated with the master fields in their kernel. 
This results in the notion of dual pictures: each picture consists of sectors labelled by boundary conditions in dual pictures, \emph{i.e.} by frozen dual moduli of type (ii) and (iii); each such sector
consists of the remaining variable moduli, \emph{i.e.} the $Z'$-moduli and the moduli of type (ii) and (iii) that remain visible in the picture in question. 

The aforementioned notion of dual pictures may be enriched by duality extensions \cite{Boulanger:2011dd}: given a quasi-FDA $\widehat{\cal A}$ with master fields $\widehat Z^{\hat i}$ of fixed degrees $p_{\hat i}\in\mathbb N$ a larger quasi-FDA $\widehat {\cal A}^{\uparrow}$ can be formed by replacing each $\widehat Z^i$ by $\widehat Z^{\uparrow i}$ consisting of master fields of degrees $\{p_i,p_i+2,p_i+4,\dots\}$, and each free parameter (real number) in $Q^i$ by a real element in the central algebra generated by the $\widehat J^r$. 

Using the above terminology, higher-spin gravities are based on universal, associative, quasi-FDAs with twisted-central terms and fermionic zero-modes in $\O(\mathfrak{C})$ associated with a fibre sub-manifold ${\cal Y}$; see Eqs. \eq{e1}--\eq{e3}. 
In the four-dimensional case, Vasiliev's original twistorial formulation is in terms of $\widehat Z^{\hat i}=(\widehat \Phi,\widehat A;\widehat J,\,\,\widehat{\!\!\bar J})$ of degrees $(0,1;2,2)$; for a duality-extended formulation and a related action principle, see \cite{Boulanger:2011dd}. 
The total manifold $\mathfrak{C}\stackrel{\rm loc}{\cong} T^\ast{\cal X}\times {\cal T}$, where $T^\ast{\cal X}$ is universal and ${\cal T}={\cal Z}\times {\cal Y}\cong \Comp^2\times \Comp^2$, has the structure of a sort of double fibration on which operates a generalized Penrose twistor transform \cite{PenroseRindler}; see also \cite{Gelfond:2008td}.  
The twisted-central elements $(\widehat J,\,\,\widehat{\!\!\bar J})\equiv (\widehat J,\,\,\widehat{\!\!\bar J})|_{{\cal T}_{p}}$ for any ${\cal T}_{p}:=\{p\in T^\ast{\cal X}\}\times {\cal Z}\times {\cal Y}$.
If the full system can be projected down to an unfolded subsystem on $\check{\mathfrak C}:\stackrel{\rm loc}{\cong}{\cal X}\times{\cal T}$ (\emph{e.g.} by imposing Eq. \eq{reduceP}), the latter describes deformations of $\C|_{{\cal T}_{p}}$ generated by $\widehat \Phi|_{{\cal T}_{p}}\star \widehat J$ and its hermitian conjugate. 
Further projection down to $\check {\mathfrak C}_4:\stackrel{\rm loc}{\cong}{\cal X}_4\times {\cal T}$, where ${\cal X}_4\subset {\cal X}$ is a four-manifold, provides the minimal type of picture on which operates a Penrose-style transform: 

\noindent Projection to ${\mathfrak C}_4 :\stackrel{\rm loc}{\cong}{\cal X}_4\times \{ Z=0\}\times {\cal Y}$, yields an exterior FDA on ${\cal X}_4$ consisting of $\Phi=\widehat \Phi|_{{\mathfrak C}_4}$ and $W=(\widehat A-\widehat K)|_{{\mathfrak C}_4}$, with $\widehat K$ containing the canonical Lorentz connection (see Eq. \eq{fieldredef}), and with soldering one-form $E=\ft12(1-\pi)W$ (see Section \ref{Sec:soldering}). 
The variable spacetime moduli, to be extracted via classical observables as discussed in Section \ref{Sec:0charges}, 
consist of initial data $\Phi'=\Phi|_{p_0}$ representing local degrees of freedom; diffeomorphism-invariant boundary data contained in the gauge function of $E$, representing global metric structures on $\partial {\cal X}_4$; and other global data contained in $W$ and the transition functions (see discussion in Section \ref{gaugefmethod}).

\noindent Projection to ${\cal T}_{p_0}$ for some fixed $p_0\in {\cal X}_4$ yields an associative quasi-FDA on ${\cal Z}$ consisting of $\widehat \Phi':=\widehat\Phi|_{{\cal T}_{p_0}}$ and $\widehat V':=\widehat A|_{{\cal T}_{p_0}}$ and $(\widehat J,\,\,\widehat{\!\!\bar J})\equiv (\widehat J,\,\,\widehat{\!\!\bar J})|_{{\cal T}_{p_0}}$.  The variable twistor-space moduli consist of $\Phi'=\widehat\Phi'|_{Z=0}$ and boundary values for $\widehat V'$ describing deformations of $\C|_{{\cal T}_{p_0}} $.

\noindent Thus, the zero-form moduli $\Phi'$ are visible in both pictures while the spacetime $W$-moduli and twistor-space $\widehat V'$-moduli label inequivalent twistor-space and spacetime pictures, respectively; these pictures are related via uplifts to ${\mathfrak C}$ where all one-form moduli become visible\footnote{On top of them, there are further moduli associated with $T^\ast{\cal X}\rightarrow {\cal X}\rightarrow {\cal X}_4$ that we leave for future investigations; for example, in the case of spherically symmetric type-D solutions, where we shall activate all moduli on ${\mathfrak C}_4$ and ${\mathfrak C}'_{p_0}$, one may consider evaluating the abelian $p$-form charges \eq{abelianpform} on $p$-spheres in generalized spacetimes ${\cal X}$.}; as far as the ``direction'' of the twistor-map is concerned, the asymmetry between ${\cal X}_4$ and ${\cal T}$ implies that standard algebraic methods facilitate starting in ${\mathfrak C}'_{p_0}$, uplifting to $\check{\mathfrak C}_4$ via gauge functions, and then reducing to ${\cal X}_4$.

\scs{Dynamical Equations in Spacetime}\label{weakfields}

In this Appendix we outline the derivation of the generally-covariant equations of motion for dynamical component fields in four-dimensional spacetime starting from the full master equations \cite{Sezgin:2002ru}.

\scss{Graded-commutative free-differential algebra on ${\cal X}$}

The master fields on $\mathfrak{C}$ consists of totally-symmetric poly-vector fields on ${\cal X}$ valued in algebras of differential forms on ${\cal X}\times {\cal Z}\times {\cal Y}$. This system can be truncated consistently to a system on ${\mathfrak C}$ by setting all such poly-vector fields of strictly positive rank equal to zero by imposing\footnote{A Maurer-Cartan form $\O^{(0)}(X^M,P_M;dX^M,dP_M;Y^{\underline\a})$ depending non-trivially on both $X^M$ and $P_M$ cannot be restricted to $\msp(4;\Comp)$; in this sense, the extension from ${\cal X}$ to $T^\ast{\cal X}$, which is natural from the point-of-view of seeking underlying first-quantized origins of second-quantized field theory, is also naturally connected to higher-spin extensions of gravity.} 
\be \widehat U^M~=~0\ ,\qquad \partial^M(\widehat \Phi,\widehat U_M,\widehat S_{\underline\a})~=~0\ .\label{reduceP}\ee
Further projection down to a graded-commutative free-differential algebra on ${\mathfrak C}:={\cal X}\times \{Z^{\underline\a}=0\}\times {\cal Y}$ can be achieved by choosing an ordering scheme $B$ and imposing  
\be \left.\big[\widehat \Phi\big]^B\right|_{Z=0}~=~\big[\Phi(X,Y)\big]^B\ ,\qquad \left.\big[\widehat U\big]^B\right|_{Z=0}~=~dX^M \big[U_M(X,Y)\big]^B\ ,\qquad \widehat V_{\underline\a}|_{\Phi=0}\ =\ 0\ ,\ee
\emph{i.e.} assuming trivial boundary conditions on ${\cal Z}$, and by assuming expansions of the form 
\be  \widehat \Phi~=~\sum_{n=1}^\infty \widehat \Phi^{(n)}\ ,\quad \widehat S_{\underline\a}~=~ Z_{\underline\a}-2i\sum_{n=1}^\infty \widehat V^{(n)}_{\underline\a}\ ,\quad 
\widehat U~=~ \sum_{n=0}^\infty \widehat U^{(n)}(X,Y,Z)\ ,\\[5pt]
\ee
where $\left(\widehat \Phi^{(n)},\widehat V^{(n)}_{\underline\a},\widehat U^{(n)}\right)$ are $n$th order in $\Phi$ (with $\widehat \Phi^{(1)}\equiv \Phi$ and $\widehat U^{(0)}\equiv U$); for generic values of $X^M$, these perturbative building blocks are assumed to have symbols in $B$-order that are real-analytic in $Y^{\underline\a}$ and $Z^{\underline\a}$ and to belong to an associative subalgebra $\widehat{\cal A}\subset \O({\cal Z})\otimes \O^{[0]}({\cal Y})$  (which requires $[\Phi]^B$ to take values in suitable classes of functions in $\O^{[0]}({\cal Y})$ \cite{Vasiliev:1990vu,Prokushkin:1998bq,Bekaert:2005vh}). 
Defining the homotopy contractor $\rho_v= i_v ({\cal L}_v)^{-1}$, where ${\cal L}_v=\{i_v,q\}$, $q=dZ^{\underline\a}\partial_{\underline\a}$, $v=v^{\underline\a}(Z)\partial_{\underline\a}$, $\partial_{\underline\a}:={\partial\over \partial Z^{\underline\a}}$ and $v^{\underline\a}(0)=0$, it follows that if $\widehat f,\widehat g\in\widehat {\cal A}$ and ${\cal L}_v\widehat f=\widehat g$ then $\widehat f=f+({\cal L}_v)^{-1}\widehat g$ where $f\in \O^{[0]}({\cal Y})$. Thus, Eqs. \eq{dSa}--\eq{INT3} are perturbatively equivalent to\footnote{For the perturbative regularization of $({\cal B}^{(n)},\overline{\cal B}^{(n)})$, see discussion at the end of Section \ref{0formcharges}.}
\be \widehat \Phi^{(n)}~=~ \delta_{n1}\Phi-\rho_v\sum_{n_1+n_2=n} [\widehat V^{(n_1)},\widehat \Phi^{(n_2)}]_\pi\ ,\label{rho1}\ee
\be \widehat V^{(n)}~=~ q\widehat \l^{(n)}-\rho_v\sum_{n_1+n_2=n} \left(\widehat V^{(n_1)}\star \widehat V^{(n_2)}+{\cal B}^{(n_1)}\star \widehat \Phi^{(n_2)} \star \widehat J+\overline{\cal B}^{(n_1)}\star \widehat \Phi^{(n_2)} \star \,\,\widehat{\!\!\bar J}\,\right)\ ,\label{rho2}\ee
\be\widehat U^{(n)}~=~ d\rho_v q \widehat \l^{(n)}-\rho_v \sum_{n_1+n_2=n} \left[\widehat V^{(n_1)},\widehat U^{(n_2)}\right]_\star\ ,\label{rho3} \ee
where $\widehat \l^{(n)}\in \widehat {\cal A}$ are gauge artifacts which can be eliminated by imposing the 
\be \mbox{twistor gauge condition:}\quad i_v \widehat V~\stackrel{!}{=}~0\ ,\label{twistorgauge}\ee
which implies $\widehat \l^{(n)}=0$ and
\be \widehat U~=~ (1+\sum_{n=1}^\infty\widehat L^{(n)})^{-1} U\ ,\qquad \widehat L^{(n)}(\cdot)~:=~\rho_v\left[\widehat V^{(n)},\cdot\right]_\star\ .\ee
The residual gauge symmetries are given by 
\be \widehat \e~=~(1+\sum_{n=1}^\infty \widehat L^{(n)})^{-1} \e \ ,\label{residualhs(4)}\ee
where in the case of the minimal bosonic model, $\e\in \mathfrak{hs}(4)$, the minimal-bosonic higher-spin Lie algebra given by arbitrary polynomials in $Y^{\underline\a}$ obeying $\t(\e)=\e^\dagger=-\e$; and  in the case of  the non-minimal-bosonic model, $\e\in \mathfrak{hs}_1(4)$, the non-minimal extension of $\mathfrak{hs}(4)$ in which $\e$ obeys the weaker conditions $\pi\bar\pi(\e)=\e$ and $\e^\dagger=-\e$.
The resulting reduced albeit perturbatively defined unfolded system on ${\cal X}$ is then given by the reduction of \eq{2.55} to $Z=0$, \emph{viz.} 
\bea &&\left.\left[\nabla\widehat W+\widehat W\star \widehat W + \frac1{4i} \left(r^{\a\b} \widehat M_{\a\b}+\bar r^{\ad\bd} \widehat {\bar M}_{\ad\bd}\right)\right]\right|_{Z=0}\ =\ 0\ ,\label{ST1}\\[5pt] &&\left.\left[\nabla\widehat \Phi+\widehat W\star\widehat \Phi-\widehat \Phi\star\pi(\widehat W)\right]\right|_{Z=0}\ =\ 0\ ,\label{ST2}\eea
with $\widehat W$ given in \eq{fieldredef}, the full Lorentz-generators in \eq{fullM}-\eq{M(S)} and the Riemann two-form and Lorentz-covariant derivatives are defined above and in Eqs. \eq{lorcovfirst}--\eq{lorcovlast}, respectively; assuming $(\omega^{\a\b},\bar\o^{\ad\bd})=(\omega^{\a\b},\bar\o^{\ad\bd})^{(0)}$, the manifest Lorentz-invariance implies that\footnote{By definition, $\widehat W=\widehat U-\widehat K=(1+\sum_{n=1}^\infty L^{(n)})^{-1}U-\omega-\sum_{n=1}^\infty \widehat K^{(n)}$ where $\o=\frac1{4i}(\o^{\a\b}y_\a\star y_\b+\bar\o^{\ad\bd} \yb_{\ad}\star\yb_{\bd})$. It follows that $\widehat W^{(0)}=U-\omega=: W$ such that $\widehat W=W+((1+\sum_{n=1}^\infty \widehat L^{(n)})^{-1}-1)(W+\omega)-\sum_{n=1}^{\infty}\widehat K^{(n)}$. Since both $\widehat W$ and $W$ consists of canonical Lorentz tensors, it follows that the terms proportional to $\omega$ must cancel, which yields \eq{widehatW}.}
\be \widehat W~=~\sum_{n=0}^\infty \widehat W^{(n)}~=~(1+\sum_{n=1}^\infty\widehat L^{(n)})^{-1}W\ ,\qquad W~=~dX^M W_M(X,Y)\ .\label{widehatW}\ee
Using $v^{\underline\a}=Z^{\underline\a}$ and the normal order defined by Eqs. \eq{Normal1}--\eq{Normal4}, Eqs. \eq{ST1} and \eq{ST2} take on the manifestly Lorentz-covariant form
\bea &&\nabla W+W\star W+r+\sum_{n=1}^\infty J^{(n)}(W,W;\Phi,\dots,\Phi)\ =\ 0\ ,\label{C8}\\[5pt]
&&\nabla \Phi+[W,\Phi]_\pi+\sum_{n=2}^\infty P^{(n)}(W,W;\Phi,\dots,\Phi) \ = \ 0\ ,\label{C9}\eea
where $r:=d\omega+\omega\star \omega$ with $\omega:=\frac1{4i} \left(\o^{\a\b} y_\a\star y_\b+\bar \o^{\ad\bd} \yb_{\ad}\star \yb_{\bd}\right)$, and the curvature corrections 
\bea J^{(n)}&=& \sum_{n_1+n_2=n} \left.(\widehat W^{(n_1)}\star \widehat W^{(n_2)}+i (r^{\a\b}\widehat V^{(n_1)}_\a\star \widehat V^{(n_2)}_\b+\bar r^{\ad\bd}\widehat{\bar V}^{(n_1)}_{\ad}\star \widehat{\bar V}^{(n_2)}_{\bd}))\right|_{Z=0}\ ,\\[5pt]
P^{(n)}&=&\sum_{n_1+n_2=n} \left.[\widehat W^{(n_1)},\widehat \Phi^{(n_2)}]_\pi\right|_{Z=0}\ .\eea
By construction, Eqs. \eq{C8} and \eq{C9} are Cartan integrable order by order in the $\Phi$-expansion and define an exterior (or graded-commutative) free differential algebra on ${\cal X}$, that can be written on standard form by imposing \eq{shiftgauge} and eliminating $(r^{\a\b},\bar r^{\ad\bd})$ from $J^{(n)}$. 

\scss{Dynamical field equations on ${\cal X}_4$}

Reducing the universal system \eq{C8}--\eq{C9} down to a four-dimensional sub-manifold ${\cal X}_4\subset{\cal X}$ with local coordinates $x^\mu$ and assuming that 
\bea e_{\mu,\a\ad}&:=&2i\l^{-1}{\partial^2\over \partial y^\a \partial {\bar y}^{\ad}} \left[W_\mu(x|Y)\right]^{\rm Weyl}|_{Y=0}\label{vierbein}\eea
is invertible yields a dynamical field content given by
\bea \mbox{non-minimal-bosonic model}&:& \phi\ ,\ a_\m\ ,\ g_{\m\n}\ , \{\phi_{\mu_1\dots \mu_s}\}_{s\ =\ 3,4,5,6,\dots}\ ,\\[5pt]
\mbox{minimal-bosonic model}&:& \phi\ ,\ g_{\m\n}\ , \{\phi_{\mu_1\dots \mu_s}\}_{s\ =\ 4,6,\dots}\ ,\eea
where the (pseudo-)scalar $\phi$, Maxwell potential $a_\mu$ and metric $g_{\m\n}$ are given by
\bea \phi&=&\Phi(x|Y)|_{Y=0}\ ,\qquad a_\mu\ =\ W_\mu(x|Y)|_{Y=0}\ ,\qquad g_{\mu\nu}\ =\ e_\mu{}^a(x) e_{\nu~a}(x)\ ,\label{c.14}\eea
and the Fronsdal fields are given by ($s\geqslant 3$)
\bea \phi_{\mu_1\dots \mu_s}&=& 2i\l^{-1} e_{(\mu_1}{}^{\a_1\ad_1}\cdots e_{\m_{s-1}}{}^{\a_{s-1}\ad_{s-1}} {\partial^2\over \partial y^{\a_1} \partial {\bar y}^{\ad_1}} \cdots {\partial^2\over \partial y^{\a_{s-1}} \partial {\bar y}^{\ad_{s-1}}} W_{\mu_s)}(x|Y)|_{Y=0}\ .\qquad\label{c.15}\eea
The dynamical equations of motion read 
\bea (\nabla^2_g+2\l^2)\phi&=&T\ ,\qquad \nabla^\mu_g f_{\m\n}~=~T_\n\ ,\label{EOM01}\\[5pt]
G_{\mu\nu}+3\l^2 g_{\m\n}&=& T_{\m\n}\ ,\qquad
G_{\m_1\dots \m_s}~=~ T_{\m_1\dots \m_s}\ ,\label{EOM2s}\eea
where $\nabla_g$ is the standard metric connection; $f_{\mu\nu}=2\partial_{[\m}a_{\n]}$; $G_{\m\n}$ is the Einstein tensor; $G_{\m_1\dots\m_s}$ are the covariantized, self-adjoint Fronsdal operators (containing the standard minimal metric couplings); and the composite sources $(s=0,1,2,\dots)$\footnote{The algorithm for calculating the composite sources is spelled out in \cite{Sezgin:2002ru}; it amounts to iterative elimination of auxiliary fields and imposition of generalized holonomic gauges which can be reached at every order in the weak-field expansion under the usual assumptions of perturbation theory.}
$T_{\m(s)}= \sum_{n=2}^\infty T^{<n>}_{\m(s)}$ 
where $T^{<n>}_{\m(s)}$ are $n$th order in the weak fields, \emph{i.e.} $(\phi,a_\m,\{\phi_{\m_1\dots\m_s}\})$. For fixed $n$, the quantities $T^{<n>}_{\mu(s)}$ and $r^{<n>}_{\mu(s)}$ have derivative expansions\footnote{For example, the nonlocal quadratic scalar-field stress-energy tensor $T^{(2)}_{\m\n}[\phi]$, that depends quadratically on $\phi$ and all its derivatives, was calculated in \cite{Kristiansson:2003xx}.} to all orders in $\l^{-1}\nabla_\m$. These dimensionless operators become large when acting on localizable weak-field fluctuations; by examining $\star$-products at $Z=0$, one can show that the derivative expansions are indeed strongly coupled and actually formally divergent for fluctuations fields belonging to lowest-weight and highest-weight spaces \cite{Iazeolla:2008ix}. 

We note that, while the reduced equations of motion on ${\cal X}_4$ are manifestly generally covariant, their invariance under spin-one and higher-spin gauge transformations, which can be read off (\'a la Cartan) from \eq{C8} and \eq{C9} and take the form
\bea \delta a_\m&=& \partial_\mu\e + r_\m\ ,\qquad \delta \phi_{\m_1\dots\m_s}\ =\ \nabla_{(\m_1} \e_{\m_2\dots \m_s)}+ r_{\m_1\dots\m_s}\ ,\\[5pt]
\delta\phi&=&r\ ,\qquad \delta g_{\m\n}\ =\ r_{\m\n}\ ,\eea
with $r_{\mu(s)}$ given by double expansions in weak fields and derivatives, is subtle in the sense that in order to verify it one would have to collect an infinite number of terms at each fixed order in the double expansion.


\scs{Spin-frames Adapted to $K$-Matrices}\label{App:L-rot}


\scss{Canonical forms of $L$-rotated $K$-matrices}

The van der Waerden symbols can be realized in a given spin-frame 
\be U~=~(u^\pm_\a,{\bar u}^\pm_{\ad})\ ,\qquad \bar u^\pm_{\ad}~=~(u^\pm_\a)^\dagger\ ,\qquad u^{+\a}u^-_\a=1=\bar u^{+\ad}\bar u^-_{\ad}\ ,\label{D.1}\ee
\be \e_{\a\b} \ = \ (u^- u^+- u^+ u^-)_{\a\b}\ ,\qquad \e_{0123}~=~1\ ,\ee
as
\bea \s_0|_U& = & u^+\bar u^++ u^-\bar u^- \ , \qquad \s_1|_U~=~u^+\bar u^-+ u^-\bar u^+ \ ,\label{sigmaupm}\\[5pt]
\s_2|_U & = & i(u^-\bar u^+- u^+\bar u^-) \ ,\qquad \s_3|_U \ = \ u^+\bar u^+- u^-\bar u^-\ ,\eea
\be \s_{01}|_U ~=~ u^+ u^+- u^- u^-\ , \quad \s_{02}|_U ~=~ -i (u^+ u^++ u^- u^-) \ ,\quad  \s_{03}|_U ~=~- (u^+ u^-+ u^- u^+) \ee\be
\s_{12}|_U~=~ i\s_{03}|_U\ ,\qquad \s_{23}|_U~=~ i\s_{01}|_U\ ,\qquad \s_{31}|_U~=~i\s_{02}|_U\ ,\label{D.2}\ee
with $\bar\s_{ab}|_U$ given by complex conjugates. In what follows we shall let $U$ denote a well-defined spin-frame at the base-point $p_0$ where $(\widehat \Phi',\widehat S'_{\underline\a})$ are evaluated.

For a given $K=E,J,iB,iP$ ($E=P_0=M_{0'0}$, $J=M_{12}$, $iB=iM_{03}$, $iP=iP_1=iM_{0'1}$) there exists an adapted spin-frame $\widetilde U\equiv \widetilde U_{(K)}$ in which $K^L\equiv L^{-1}\star K\star L\equiv \frac12 Y^{\underline\a}K^L_{\underline{\a\b}} Y^{\underline\b}$ assumes the following canonical form ($\Upsilon\in \Real$ for $E^L$ and $J^L$, while it can be real or imaginary, depending on the spacetime region, for $iB^L$ and $iP^L$, see Table \ref{Table1}):
\be\ba{lclcl}  E^L&:&\left[\ba{cc} 2 \Upsilon \tilde u^+ \tilde u^-& \sqrt{1+\Upsilon^2} (\tilde u^+\tilde {\bar u}^++\tilde u^-\tilde {\bar u}^-)\\ \sqrt{1+\Upsilon^2} (\tilde {\bar u}^+\tilde u^++\tilde {\bar u}^-\tilde u^-)& 2\Upsilon \tilde {\bar u}^+ \tilde {\bar u}^-\ea\right]&=&\Upsilon\,\Gamma^{03}|_{{\widetilde U}}- \sqrt{1+\Upsilon^2}\, \Gamma^0|_{{\widetilde U}}\ ,\\[20pt]    
J^L&:&\left[\ba{cc} 2i \sqrt{1+\Upsilon^2}  \tilde u^+ \tilde u^-&\Upsilon (\tilde u^+\tilde {\bar u}^-+\tilde u^-\tilde {\bar u}^+)\\ \Upsilon (\tilde {\bar u}^+\tilde u^-+\tilde {\bar u}^-\tilde u^+)& -2i\sqrt{1+\Upsilon^2} \tilde {\bar u}^+ \tilde {\bar u}^-\ea\right]&=&-\sqrt{1+\Upsilon^2}\,\Gamma^{12}|_{{\widetilde U}}+\Upsilon \, \Gamma^1|_{{\widetilde U}}\ ,\\[20pt]
iB^L&:&\left[\ba{cc} 2i \sqrt{1+\Upsilon^2}  \tilde u^+ \tilde u^-&i\Upsilon (\tilde u^+\tilde {\bar u}^++\tilde u^-\tilde {\bar u}^-)\\ i\Upsilon (\tilde {\bar u}^+\tilde u^++\tilde {\bar u}^-\tilde u^-)& 2i\sqrt{1+\Upsilon^2} \tilde {\bar u}^+ \tilde {\bar u}^-\ea\right]&=&-i\left(\sqrt{1+\Upsilon^2}\,\Gamma^{03}|_{{\widetilde U}}+\Upsilon \, \Gamma^0|_{{\widetilde U}}\right)\ ,\\[20pt] 
iP^L&:&\left[\ba{cc} 2 \Upsilon \tilde u^+ \tilde u^-& i\sqrt{1+\Upsilon^2} (\tilde u^+\tilde {\bar u}^-+\tilde u^-\tilde {\bar u}^+)\\ i\sqrt{1+\Upsilon^2} (\tilde {\bar u}^+\tilde u^-+\tilde {\bar u}^-\tilde u^+)& -2\Upsilon \tilde {\bar u}^+ \tilde {\bar u}^-\ea\right]&=&i\left(\Upsilon\,\Gamma^{12}|_{{\widetilde U}}- \sqrt{1+\Upsilon^2}\, \Gamma^1|_{{\widetilde U}}\right)\ .\ea\ee
The $K$-adapted spin-frames $\widetilde U_{(K)}$ may have ill-defined limits at $p_0$ while 
\be (\Gamma_A)_{\underline{\a\b}}|_{\widetilde U_{(K)}}|_{p\rightarrow p_0}~=~ (\Gamma_A)_{\underline{\a\b}}\ .\ee
In the remainder of this Appendix we collect the transformations from the fixed spin-frame $U$ to $\widetilde U_{(E)}$ and $\widetilde U_{(J)}$.

\scss{$E$-adapted and $J$-adapted spin-frames}

The decomposition \eq{kv} of the matrix 
\be E^L_{\underline{\a\b}}~:=~L_{\underline\a}{}^{\underline\a'} L_{\underline\b}{}^{\underline\b'} (\Gamma_0)_{\underline{\a'\b'}}~=~\left[\ba{cc}\vark^L_{(E)\a\b}&v^L_{(E)\a\bd}\\ \bar v^L_{(E)\ad\b}& \bar\vark^L_{(E)\ad\bd}\ea\right]\ ,\ee
with $L_{\underline{\a\b}}(x^\mu)$ given in stereographic coordinates by \eq{3.20}, takes the following form in the global embedding coordinates $X^A$ defined in \eq{A.15} ($q(X):=\sqrt{1+X^\m X_\m}$):
\bea \vark^L_{(E)\a\b} & = & X_{(\a}{}^{\ad}\left(u^{+}\bar u^{+}+u^{-}\bar u^{-}\right)_{\b)\ad}\nn\\[5pt]
& = & X_3(u^{+}u^{-}+u^{-}u^{+})_{\a\b}+(X_1+iX_2)(u^{-}u^{-})_{\a\b}-(X_1-iX_2)(u^{+}u^{+})_{\a\b} \nn\\[5pt]
v^L_{(E)\a\bd} &=&  \frac{1+q}{2}\left[\left(1+\frac{X_0^2+X_i X_i}{(1+q)^2}\right)(u^{+}\bar u^{+}+u^{-}\bar u^{-})_{\a\bd}\right. \nn\\[5pt]
&&- \left.2\frac{X_0}{(1+q)^2}\left(X_3(u^{+}\bar u^{+}- u^{-}\bar u^{-})_{\a\bd}+X_1(u^{+}\bar u^{-}+ u^{-}\bar u^{+})_{\a\bd}+iX_2(u^{-}\bar u^{+}- u^{+}\bar u^{-})_{a\bd}\right)\right]\ ,\qquad \eea
obeying $\vark^L_{(E)\a\b}|_{X^i=0}=0$ and $v^L_{(E)\a\bd}|_{X^i=0}=(\s_0)_{\a\bd}$. In global spherical coordinates $(t,r,\theta,\phi)$, as defined in \eq{AdSspherical}, one has
\be
\vark^L_{(E)\a\b}~=~ r\left[\cos\th(u^{+}u^{-}+u^{-}u^{+})_{\a\b}+\sin\th e^{i\phi}(u^{-}u^{-})_{\a\b}-\sin\th e^{-i\phi}(u^{+}u^{+})_{\a\b}\right] \ .\ee
For $r>0$, the canonical form with $\Upsilon=r$ is assumed on the $E$-adapted spin-frame 
\bea \tilde u^{+}_{(E)\a} & = & \frac{p}{\sqrt{2}}\left[\sqrt{1+\cos\th}\,u^{+}_\a+\sqrt{1-\cos\th}\,e^{i\phi}\,u^{-}_\a\right] \\[5pt]
\tilde u^{-}_{(E)\a} & = & \frac{p^{-1}}{\sqrt{2}}\left[-\sqrt{1-\cos\th}\,e^{-i\phi}\,u^{+}_\a+\sqrt{1+\cos\th}\,u^{-}_\a\right] \ ,\eea
where
\bea p(x)\ = \ \left(\frac{\sqrt{1+r^2}+|\cos t| +r\sin t}{\sqrt{1+r^2}+|\cos t| -r\sin t}\right)^{1/4} \ , \eea
leading to \eq{kLE} and \eq{vLE}.

Analogously, the $L$-rotation of $K=J=M_{12}$ yields
\bea v^L_{(J)\a\bd} & = & -i\left[(X_1+iX_2)u^{-}_\a\bar u^{+}_{\bd}-(X_1-iX_2)u^{+}_\a\bar u^{-}_{\bd}\right] \ , \\[5pt]
 \vark^L_{(J)\a\b} & = & i\left[\left(1+\frac{X^2_1+X^2_2}{1+q}\right)(u^{+}_{\a}u^{-}_{\b}+u^{-}_{\a}u^{+}_{\b}) \right.\nn\\[5pt]
 &+&\left.\frac{(X_0-X_3)(X_1+iX_2)}{1+q}u^{-}_\a u^{-}_\b + \frac{(X_0+X_3)(X_1-iX_2)}{1+q}u^{+}_\a u^{+}_\b \right] \ ,\eea
which obey $\vark^L_{(J)\a\b}|_{X^i=0}=-(\s_{12})_{\a\b}$ and $v^L_{(E)\a\bd}|_{X^i=0}=0$, and that acquire the canonical form with $\Upsilon=\sqrt{1+r^2\sin^2\th}$ on the $J$-adapted spin-frame
\bea \tilde u^{+}_{(J)\a}& = & \frac{e^{i\ft{\pi}4}}{\sqrt{2}}\left[f_+\,e^{-i\phi/2}\,\tilde u^{+}_{(E)\a}-f_-\,\,e^{i\ft{\phi}2}\,\tilde u^{-}_{(E)\a}\right] \\[5pt]
\tilde u^{ -}_{(J)\a}& = & \frac{e^{-i\ft{\pi}4}}{\sqrt{2}}\left[f_-\,e^{-i\ft{\phi}2}\,\tilde u^{+}_{(E)\a}+f_+\,e^{i\ft{\phi}2}\,\tilde u^{-}_{(E)\a}\right] \ , \eea
where $\tilde u^{\pm}_{(E)\a}$ is the $E$-adapted spin-frame and
\be f_\pm(r,\th)\  = \ \sqrt{1\pm\frac{\cos\th}{\sqrt{1+r^2\sin^2\th}}} \ .\ee
%


\scs{ Weyl-ordered and Regular Presentation of Projectors}\label{AppProj}


Given the complexified Heisenberg algebra \eq{HA}, a set of (diagonal) projectors $P_{\bf n}\equiv P_{{\bf n},{\bf n}}$ obeying the orthonormality condition \eq{3.31} and with eigenvalues ${\bf n}=(n_1,n_2)$, $n_i\in\mathbb{Z}+\ft12$, as in \eq{3.8}, is (for details, see for example \cite{Iazeolla:2008ix}) 
\be P_{\mathbf n} ~=~P_{n_1}(w_1)\star P_{n_2}(w_2)\ ,\label{diagproj}\ee
\be w_i ~=~ y^+_i y^-_i\ =\ y^+_i\star y^-_i+\ft12\ =\ y^-_i\star y^+_i-\ft12\qquad
\mbox{(no sum over $i$)}\ ,\ee
with single-Fock-space projectors having the following Weyl-ordered regular presentation ($n\in\mathbb{Z}+\ft12$, $\ve:=n/|n|$, $w:=y^- y^+$) 
\be P_n(w) ~=~ 2(-1)^{|n|-\ft12}\,\oint_{C(\ve)} \frac{ds}{2\pi  i}\frac{(s+1)^{n-\ft12}}{(s-1)^{n+\ft12}}\,\left[e^{-2sw}\right]_{\rm Weyl}\ ,
\ee
where $C(\ve)$ is prescribed to be a small contour encircling $\ve$. 
With this prescription, and writing $P_n\equiv\oint_{C(\ve)} ds f_n(s|w)$, the $\star$-product composition $P_n\star P_{n'}\equiv (\oint_{C(\ve)} ds f_n(s|w))\star (\oint_{C(\ve')} ds' f_{n'}(s'|w))$ is, by the very definition of regular presentations, performed by exchanging the auxiliary integrals with the $\star$-product, performing the latter and then performing the auxiliary integrations one after the other, say the $s'$ integral while keeping $s$ fixed, \emph{viz.}
\be P_n\star P_{n'}~:=~ \oint_{C(\ve)} ds \left[\oint_{C_s(\ve')} ds' \left(f_{n}(s|w) \star f_{n'}(s'|w)\right)\right]\ ,\ee
where thus $C_s(\ve')=\left\{s':|s'-\ve'|\ll|s-\ve|\ll1\right\}$. Using the $\star$-product lemma 
\bea \left[e^{-2sw\phantom{'}}\right]_{\rm Weyl}\star \left[e^{-2s'w}\right]_{\rm Weyl}&=& {1\over
1+ss'}\left[\exp\left(-2{\frac{s+s'}{1+ss'}\,w}\right)\right]_{\rm Weyl}\ ,\label{swtw}\eea
which holds by analytical continuation for all $s$ and $s'$ such that $ss'\neq -1$, and changing variables from $s'$ to 
\be u(s')~=~(s+s')(1+ss')^{-1}\quad \mbox{ at fixed $s\in C(\ve)$}\ ,\label{changeofvariables}\ee
it follows that\footnote{If $\ve=\ve'$ then $u\in C(\ve')$ provided $|s'-\ve'|\ll 1$ and $|s-\ve|\ll1$ while if $\ve=-\ve'$ then $u\in C(\ve')$ provided $|s'-\ve'|\ll |s-\ve|\ll1$.} $u\in C(\ve')$ for all $s\in C(\ve)$; as a result, the auxiliary $s$-integral factorizes out and one has 
\bea P_n\star P_{n'}&=& \left[2(-1)^{|n|+n'-1} \oint_{C(\ve)}{ds\over 2\pi i} \frac{(s+1)^{n-n'-1}}{(s-1)^{n-n'+1}}\right] 2(-1)^{|n'|-\ft12} \oint_{C(\ve')}{du\over 2\pi i} \frac{(u+1)^{n-\ft12}}{(u+1)^{n+\ft12}}\,\left[e^{-2uw}\right]_{\rm Weyl}\qquad \nn\\[5pt]
&=& \delta_{n,n'} P_{n'}\ .\label{recort}\eea
We note that this orthonormalization is consistent with associativity, which requires $P_{-\ft12}\star P_{\ft12}=2P_{-\ft12}\star(w\star P_{\ft12})=-P_{-\ft12}\star P_{\ft12}=0$, as can be seen from $y^{\ve} \star P_{-\ft\ve 2}=0=P_{-\ft\ve 2}\star y^{-\ve}$ ($\ve=\pm$). 
Moreover, it can be seen from the above calculation that the order in which the auxiliary integrals is performed is immaterial for the final result and that one may also choose to replace one of the two auxiliary integrals by its residues prior to performing the $\star$-product, which is equivalent to collapsing the closed contour $C_s(\ve')=\left\{s':|s'-\ve'|\ll|s-\ve|\ll1\right\}$ above. 
Using the latter, slightly simplified, prescription, the space ${\cal A}:={\cal A}^+\oplus {\cal A}^-$ spanned by the generalized Fock-space ($+$) and anti-Fock-space $(-)$ projectors 
\be P_{n|n'}~:=~\ft1{\sqrt{(|n|-\ft12)!(|n'|-\ft12)!}} (\ve y^\ve)^{\star (|n|-\ft12)}\star P_{\ft{\ve}2}\star (y^{-\ve'})^{\star (|n'|-\ft12)}\ee
(which are non-trivial only if $\ve=\ve'$ and inherit a regular presentation from $P_{\ft{\ve}2}$)
can be equipped with a $\star$-product rule whereby one auxiliary integral is introduced for each composition.
Thus, given ${\cal O}_i\in {\cal A}$ ($i=1,\dots,N$), their $N$-fold $\star$-products, namely $(\cdots ({\cal O}_1\star {\cal O}_2)\star \cdots \star {\cal O}_{N-1})\star {\cal O}_N$ and all other arrangements obtained by permuting the composition order, are to be evaluated by introducing one auxiliary integral prior to each $\star$-product; the former integrals can then be factored out, one after the other, as in Eq. \eq{recort}. By induction, it follows that all nestings yield the same answer, \emph{viz.} 
\be (\cdots ({\cal O}_1\star {\cal O}_2)\star \cdots \star {\cal O}_{N-1})\star {\cal O}_N~=~ \sum_{n,n'} P_{n|n'} (\check {\cal O}_{1}\cdots \check{\cal O}_N)_{n|n'}\ ,\ee
using the notation ${\cal O}_i:= \sum_{n,n'} P_{n|n'} \check {\cal O}_{i;n|n'}$ and 
$(\check {\cal O}\check{\cal O}')_{n|n'}:=\sum_{m} \check {\cal O}^{\phantom{'}}_{n|m}\check{\cal O}'_{m|n'}$ and the lemma
\be P_{n|n'}\star P_{m|m'}~=~\delta_{n',m} P_{n|m'}\ .\ee
Hence, in particular, one has  
$({\cal O}\star  {\cal O}')\star{\cal O}''={\cal O}\star ( {\cal O}'\star{\cal O}'')$ manifesting associativity. 

As for the double-Fock-space projectors in \eq{diagproj}, their regular presentation  reads ($\ve_i:=n_i/|n_i|$)
\bea P_{n_1 ,n_2} & = & 4(-1)^{\sum_i |n_i|-1}\,\oint_{C(\ve_1)} \frac{ds_1}{2\pi i} \frac{(s_1+1)^{n_1-\ft12}}{(s_1-1)^{n_1+\ft12}}\oint_{C(\ve_2)}\frac{ds_2}{2\pi i}\frac{(s_2+1)^{n_2-\ft12}}{(s_2-1)^{n_2+\ft12}}\,\left[e^{-2\sum_i s_i w_i}\right]_{\rm Weyl}\ . \qquad\label{intproj}\eea
As each Fock-space projector has the form \eq{singleproj} and $Y_{\underline{\a}}\star \k_y\bar{\k}_{\yb} = - \k_y\bar{\k}_{\yb}\star Y_{\underline{\a}}$, one has
\bea P_{n_1, n_2} \star \k_y \bar \k_{\bar y } \ = \ \kappa(K_{(\ve_1\ve_2)})\,(-1)^{| n_1|+| n_2|-1}\,P_{n_1, n_2}  \ ,\label{Pkbark}\eea
where the sign factors $\kappa(K)$ for $K=E,J,iB,iP$, which are collected in Table \ref{Table1}, are determined using Gaussian integration. In the two $\pi$-even cases, $J=M_{12}$ and $iB=iM_{03}$, the integrals factorize into holomorphic and anti-holomorphic pieces as follows (assuming Weyl order):
\be e^{\mp 4J}\star \k_y\bar\k_{\yb}~=~(e^{\mp i y^+ y^-}\star \k_y)\star(e^{\pm i \bar y^+ \bar y^-}\star \bar\k_{\yb})\ ,\ee
\be e^{\mp 4iB}\star \k_y\bar\k_{\yb}~=~(e^{\mp i y^+ y^-}\star \k_y)\star(e^{\mp i \bar y^+ \bar y^-}\star \bar\k_{\yb})\ ,\ee
using the spin-frames in Eqs. \eq{D.1}--\eq{D.2} and $y^\pm:=u^\pm y$ and $\bar y^\pm:=\bar u^\pm \bar y$. From the chiral $\star$-product 
\be e^{\mp i y^+ y^-}\star \k_y~=~i\,e^{\mp i y^+ y^-}\ ,\ee
it follows that
\be e^{\mp 4J}\star \k_y\bar\k_{\yb}~=~i\,(-i)\, e^{\mp 4J}\quad \Rightarrow\quad \kappa(J)~=~+1\ ,\ee
\be e^{\mp 4iB}\star \k_y\bar\k_{\yb}~=~i^2\,e^{\mp 4iB}\quad \Rightarrow\quad \kappa(iB)~=~-1\ .\ee
In the two $\pi$-odd cases, $E=P_0$ and $iP=iP_3$, say, the integrals factorize in a similar fashion as follows:
\be e^{\mp 4E}\star \k_y\bar\k_{\yb}~=~(e^{\mp y^+ \bar y^+}\star \k_+)\star(e^{\mp y^- \bar y^-}\star \k_-)\ ,\ee
\be e^{\mp 4iP}\star \k_y\bar\k_{\yb}~=~(e^{\mp i y^+ \bar y^+}\star \k_+)\star(e^{\pm i y^- \bar y^-}\star \k_-)\ ,\ee
where we have defined $\k_\pm:=2\pi [\d(y^\pm)\d(\bar y^\pm)]_{\rm Weyl}$. It follows that 
\be \kappa(iP)~=~\kappa(J)~=~+1\ ,\ee
while $\kappa(E)$ can be brought back to the case of $\kappa(iB)$ by analytical continuation leading to
\be \kappa(E)~=~i^2 \kappa(iB)~=~-1\ .\ee
Concerning the rank-$|n|$ projectors ($K_{(q)}:=\ft12(qw_1+w_2)$, $q=\pm 1$)
\be {\cal P}_{n}(K_{(q)}) := \sum_{\tiny \ba {c}n_2+qn_1=n\\[-3pt] \e_1\e_2=q\ea}P_{n_1, n_2}\ ,\qquad n\in\{\pm 1, \pm 2, \dots\}\ ,\label{ranknproj}\ee
they are invariant under the centralizer $\mathfrak{c}_{\msp(4;\Comp)}(K_{(q)})$. In particular,  the ground-state projectors 
\be P_{\ft{\ve_1} 2,\ft{\ve_2}2}~=~ {\cal P}_{\ve_2}(K_{(\ve_1\ve_2)})\ .\ee
To perform the sum in \eq{ranknproj} one may first perform the closed-contour integrals in the single-Fock-space projectors, which yields
\be P_n(w) ~=~
\frac{1}{(|n|-\ft12)!}\,(\ve y^{\ve})^{|n|-\ft12}\,\star\,P_{\ft{\ve}2}\,\star\,(y^{-\ve})^{|n|-\ft12}~=~ 2(-1)^{|n|-\ft12} \left[e^{-2w}L_{n-\ft12}(4w)\right]_{\rm Weyl}\ ,  \label{singleproj}\ee
where $L_{n-\ft12}(x)\equiv L_{n-\ft12}^{(0)}(x)$ and 
\be L^{(\a)}_{n-\ft12}(x)~=~{x^{-\a}\,e^x \over (n-\ft12)!}{d^{n-\ft12}\over
dx^{n-\ft12}}(e^{-x}x^{n-\ft12+\a})\ee
are the generalized Laguerre polynomials with $n\geqslant \ft12$. We note that \eq{singleproj} holds for all $n$ 
by virtue of Kummer's transformation
\bea  L^{(\a)}_{n-\ft12}(x) \ = \ \frac{e^x\sin((n-\ft12)\pi)}{\sin((n-\ft12+\a)\pi)}\,L^{(\a)}_{-n-\ft12-\a}(-x) \ .\eea
From the recurrence relation $\sum_{p+q=r} L_p^{(\a)}(x)\,L_q^{(\b)}(y) = L^{(\a+\b+1)}_r(x+y)$ for $p,q,r\,\in\,\mathbb{N}$ and Kummer's transformation
it follows that ($n=\pm 1, \pm 2, \dots$; $\ve:=n/|n|=\ve_2$ using $n_2+qn_1=\ve_2(|n_1|+|n_2|)$)
\bea {\cal P}_{n}(K_{(q)}) &=& \sum_{\tiny \ba {c}n_2+qn_1=n\\[-3pt] \e_1\e_2=q\ea}4(-1)^{|n|-1}\left[e^{-2(w_1+w_2)}L_{n_1-\ft12}(4w_1)L_{ n_2-\ft12}(4 w_2)\right]_{\rm Weyl}\\[5pt] &=&4(-)^{n-\ft{1+\ve}2} \,\left[e^{-4K_{(q)}}L^{(1)}_{n-1}(8K_{(q)})\right]_{\rm Weyl}\label{enhanced2}\\[5pt]& = & 2(-)^{n-\ft{1+\ve}2}\,\oint_{C(\ve)} \frac{ds}{2\pi i}\,\left(\frac{s+1}{s-1}\right)^{n}\,\left[e^{-4sK_{(q)}}\right]_{\rm Weyl}\ .\label{enhanced}\eea

\scs{Details of the Deformed Oscillators}\label{App:D}

In this Appendix, we spell out various details of solving the deformed-oscillator problem in Eqs. \eq{defosc1} and \eq{defosc2} by casting it via the Laplace transformation \eq{WSigmaansatz} into the solvable $\circ$-product equation \eq{ringeq} \cite{Prokushkin:1998bq} leading to the solution given in Eqs. \eq{j2} and \eq{j}. 
In the first part, we retrieve the $\circ$-product equation \eq{ringeq} by a different route than that taken in Section \ref{Sec:defosc}, namely by Laplace transforming the deformed oscillators in (anti-)normal-ordered bases rather than in the Weyl-ordered basis used in \eq{WSigmaansatz}. In the second part, we solve Eq.  \eq{ringeq} taking into account the splitting into even and odd Laplace transforms on $[-1,1]$ as well as non-trivial contributions distributions supported at the mid-point corresponding to Fock-space projectors \cite{Sezgin:2005pv,Iazeolla:2007wt}.

\scss{Laplace transforming in (anti-)normal-ordered bases}

Instead of relying on the limit \eq{deltalimit}, the deformed-oscillator equations \eq{defosc1} and \eq{defosc2} can be mapped to the $\circ$-product equation \eq{ringeq} by performing the Laplace transformation using either normal order $:\cdot :_+$ or anti-normal order $:\cdot :_-$ with respect to the complexified Heisenberg algebra $z^\pm=u^{\pm\a} z_\a$ obeying $[z^-,z^+]=-2i$, where the $\star$-product is represented by 
\bea :\,f(z^-,z^+)\,:_+ \star :\,g(z^-,z^+)\,:_+ \ = \ \int\frac{d\xi^-d\eta^+}{4\pi}\,e^{\ft{i}2\xi^-\eta^+}:f(z^+,z^-+\xi^-)\,g(z^+-\eta^+,z^-) :\ ,\label{normstar}\eea
\bea :\,f(z^-,z^+)\,:_- \star :\,g(z^-,z^+)\,:_- \ = \ \int\frac{d\xi^+d\eta^-}{4\pi}\,e^{-\ft{i}2\xi^+\eta^-}:f(z^++\xi^+,z^-)\,g(z^+,z^--\eta^-): \ .\label{normstar-}\eea
The (anti-)normal-ordered form of the inner Klein operator $\k_z$ is given by
\bea \k_z ~=~ \ (-1)_\star^{N_z} ~=~  :\,e^{-2\s N_z}\,:_\s \ , \qquad N_z~:=~\ft{i}2 z^+\star z^-\ ,\label{kznorm}\eea
which breaks manifest $SL(2;\Comp)$-covariance as well as $\tau$-covariance in the sense that $\t(\k_z)=-\k_z$ while $\tau(:e^{-2 \s N_z}:_\s)=:e^{2\s N_z}:_{-\s}$. The deformed oscillator problem gets the form ($N_z=\ft i2 w_z-\ft12$; $w_z=\ft12\{z^+,z^-\}_\star$)
\bea [\Sigma_\a\,,\,\Sigma_\b]_\star \ = \ -2i\e_{\a\b}\left(1-{\cal B}\nu\,:e^{-i\sigma w_z}:_\sigma \right) \ ,\label{defsigmanorm}\eea
where the index $\mathbf{n}$ has been suppressed. One proceeds by making the Ansatz
\bea \Sigma^{\pm}_{\s}  :=  u^{\pm\,\a}\Sigma_\a &= & \int_{-1}^1 dt\, f^{\pm}_\s(t)\,: z^{\pm} e^{\ft{i\s}2\,(t-1)w_z} :_\s \label{Sigmaansatz}\\[5pt] &=& -2i\frac{\partial}{\partial\r^\pm}\left.\int_{-1}^1 dt \,:e^{\ft{i}2\,\left(\s(t-1)w_z+\r^+ z^++\r^-z^-\right)}:_\s f_{\pm}(t)\right|_{\r^{\pm}=0} \ . \label{normSigma} \eea
chosen such that $f^{\pm}_\s(t)$ shall turn out to be the same as in \eq{WSigmaansatz}. 
From the lemma
\bea && :\exp \ft{i}2\left(\s(t-1)w_z+\r^+z^++\r^-z^-\right):_\s\star :\exp\ft{i}2\left(\s(t'-1)w_z+\r^{\prime\,+}z^++\r^{\prime\,-}z^-\right): _\s\nn\\[5pt]
&=& :\exp \ft{i}2\left(\s(tt'-1)w_z+\tilde \r^+ z^++\tilde \r^- z^-+\ft12(1+\s)\r^-\r^{\prime\,+}-\ft12(1-\s)\r^{\prime-}\r^+\right):_\s\hspace{1cm}\ ,\eea
\be \tilde \r^\pm~:=~\ft12(1\pm\s)(\r^\pm+t \r^{\prime\pm})+\ft12(1\mp\s)(\r^{\prime \pm}+t' \r^\pm)\ ,\ee
one gets
\bea \left[:z^-\,e^{\ft{i\s}2\,(t-1)w_z}:_\s\,,\,:z^+\,e^{\ft{i\s}2 (t'-1)w_z}:_\s\right]_\star \ = \ -2i:\left(1+\ft{i\s}2(tt'-1)w_z\right)\, e^{\ft{i\s}2(tt'-1)w_z}:_\s\ ,\eea
that is, 
\bea \int_{-1}^1 dt\int_{-1}^1 dt'f^-_\s(t)f^+_\s(t')\left[ 1+\ft{i\s}2(tt'-1)w_z \right]\,:e^{\ft{i\s}2(tt'-1)w_z}:_{\s} \ = \ 1-{\cal B}\nu \,:e^{-i\s w_z}:_\s \ .\eea
Inserting $1=\int_{-1}^1 du\,\d(tt'-u)$ into left-hand side and using $h_\s(u):=(f^-_\s\circ f^+_\s)(u)$ with $\circ$ given by the convolution defined in \eq{ringo}, one obtains the integral equation 
\bea  \int_{-1}^1 du \,h(u)\,\left[(u-1)\frac{\partial}{\partial u}+1\right]:e^{\ft{i\s}2(u-1)w_z}:_\s \ = \  1-{\cal B}\nu \,:e^{-i\s w_z}:_\s \ , \eea
with the unique solution $h_\s(u)=\d(u-1)-\ft{\s{\cal B}\nu}2$. The original problem is therefore mapped to the $\circ$-product equation \eq{ringeq}. Going back to Weyl order using\footnote{The $\r^+\r^-$-term does not contribute to \eq{normSigma}. }
\bea :e^{\ft{i}2(\s(t-1)z^+z^- +\r^+z^+ +\r^-z^-)}:_\s \ = \ \frac{2}{t+1}\,\left[e^{\ft{i}{t+1}(\s(t-1)z^+z^- +\r^+z^+ +\r^-z^- -\ft\s 2\r^+\r^-)}\right]_{\rm Weyl} \ ,\label{normtoWeyl}\eea
one then retrieves the Ansatz given in \eq{WSigmaansatz}.  
In order to derive the re-ordering formula above, it is convenient to Fourier-transform the left-hand side as
\bea :e^{\ft{i}2( s z^+ z^-+\r^+z^++\r^- z^-)}:_+ \ = \ \int\frac{dkd\bar k}{4\pi} :e^{-\ft12(s kz^+ +\bar k z^-)+\ft{i}2(k\bar k+\r^+z^++\r^- z^-)}:_+ \ .\label{FT}\eea
Then, the Baker-Campbell-Hausdorff formula can be used to write the normal-ordered, $z$-dependent part of the integrand as
\bea & :e^{-\ft12(s kz^+ +\bar k z^-)+\ft{i}2(\r^+z^++\r^- z^-)}:_+  \ = \ e_\star^{\ft{i}2(\r^++isk)z^+}\star e_\star^{\ft{i}2(\r^-+i\bar k)z^-} &\nn\\[5pt]
  & \ = \ e_\star^{\ft{i}2(\r^++isk)z^+ + \ft{i}2(\r^-+i\bar k)z^- + \ft{i}4 (sk\bar k-\r^+\r^-)+\ft14(sk\r^-+\bar k\r^+)}&\nn\\[5pt]
  & \ = \ \left[e^{\ft{i}2(\r^++isk)z^+ + \ft{i}2(\r^-+i\bar k)z^- + \ft{i}4 (sk\bar k-\r^+\r^-)+\ft14(sk\r^-+\bar k\r^+)}\right]_{\rm Weyl}\ , &\label{Weylexp}\eea
where the last equality follows from the fact that  $e_\star^{f(z)}=\left[e^{f(z)}\right]_{\rm Weyl}$ if $f$ is linear in $z$ (as any linear combination $A^i z_i$, with $i=-,+$, satisfies $A^{i_1} z_{i_1}\star\cdots\star A^{i_n} z_{i_n}=A^{i_1} z_{i_1}\cdots A^{i_n} z_{i_n}$ since any contraction is proportional to $\e_{ij}$ with $\e_{-+}=1=-\e_{+-}$, $\e_{++}=0=\e_{--}$). Inserting now the Weyl-ordered result \eq{Weylexp} in \eq{FT} and performing the integration one obtains \eq{normtoWeyl} for $\s=+$. The case of $\s=-$ is treated analogously.

\scss{Solving the $\circ$-product equation}

In order to solve the $\circ$-product equation \eq{ringeq}, \emph{i.e.}
\bea  (f^+_\s\circ f^-_\s)(u) \ = \ \d(u-1)-\frac{\s{\cal B}\nu}{2} \ ,\label{ringprobl}\eea
one begins by observing that the space of functions on the interval $[-1,1]$ decompose under the $\circ$-product into even and odd functions, \emph{viz.}
\bea f^{(\pi)}\circ g^{(\pi')} ~=~ \d_{\pi\pi'} f^{(\pi)}\circ g^{(\pi')} \ , \qquad f^{(\pi)}(-s) \ = \ \pi f^{(\pi)}(s) \ , \qquad \pi,\pi' \ = \ \pm 1 \ .\eea
Therefore \eq{ringprobl}  separates into the following two independent equations:
\bea  (f^{-(+)}_\s\circ f^{+(+)}_\s)(u) & = & I_0^{(+)}(u)-\frac{\s{\cal B}\nu}{2} \ ,\label{ringprobl+}\\[5pt]
(f^{-(-)}_\s\circ f^{+(-)}_\s)(u) & = & I_0^{(-)}(u) \ ,\label{ringprobl-}\eea
where
\bea I_0^{(\pm)} \ := \ \frac{1}{2}\left[\d(u-1)\pm\d(u+1)\right] \ ,\eea
acts as the identity in the $\circ$-product algebra.
Equations \eq{ringprobl+} and \eq{ringprobl-} can be cast into algebraic equations by expanding ($t\in[-1,1]$)
\bea f^{\pm(\pi)}_\s(t)&:=& m^{\pm(\pi)}_\s(t)+\sum_{k=0}^\infty \l^\pm_{\s,k} p^{(\pi)}_{k}(t)\
,\qquad m^{\pm(\pi)}_\s ~:=~ \sum_{k=0}^\infty \mu^{\pm}_{\s,k}\,I^{(\pi)}_k \ ,\label{mexp}\eea
in terms of $I^{(\pm)}_0$ and ($k\geqslant 1$)
 \bea
 I^{(\pi)}_k(u)&:=&\left[{\rm sign}(u)\right]^{\frac12(1-\pi
 )}~\int_{-1}^1 ds_1 \cdots \int_{-1}^1 ds_k~\delta(u-s_1\cdots
 s_k)\nn\\[5pt]
 &=&\left[{\rm sign}(u)\right]^{\frac12(1-\pi)}{\left(\log
 \frac1{u^2}\right)^{k-1}\over (k-1)!}\ ,
 \eea
obeying the algebra ($k,l\geqslant 0$)
 \be
 I^{(\pi)}_k\circ I^{(\pi)}_l\ =\ I^{(\pi)}_{k+l}\ ,
 \label{ring}
 \ee
and $p^{(\pi)}_k(t)$ ($k\geqslant 0$) are the $\circ$-product projectors
\bea p^{(\pi)}_k(s)&:=& {(-1)^k\over k!} \d^{(k)}(s)\ ,\qquad \pi =\
(-1)^k\ ,\label{pk}\eea
obeying
\bea p^{(\pi)}_k\circ f&=& L_k[f] p^{(\pi)}_k\ ,\qquad L_k[f]\ =\ \int_{-1}^1
ds~ s^k f(s)\ .\label{proj1}\eea
The property \eq{ring} implies that $m^{(\pi)}_-\circ m^{(\pi)}_+$ can be mapped to the algebraic product $\widetilde{m}^{(\pi)}_- (\xi) \widetilde{m}^{(\pi)}_+(\xi)$ between the symbols ($\xi\in\Comp$)
\bea \widetilde{m}^{\pm(\pi)}_\s (\xi)\ := \ \sum_{k=0}^\infty \mu^{\pm(\pi)}_{\s,k}\,\xi^k \ . \eea
Therefore, substituting \eq{mexp} into \eq{ringprobl+} and \eq{ringprobl-} and using \eq{ring} and \eq{proj1} one is left with the algebraic equations
\be  \widetilde{m}^{-(+)}_\s\, \widetilde{m}^{+(+)}_\s ~=~ 1-\frac{\s{\cal B}\nu}{2}\,\xi \ , \qquad
\widetilde{m}^{-(-)}_\s\, \widetilde{m}^{+(-)}_\s ~=~ 1 \ ,\label{meq2}\ee
and the following condition on the projector part of the expansion \eq{mexp}:
\bea \l^{-(\pi)}_{\s,k}\, L_{n}[m^{+(\pi)}_\s]+ \l^{+(\pi)}_{\s,k}\, L_{n}[m^{-(\pi)}_\s]+\l^{-(\pi)}_{\s,k}\l^{+(\pi)}_{\s,k}  \ = \ 0 \ .\label{proj2}\eea
The solution space to \eq{meq2} is parameterized by an undetermined function $\widetilde{g}_\s^{(\pi)}$ as follows\footnote{Note that, differently from the Lorentz-covariant solutions in \cite{Prokushkin:1998bq,Sezgin:2005pv,Iazeolla:2007wt}, the algebraic equations  involve the product of two different functions rather than the square of a single one.}:
\be \widetilde{m}^{\pm(+)}_\s~=~ (\widetilde{g}_\s^{(+)})^{\pm 1} \sqrt{1-\frac{\s{\cal B}\nu}{2}\,\xi}\ ,\qquad\widetilde{m}^{\pm(-)}_\s~=~ (\widetilde{g}_\s^{(-)})^{\pm 1}\ .\ee
Likewise, the solution space to \eq{proj2} contains an undetermined set of coefficients, say $\l^{+(\pi)}_{\s,k}$. One can show that these undetermined quantities are gauge artifacts. One natural gauge choice is to work with symmetric solutions 
\be f^{\pm}_\s~=~f_\s\quad \Rightarrow\quad \mu^{\pm(\pi)}_{\s,k}~=~\mu^{(\pi)}_{\s,k}\ ,\qquad \l^{\pm(\pi)}_{\s,k}~=~\l^{(\pi)}_{\s,k}\ .\ee
and we shall hence drop the $\pm$ referring to the spin-frame henceforth. Thus 
\be \widetilde m^{(+)}_\s~=~\varepsilon^{(+)}_\s\sqrt{1-\ft{\s{\cal B}\n}{2}\xi}\ ,\qquad \widetilde{m}^{\pm(-)}_\s~=~\varepsilon^{(-)}_\s\ ,\ee
where $(\varepsilon^{(\pm)}_\s)^2=1$, and ($\pi=(-1)^k$)
\bea \l^{(\pi)}_{\s,k}\left(\l^{(\pi)}_{\s,k}+2L_k[m^{(\pi)}_\s]\right) \ = \ 0 \ .\eea
It follows that
\bea m^{(+)}_\s~=~\ \varepsilon^{(+)}(I^{(+)}_0(s)+q^{(+)}_\s(s)) \ , \qquad m^{(-)}_\s ~=~ \varepsilon^{(-)} I^{(-)}_0(s)\ ,\eea
and that either $\l^{(\pi)}_{\s,k}=0$, and the projectors $p_k$ do not contribute to the internal connection, or $\l^{(\pi)}_{\s,k} \ = \ -2L_k[m^{(\pi)}_\s] \ = \ -2L_k[m^{(\pi)}_\s] $, \emph{i.e.},
\bea \l^{(\pi)}_{\s,k} \ = \ -2\th_{\s,k} L_k[m^{(\pi)}_\s] \ , \qquad \th_{\s,k} \ = \ \{0,1\} \ . \label{lambdak}\eea
Requiring that $\S_\a=z_\a$ for $\nu=0$ and $\th_{\s,k}=0$, that is, $f^\pm_\s (s)|_{\nu=0=\theta_{\s,k}}=\d(s-1)=I^{(+)}_0(s)+I^{(-)}_0(s)$, which implies that 
\be \varepsilon^{(\pm)}_\s~=~1\ ,\qquad \mu^{(\pi)}_{\s,0}~=~1\ ,\qquad \widetilde q^{(+)}_\s~ =~\widetilde m^{(+)}_\s-1~=~\sum_{k=1}^{\infty}\m^{(\pi)}_{\s,k} \xi^k~=~\sqrt{1-\ft{\s{\cal B}\nu}2\,\xi}-1\ .\ee
The result is the confluent hypergeometric function
\bea q^{(+)}_\s(s) \ = \ \sum_{k=1}^\infty {\ft12 \choose k}\left(-\frac{\s{\cal B}\nu}{2}\right)^{k}\frac{\left(\log\ft1{s^2}\right)^{k-1}}{(k-1)!} \ = \ -\frac{\s{\cal B}\nu}{4}\,{}_1F_1\left[\frac{1}{2};2;\frac{\s{\cal B}\nu}{2}\log\frac{1}{s^2}\right] \ .\label{1F1}\eea
In order to determine the projector-dependent part of \eq{mexp}, all one has to do at this point is to compute the expansion coefficients $\l^{(\pi)}_{\s,k}$ from \eq{lambdak}, \emph{i.e.} to calculate 
\be L_k[m_\s]~=~L_k\left[\d(s-1)+q^{(+)}_\s(s)\right]~=~1+L_k[q^{(+)}_\s]\ ,\qquad L_k[q^{(+)}_\s]~=~ -{1+(-1)^k\over
2}\left(1-\sqrt{1-{\s{\cal B}\nu\over 1+k}}\right)\ ,\label{Lkq}\ee
which shows that $\l^{(\pi)}_{\s,k}$ are $\nu$-dependent only for even $k$. The holomorphic solutions can thus be given in normal order as follows:
\be \S^\pm_\s~=~ z^\pm -2i V^\pm_\s \ , \qquad V^\pm_\s ~=~  V^{\pm(part)}_\s+ V^{\pm(proj)}_\s  \ ,\label{Sigma}\ee\be
 V^{\pm(part)}_\s ~=~\frac{i}2  \int_{-1}^1 ds~ q^\pm_\s(s)\,:z^\pm\,
 e^{\ft{i\s}2(s-1)z^+ z^-}:_\s\ ,\label{part}\ee\be  V^{\pm(proj)}_\s ~=~
 -i \sum_{k=0}^\infty \theta_{\s,k} L_k[m^\pm_\s] :z^\pm P_{\s,k}(z^+ z^-):_\s\ ,\label{hom}\ee
where
\bea P_{\s,k}(z^+ z^-)&=&\int_{-1}^1 ds ~:e^{\ft{i\s}2(s-1)z^+ z^-}:_\s p_k(s)\ =\ {(i\s/2)^k\over
k!}:(z^+ z^-)^k e^{-\ft{i\s}2z^+ z^-}:_\s\ ,\label{Pk}\eea
are projectors in the $\star$-product algebra \eq{normstar}, \emph{viz.}
\bea P_{\s,k}\star P_{\s,l}\ =\
\delta_{kl}P_{\s,k}\ ,\eea
and the symmetric gauge is reached by taking
\be m^\pm_\s(s)~:=~\delta(s-1)+q^\pm_\s(s)~=~\delta(s-1)+q_\s(s)\ ,\ee
with $q_\s$ given by \eq{1F1}.
The anti-holomorphic solution $\bar\S^\pm_{\s}=(\S^\pm_{\s})^\dagger$ implying that $\bar\th_{\s,k}=\th_{\s,k}$. 
In the symmetric case, the projector part \eq{hom} of the internal connection can be
written as
\bea
A^{\pm (proj)}_\s&=&-i:z^\pm\sum_{k=0}^\infty\left[\theta_{\s,k}
P_{\s,k}-\left(1-\sqrt{1-{\s{\cal B}\nu\over
1+2k}}\right)\theta_{\s,2k}P_{\s,2k}\right]: \ .\eea
Independently of the values of $\theta_{\s,k}$, the branch-cut can be chosen such that the internal connection is analytic for ${\rm Re}(\s{\cal B}\nu)<1$, where also the particular solution can be shown to be real analytic \cite{Sezgin:2005pv}.
In particular, at $\nu=0$ one has the undeformed oscillators
\be \S^{\pm(proj)}_\s~=~:z^\pm(1-2\sum_k \theta_{\s,k} P_{\s,k}):_\s\ ,\qquad \left[\S^{-(proj)}_\s,\S^{-(proj)}_\s\right]_\star~=~-2i\ ,\ee
as can be seen by defining $P_\s:=\sum_k \theta_{\s,k} P_{\s,k}$ and using $(1-2P_\s)^{\star 2}=1$ and $[z^+\star z^-,P_\s]_\star=0$.

\end{appendix}



\begin{thebibliography}{99}




\bibitem{Didenko:2009td}
  V.~E.~Didenko and M.~A.~Vasiliev,
  ``Static BPS black hole in 4d higher-spin gauge theory,''
  Phys.\ Lett.\  B {\bf 682} (2009) 305
  [arXiv:0906.3898 [hep-th]].

\bibitem{vasiliev} M.A. Vasiliev,
``Consistent equations for
interacting gauge fields of all spins in $3+1$ dimensions,'' Phys.
Lett. {\bf B243} (1990) 378.

\bibitem{Vasiliev:2003ev}
  M.~A.~Vasiliev,
  ``Nonlinear equations for symmetric massless higher spin fields in
  (A)dS(d),''
  Phys.\ Lett.\  B {\bf 567} (2003) 139
  [arXiv:hep-th/0304049].

\bibitem{Vasiliev:1999ba}
  M.~A.~Vasiliev,
  ``Higher spin gauge theories: Star-product and AdS space,''
  arXiv:hep-th/9910096.

\bibitem{Bekaert:2005vh}
  X.~Bekaert, S.~Cnockaert, C.~Iazeolla and M.~A.~Vasiliev,
  ``Nonlinear higher spin theories in various dimensions,''
  arXiv:hep-th/0503128.

\bibitem{Iazeolla:2008bp}
  C.~Iazeolla,
  ``On the Algebraic Structure of Higher-Spin Field Equations and New Exact
  Solutions,''
  arXiv:0807.0406 [hep-th].

\bibitem{Vasiliev:1995dn}
  M.~A.~Vasiliev,
  ``Higher spin gauge theories in four-dimensions, three-dimensions, and
  two-dimensions,''
  Int.\ J.\ Mod.\ Phys.\  D {\bf 5} (1996) 763
  [arXiv:hep-th/9611024].

\bibitem{Boulanger:2011dd}
  N.~Boulanger and P.~Sundell,
  ``An action principle for Vasiliev's four-dimensional higher-spin gravity,''
  arXiv:1102.2219 [hep-th].
  
\bibitem{Gubser:2002tv}
  S.~S.~Gubser, I.~R.~Klebanov and A.~M.~Polyakov,
  ``A Semiclassical limit of the gauge / string correspondence,''
  Nucl.\ Phys.\  B {\bf 636}, 99 (2002)
  [arXiv:hep-th/0204051].
  
\bibitem{Kruczenski:2004wg}
  M.~Kruczenski,
  ``Spiky strings and single trace operators in gauge theories,''
  JHEP {\bf 0508}, 014 (2005)
  [arXiv:hep-th/0410226].

\bibitem{Kristiansson:2003xx}
  F.~Kristiansson and P.~Rajan,
  ``Scalar field corrections to AdS(4) gravity from higher spin gauge
  theory,''
  JHEP {\bf 0304} (2003) 009
  [arXiv:hep-th/0303202].

\bibitem{Boulanger:2008tg}
  N.~Boulanger, S.~Leclercq and P.~Sundell,
  ``On The Uniqueness of Minimal Coupling in Higher-Spin Gauge Theory,''
  JHEP {\bf 0808} (2008) 056
  [arXiv:0805.2764 [hep-th]].

\bibitem{Bekaert:2010hw}
  X.~Bekaert, N.~Boulanger and P.~Sundell,
  ``How higher-spin gravity surpasses the spin two barrier: no-go theorems
  versus yes-go examples,''
  arXiv:1007.0435 [hep-th].

\bibitem{Vasiliev:1988xc}
  M.~A.~Vasiliev,
  ``Equations of motion of interacting massless fields of all spins as a free differential algebra,''
  Phys.\ Lett.\  B {\bf 209} (1988) 491.

\bibitem{Vasiliev:1988sa}
  M.~A.~Vasiliev,
  ``Consistent equations for interacting massless fields of all spins in the first order in curvatures,''
  Annals Phys.\  {\bf 190}, 59 (1989).
  
\bibitem{Vasiliev:1990vu}
  M.~A.~Vasiliev,
  ``Properties of equations of motion of interacting gauge fields of all spins
  in (3+1)-dimensions,''
  Class.\ Quant.\ Grav.\  {\bf 8} (1991) 1387.
  
\bibitem{more}
  M.~A.~Vasiliev,
  ``More on equations of motion for interacting massless fields of all spins in
  (3+1)-dimensions,''
  Phys.\ Lett.\  B {\bf 285} (1992) 225.

\bibitem{Engquist:2005yt}
  J.~Engquist and P.~Sundell,
  ``Brane partons and singleton strings,''
  Nucl.\ Phys.\  B {\bf 752} (2006) 206
  [arXiv:hep-th/0508124].

  
\bibitem{Flato:1978qz}
  M.~Flato and C.~Fronsdal,
  ``One Massless Particle Equals Two Dirac Singletons: Elementary Particles In
  A Curved Space. 6,''
  Lett.\ Math.\ Phys.\  {\bf 2} (1978) 421.

\bibitem{Sezgin:2002rt}
  E.~Sezgin and P.~Sundell,
  ``Massless higher spins and holography,''
  Nucl.\ Phys.\  B {\bf 644} (2002) 303
  [Erratum-ibid.\  B {\bf 660} (2003) 403]
  [arXiv:hep-th/0205131].

\bibitem{Sezgin:2003pt}
  E.~Sezgin and P.~Sundell,
  ``Holography in 4D (super) higher spin theories and a test via cubic scalar
  couplings,''
  JHEP {\bf 0507} (2005) 044
  [arXiv:hep-th/0305040].

\bibitem{Giombi:2009wh}
  S.~Giombi and X.~Yin,
  ``Higher Spin Gauge Theory and Holography: The Three-Point Functions,''
  JHEP {\bf 1009} (2010) 115
  [arXiv:0912.3462 [hep-th]].

\bibitem{Giombi:2010vg}
  S.~Giombi and X.~Yin,
  ``Higher Spins in AdS and Twistorial Holography,''
  JHEP {\bf 1104} (2011) 086
  [arXiv:1004.3736 [hep-th]].

\bibitem{Klebanov:2002ja}
  I.~R.~Klebanov and A.~M.~Polyakov,
  ``AdS dual of the critical O(N) vector model,''
  Phys.\ Lett.\  B {\bf 550} (2002) 213
  [arXiv:hep-th/0210114].

\bibitem{Douglas:2010rc}
  M.~R.~Douglas, L.~Mazzucato and S.~S.~Razamat,
  ``Holographic dual of free field theory,''
  Phys.\ Rev.\  D {\bf 83} (2011) 071701
  [arXiv:1011.4926 [hep-th]].

\bibitem{Koch:2010cy}
  R.~d.~M.~Koch, A.~Jevicki, K.~Jin and J.~P.~Rodrigues,
  ``$AdS_4/CFT_3$ Construction from Collective Fields,''
  Phys.\ Rev.\  D {\bf 83} (2011) 025006
  [arXiv:1008.0633 [hep-th]].

\bibitem{Prokushkin:1998bq}
  S.~F.~Prokushkin and M.~A.~Vasiliev,
  ``Higher-spin gauge interactions for massive matter fields in 3D AdS
  spacetime,''
  Nucl.\ Phys.\  B {\bf 545} (1999) 385 [arXiv:hep-th/9806236].
  
\bibitem{Didenko:2006zd}
  V.~E.~Didenko, A.~S.~Matveev and M.~A.~Vasiliev,
  ``BTZ black hole as solution of 3d higher spin gauge theory,''
  arXiv:hep-th/0612161.

\bibitem{Gutperle:2011kf}
  M.~Gutperle and P.~Kraus,
  ``Higher Spin Black Holes,''
  JHEP {\bf 1105} (2011) 022
  [arXiv:1103.4304 [hep-th]].

\bibitem{Ammon:2011nk}
  M.~Ammon, M.~Gutperle, P.~Kraus and E.~Perlmutter,
  ``Spacetime Geometry in Higher Spin Gravity,''
  arXiv:1106.4788 [hep-th].

\bibitem{Sezgin:2005pv}
  E.~Sezgin and P.~Sundell,
  ``An exact solution of 4D higher-spin gauge theory,''
  Nucl.\ Phys.\  B {\bf 762}, 1 (2007)
  [arXiv:hep-th/0508158].

\bibitem{Iazeolla:2007wt}
  C.~Iazeolla, E.~Sezgin and P.~Sundell,
  ``Real Forms of Complex Higher Spin Field Equations and New Exact
  Solutions,''
  Nucl.\ Phys.\  B {\bf 791} (2008) 231
  [arXiv:0706.2983 [hep-th]].

\bibitem{Mars}
  M.~Mars,
  ``A Space-time characterization of the Kerr metric,''
  Class.\ Quant.\ Grav.\  {\bf 16} (1999) 2507
  [arXiv:gr-qc/9904070].

\bibitem{Didenko:2008va}
  V.~E.~Didenko, A.~S.~Matveev and M.~A.~Vasiliev,
  ``Unfolded Description of $AdS_4$ Kerr Black Hole,''
  Phys.\ Lett.\  B {\bf 665} (2008) 284
  [arXiv:0801.2213 [gr-qc]].

\bibitem{Didenko:2009tc}
  V.~E.~Didenko, A.~S.~Matveev and M.~A.~Vasiliev,
  ``Unfolded Dynamics and Parameter Flow of Generic AdS(4) Black Hole,''
  arXiv:0901.2172 [hep-th].


\bibitem{berezin}
  F.~A.~Berezin and M.~A.~Shubin,
  ``The Schr\"{o}dinger Equation,''
  Moscow University Press, (Moscow, 1983)
  
\bibitem{Petrov:2000bs}
  A.~Z.~Petrov,
  ``The classification of spaces defining gravitational fields,''
  Gen.\ Rel.\ Grav.\  {\bf 32} (2000) 1661.


\bibitem{PenroseRindler}
  R.~Penrose and W.~Rindler,
  ``Spinors And Space-Time. 1. Two Spinor Calculus And Relativistic Fields,''
{\it  Cambridge, Uk: Univ. Pr. ( 1984) 458 P. ( Cambridge Monographs On Mathematical Physics)} ,
  ``Spinors And Space-Time. 2.: Spinor And Twistor Methods In Space-Time Geometry,''
{\it  Cambridge, Uk: Univ. Pr. ( 1986) 501p}


\bibitem{Sezgin:2002ru}
  E.~Sezgin and P.~Sundell,
  ``Analysis of higher spin field equations in four dimensions,''
  JHEP {\bf 0207} (2002) 055 [arXiv:hep-th/0205132].

\bibitem{Vasiliev:1990bu}
  M.~A.~Vasiliev,
  ``Algebraic aspects of the higher spin problem,''
  Phys.\ Lett.\  B {\bf 257} (1991) 111.

\bibitem{Bolotin:1999fa}
K.~I.~Bolotin and M.~A.~Vasiliev, ``Star-product and massless free
field dynamics in AdS(4)'', Phys.\ Lett.\ B {\bf 479} (2000) 421
[arXiv:hep-th/0001031].

\bibitem{Iazeolla:2008ix}
  C.~Iazeolla and P.~Sundell,
  ``A Fiber Approach to Harmonic Analysis of Unfolded Higher-Spin Field
  Equations,''
  JHEP {\bf 0810} (2008) 022
  [arXiv:0806.1942 [hep-th]].
  

\bibitem{Colombo:2010fu}
  N.~Colombo and P.~Sundell,
  ``Twistor space observables and quasi-amplitudes in 4D higher spin gravity,''
  arXiv:1012.0813 [hep-th].

\bibitem{Sezgin:2011hq}
  E.~Sezgin and P.~Sundell,
  ``Geometry and Observables in Vasiliev's Higher Spin Gravity,''
  arXiv:1103.2360 [hep-th].

\bibitem{wald}
  R.~M.~Wald,
  ``General Relativity,''
{\it  Chicago, Usa: Univ. Pr. ( 1984) 491p}

\bibitem{Melvin}
  M.~A.~Melvin,
  ``Pure magnetic and electric geons,''
  Phys.\ Lett.\  {\bf 8} (1964) 65.

\bibitem{Gelfond:2008td}
  O.~A.~Gelfond and M.~A.~Vasiliev,
  ``Sp(8) invariant higher spin theory, twistors and geometric BRST formulation
  of unfolded field equations,''
  JHEP {\bf 0912} (2009) 021
  [arXiv:0901.2176 [hep-th]].










\end{thebibliography}
\end{document}